\documentclass[pra,twocolumn,superscriptaddress]{revtex4}
\usepackage{hyperref}
\usepackage{amssymb, amsmath}
\usepackage{graphicx}
\usepackage{color}
\usepackage{ctable}
\usepackage{empheq}
\usepackage{lipsum} % for dummy text
\usepackage{enumitem}
\setlist{nosep}

\newcommand{\ip}[2]{\left\langle {#1} | {#2} \right\rangle}

\newcommand{\beq}{\begin{equation}}
\newcommand{\eeq}{\end{equation}}
\newcommand{\bea}{\begin{eqnarray}}
\newcommand{\eea}{\end{eqnarray}}

\newcommand{\comment}[1]{}

\renewcommand{\d}{{\rm d}}
\newcommand{\sig}{{\bf s}}
\newcommand{\idl}{{\bf i}}

\begin{document}

\title{Reduced models and design principles for half-harmonic generation in synchronously-pumped optical parametric oscillators}

\author{Ryan Hamerly}\email{rhamerly@stanford.edu}
\affiliation{Edward L.\ Ginzton Laboratory, Stanford University, Stanford, CA 94305}
\author{Alireza Marandi}
\affiliation{Edward L.\ Ginzton Laboratory, Stanford University, Stanford, CA 94305}
\author{Marc Jankowski}
\author{M.~M.~Fejer}
\affiliation{Edward L.\ Ginzton Laboratory, Stanford University, Stanford, CA 94305}
\author{Yoshihisa Yamamoto}
\affiliation{Edward L.\ Ginzton Laboratory, Stanford University, Stanford, CA 94305}
\affiliation{ImPACT Program, Japan Science and Technology Agency, 7 Gobancho, Chiyoda-ku, Tokyo 102-0076, Japan}
\author{Hideo Mabuchi}
\affiliation{Edward L.\ Ginzton Laboratory, Stanford University, Stanford, CA 94305}

\date{\today}
	
\begin{abstract}
	We develop reduced models that describe half-harmonic generation in a synchronously-pumped optical parametric oscillator above threshold, where nonlinearity, dispersion, and group-velocity mismatch are all relevant.  These models are based on (1) an eigenmode expansion for low pump powers, (2) a simulton-like sech-pulse ansatz for intermediate powers, and (3) dispersionless box-shaped pulses for high powers.  Analytic formulas for pulse compression, degenerate vs.\ nondegenerate operation, and stability are derived and compared to numerical and experimental results.
\end{abstract}

\maketitle

The optical parametric oscillator (OPO) is an indispensable tool in nonlinear optics.  As a lightsource, it benefits from the broadband $\chi^{(2)}$ nonlinearity, allowing it to produce light at near- and mid-IR frequencies \cite{Marandi2016}, an essential resource for molecular spectroscopy \cite{Parker2012}, high-harmonic generation \cite{Popmintchev2012} and dielectric laser accelerators \cite{Peralta2013}.  From an optical logic standpoint, since the $\chi^{(2)}$ effect is much stronger than the $\chi^{(3)}$ effect, nonlinearity (and thus computation) can be achieved with much lower powers.  Recently, networks of OPOs have been proposed as tools for combinatorial optimization \cite{Wang2013, Marandi2014} and machine learning \cite{Tezak2015}.  Integrated $\chi^{(2)}$ photonics is rapidly maturing and recent success with LiNbO$_3$ waveguides \cite{Jackel1991, Korkishko1998, Iwai2003, Roussev2004, Chang2016} and microstructures \cite{Poberaj2012, Rabiei2014, Guarino2007, Lin2015} in particular suggest that large-scale, integrated OPO systems are feasible in the near future.

Since optical nonlinearities are most pronounced at strong field intensities, and field intensity is enhanced in pulsed mode, there has been a growing interest in the synchronously-pumped OPO (SPOPO), in which the pump is a train of ultrashort pulses synchronized to the round-trip time of the cavity \cite{VanDriel1995}.  Highly nonlinear effects can take place at modest average powers.  SPOPOs are used for numerous applications including pulse compression \cite{Khaydarov1994, Marandi2015}, frequency-domain entanglement generation \cite{Roslund2014}, cluster-state preparation \cite{Yokoyama2013} and coherent computing \cite{Marandi2014, KentaThesis}.  On the other hand, SPOPOs have far more degrees of freedom than their continuous-wave counterparts, so modeling them and predicting their behavior is a challenge.

This paper discusses computationally efficient schemes for modeling degenerate SPOPOs.  Pulse dynamics in a SPOPO is a competition between three effects: $\chi^{(2)}$ nonlinearity, dispersion, and group-velocity mismatch (temporal walkoff).  Section \ref{sec:10-intro} introduces the physical system and its equations of motion.  These equations can be solved numerically using a split-step Fourier method (which can easily be scaled to multicore / GPU architectures for performance), giving rise to a discrete round-trip Ikeda-like map for the pulse amplitude \cite{Ikeda1979}.  While this numerical model is accurate and agrees with experiments, it is computationally costly to run, particularly for guided-wave systems with large temporal walkoff.

Sections \ref{sec:10-linear}-\ref{sec:10-boxpulse} derive approximate, physically-motivated {\it reduced models} for the SPOPO system.  These models reduce the OPO simulation time by several orders of magnitude, but within their respective regimes of operation, give steady-state pulse shapes and dynamical behavior that match the full numerical model.  The resulting computational speedup is particularly useful for large simulations of many OPOs in parallel -- for example, large-scale Ising or XY machines based on time-multiplexed OPO networks \cite{Takata2016, Inagaki2016, Hamerly2016-2}.  Moreover, these models facilitate device optimization and robustness studies, by allowing the designer to simulate a SPOPO with a wide range of test parameters.  Finally, these models shed analytic and physical insight into the dynamics of SPOPOs.

In Section \ref{sec:10-linear}, we derive a linearized model based on an eigenmode expansion.  The eigenmodes and their eigenvalues are computed, and related to analytic formulae that reveal a power-law scaling in the steady-state signal pulse width as a function of pump pulse width, dispersion and single-pass gain.  Section \ref{sec:10-nonlinear} extends this model by treating pump depletion to first order in perturbation theory, leading to equations with cubic terms that resemble the Langevin equations for continuous-wave OPOs \cite{Kinsler1991}.  This model accurately predicts the oscillation threshold, power efficiency, signal pulse shape, and stability for the SPOPO near threshold.

An ansatz based on the simulton solution in a $\chi^{(2)}$ waveguide \cite{Akhmanov1968, Trillo1996} is presented in Section \ref{sec:10-simulton}.  By postulating a sech-shaped signal pulse, effects of the pump shape, dispersion, and nonlinearity all map onto a set of ODE's for the amplitude, centroid and width of the sech pulse.  This ansatz restricts the range of validity compared to Sec.~\ref{sec:10-nonlinear} (although it can also be valid well above threshold, where the eigenmode treatment fails \cite{Jankowski2016}), but it is physically more intuitive and sheds more light into the pulse dynamics.

In the opposite regime well above threshold, Section \ref{sec:10-boxpulse} obtains an analytic form by ignoring dispersion.  The result is a box-shaped pulse whose width is a function of the pump amplitude and whose spectrum approximates a sinc-function.  We note that this section is a generalization of \cite{Becker1974} to the case of nonzero walkoff.

While the results of this paper are general and apply to any degenerate SPOPO with dispersion and temporal walkoff, for concreteness we consider a guided-wave PPLN OPO with a fiber cavity, implemented in \cite{Marandi2015, McMahon2016}, as an example system.

\section{The Synchronously Pumped OPO}
\label{sec:10-intro}

\begin{figure}[tbp]
\begin{center}
\includegraphics[width=0.80\columnwidth]{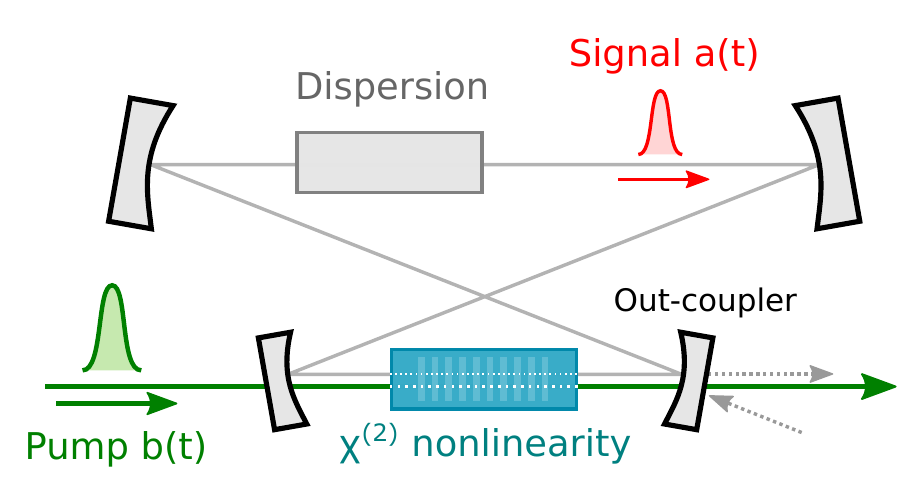}
\caption{Typical synchronously pumped OPO design.  (A PPLN waveguide OPA with optical-fiber feedback loop is considered in the text.)}
\label{fig:10-f1}
\end{center}
\end{figure}

Figure \ref{fig:10-f1} sketches the design.  The degenerate, synchronously pumped OPO consists of a cavity with a nonlinear $\chi^{(2)}$ medium, an output coupler, and a lumped dispersion element (for all dispersion excluding the $\chi^{(2)}$ medium).  In isolation, the $\chi^{(2)}$ medium is an amplifier, and the feedback loop created by the cavity turns it into an oscillator.  As a concrete example, in the fiber-coupled OPO in \cite{Marandi2015}, the $\chi^{(2)}$ medium is a PPLN waveguide and the dispersive element is the optical fiber.

\subsection{Equations of Motion}

Propagation through the OPO is a two-step process: (1) nonlinear $\chi^{(2)}$ medium and (2) linear dispersion element.  The waveguide dynamics are governed by a pair of PDE's.  To derive these equations, first write the electric field in terms of slowly-varying amplitudes \cite{BoydBook, AgrawalBook}
\bea
    \vec{E}(z, t) & = & \mathcal{E}_a \vec{E}_{T,a}(x, y) e^{i(\bar{\beta}_a z - \bar{\omega} t)} a(z, t) \nonumber \\
    & & - i\mathcal{E}_b\vec{E}_{T,b}(x, y) e^{i(\bar{\beta}_b z - 2\bar{\omega} t)} b(z, t) + \mbox{c.c.} \label{eq:10-svea}
\eea
where $a(z,t)$ and $b(z,t)$ are the envelope functions for the pump and signal.  Here $z$ is the propagation direction and $\vec{E}_{T,a}$, $\vec{E}_{T,b}$ are normalized transverse mode profiles.  The constants $\mathcal{E}_{a,b} = \sqrt{\hbar\omega_{a,b}/2n(\omega_{a,b}) \epsilon_0 c}$ are chosen so that $\int{|a|^2 \d t}$, $\int{|b|^2 \d t}$ correspond to the pump and signal photon number.  Applying Maxwell's equations to (\ref{eq:10-svea}) and adding dispersion and a $\chi^{(2)}$ nonlinearity, the envelope functions evolve as follows:\bea
    \!\!\frac{\partial a}{\partial z} & \!=\! & \left[-\frac{\alpha_a}{2} - \frac{i\beta_2^{(a)}}{2!} \frac{\partial^2}{\partial t^2} + \frac{\beta_3^{(a)}}{3!} \frac{\partial^3}{\partial t^3} + \ldots\right] a + \epsilon\,a^* b \label{eq:10-at} \\
    \!\!\frac{\partial b}{\partial z} & \!=\! & \left[-\frac{\alpha_b}{2} - u \frac{\partial}{\partial t} - \frac{i\beta_2^{(b)}}{2!} \frac{\partial^2}{\partial t^2} + \frac{\beta_3^{(b)}}{3!} \frac{\partial^3}{\partial t^3} + \ldots\right] b - \frac{1}{2}\epsilon\,a^2 \nonumber \\ \label{eq:10-bt}
\eea
where $\alpha_{a,b}$ are the waveguide power loss coefficients, $u = (\beta_1^{(b)} - \beta_1^{(a)}) = (v_a - v_b)/v_a v_b$ is the walkoff (group-velocity mismatch), $\beta_2^{(a, b)}$ and $\beta_3^{(a, b)}$ are the dispersion coefficients, and $\epsilon = \bigl(2\omega \mathcal{E}_b d_{\rm eff}/n(\omega) c\bigr) \int {E_{T,a}^2 E_{T,b} \d x\,\d y}$ is the nonlinear coupling term.  Equations (\ref{eq:10-at}-\ref{eq:10-bt}) reveal that the dynamics is a competition between three effects:

\begin{enumerate}
	\item Nonlinearity: second-harmonic generation and, when pulses overlap in time, parametric gain
	\item Dispersion: short pulses are spread out and chirped
	\item Walkoff (group velocity mismatch): pump and signal move with respect to each other, limiting the duration of their overlap
\end{enumerate}

Previous studies of this problem have either ignored the walkoff or treated it as a perturbation \cite{Becker1974, Cheung1990}, or have focused on the high-finesse limit when the single-pass PPLN gain is small \cite{Patera2010, DeValcarcel2006, Roslund2014}.  Equations (\ref{eq:10-at}-\ref{eq:10-bt}) generalize these results to the high-gain, large-walkoff case that is more commonplace when long $\chi^{(2)}$ crystals and/or ultrashort pulses are used \cite{Marandi2014, Marandi2016}.

Similar equations can be derived from a quantum model for the $\chi^{(2)}$ system \cite{Raymer1991, Werner1997}.  The procedure is similar to that used for optical fibers \cite{Drummond2001}, but in the resulting equations, the roles of $z$ and $t$ are swapped.  These quantum equations are equivalent to (\ref{eq:10-at}-\ref{eq:10-bt}) under reasonable assumptions.

For very short or high-power pulses, (\ref{eq:10-at}-\ref{eq:10-bt}) become inaccurate and higher-order effects such as $\chi^{(3)}$ and Raman scattering must be included.  Moreover, pulses spanning more than one octave merit special treatment as the slowly-varying envelope approximation breaks down \cite{Phillips2011, PhillipsThesis}; these are beyond the scope of this work.

To solve Eqs.~(\ref{eq:10-at}-\ref{eq:10-bt}), we employ the split-step Fourier method \cite{AgrawalBook}.  First, a sampling window $[0, T]$ is defined, with $T$ is large enough that all of the dynamics happens inside the window.  One can express the field in terms of a Fourier series $a(z, t) = T^{-1/2} \sum_m a_m(z) e^{-im\Omega t}$ (and likewise for $b$), where $\Omega = 2\pi/T$ and $m$ is the Fourier index.  The dispersive terms in (\ref{eq:10-at}-\ref{eq:10-bt}) are propagated in the frequency domain, while the nonlinear terms are propagated in the time domain.  Since most of the computation time is spent performing FFT's to go between time and frequency domains, we implemented the solver in CUDA \cite{CUDAGuide} because of the substantial FFT speedup afforded by modern GPUs \cite{Moreland2003, Sreehari2012}.

The second step, propagation through the dispersive element, is trivial because it is linear.  Since only the signal resonates in the setup (Fig.~\ref{fig:10-f1}), each Fourier component acquires a constant loss and phase shift $a_m \rightarrow G_0^{-1/2} e^{i\phi_m} a_m$, with $\phi_m = \phi_0 + \tfrac{\ell\lambda_a}{2c} \Omega m + \tfrac{\phi_2}{2!} (\Omega m)^2 + \ldots$, where $\phi_m$ is the signal phase measured relative to a degenerate signal whose round trip time is synchronized to the pump repetition rate.

\ctable[caption=Parameters for PPLN waveguide OPO \cite{Marandi2015, McMahon2016} used as example in this paper,
        label=tab:10-t1,
        pos=t]{ccc}{
\tnote[a]{0.3 dB/cm}
\tnote[b]{LiNbO$_3$, extraordinary polarization}
\tnote[c]{$T_p = L u$, matched to crystal walkoff length}
\tnote[d]{$\epsilon = \sqrt{2\hbar\omega\,\eta}$, where $\eta = 1.0$ W$^{-1}$cm$^{-2}$ is the normalized conversion efficiency \cite{Parameswaran2002, Langrock2007}}
\tnote[e]{$(1 - G_0^{-1})$ is total cavity loss, absorption plus out-coupling.  Here we take 5-dB loss per round trip.}
\tnote[f]{$N_{b,0} = \left[(\alpha_b/4\epsilon)(e^{\alpha_b L/2}-1)^{-1} \log(G_0 e^{\alpha_a L})\right]^2$}
\tnote[g]{$N_{b,0} = T_p b_0^2$, threshold for CW operation}}{
\hline\hline
Term & Meaning & Value \\ \hline
$\lambda_a$, $\lambda_b$  & Signal, Pump $\lambda$ & 1.5 $\mu$m, 0.75 $\mu$m \\
$L$ & Waveguide Length & 40 mm \\
$\alpha_a$, $\alpha_b$ & Waveguide Loss & 0.00691 mm$^{-1}$ \tmark[a] \\
$u$ & Walkoff & $0.329$ ps/mm\tmark[b] \\
$T_p$ & Pump Length & 13.2 ps\tmark[c] \\
$\beta_2^{(a)}$ & Signal GVD & $1.12 \times 10^{-4}$ ps$^2$/mm \\
$\beta_3^{(a)}$ & Signal TOD & $3.09 \times 10^{-5}$ ps$^3$/mm \\
$\beta_2^{(b)}$ & Pump GVD & $4.06 \times 10^{-4}$ ps$^2$/mm \\
$\beta_3^{(b)}$ & Pump TOD & $2.51 \times 10^{-5}$ ps$^3$/mm \\
$\epsilon$ & Nonlinearity & $5.16 \times 10^{-5}$ ps$^{1/2}$/mm\tmark[d] \\
$G_0$ & Power gain at threshold & 3.33\tmark[e] \\
$N_{b,0}$ & Threshold Photons & $1.94 \times 10^6$ \tmark[f] \\
$b_0$ & Threshold Amplitude & $3.84 \times 10^2$ ps$^{-1/2}$ \tmark[g] \\
\hline\hline
}

The out-coupling loss $(1-G_0^{-1})$ is the same for all modes, while the dispersion and walkoff terms give different modes different phases.  Here $\ell$ is the cavity length detuning (in units of vacuum half-wavelengths) from matching the cavity roundtrip time to the pump repetition period; $\phi_0$ and $\ell$ are not independent: $\phi_0 = \pi\ell + \mbox{const}$.  The constant reflects the fact that zero round-trip-time detuning (hereafter referred to simply as ``detuning'') may not correspond to round-trip phase equal to an integer times $\pi$ at degeneracy (a ``resonance peak'').  For signal pulses much longer than an optical cycle, this constant can be neglected because it corresponds to a small, sub-optical-cycle mismatch between the cavity roundtrip time and the pump repetition period.%small, subwavelength repetition-rate mismatch.

\subsection{Numerical Results}

\begin{figure}[t!]
\begin{center}
\includegraphics[width=1.0\columnwidth]{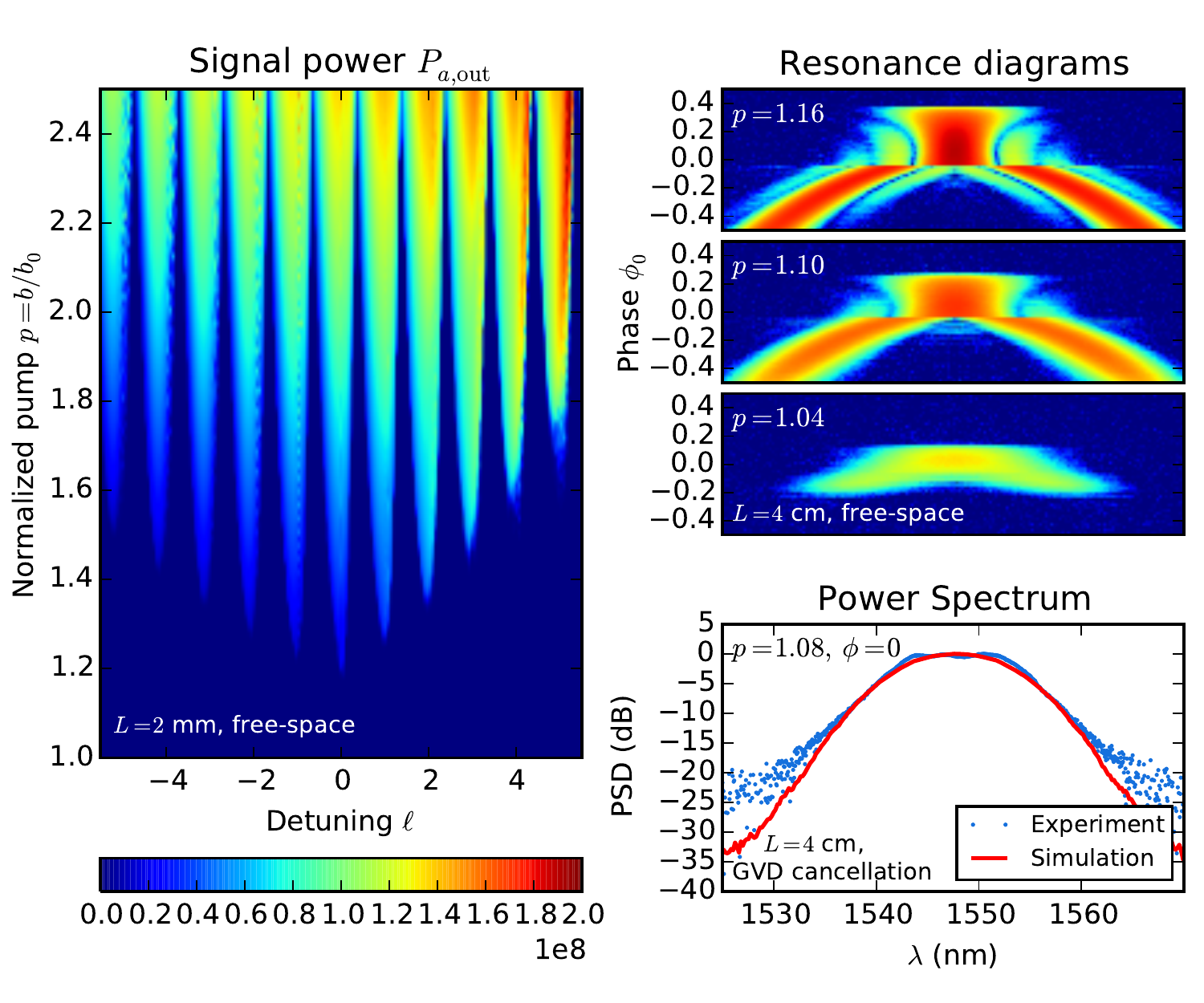}
\caption{Left: Plot of output signal power $P_{a,\rm out}$ (in photons per round-trip) for 2-mm crystal, no dispersion compensation (``free-space'').  Right: Resonance plots of the power spectrum $P(\lambda)$ for 4-cm crystal, no dispersion compensation, with normalized pump values $p \equiv b/b_0 = 1.16, 1.10, 1.04$ (top), and spectrum for GVD-compensated cavity at $p = 1.08$ (bottom).}
\label{fig:10-f2}
\end{center}
\end{figure}

Figure \ref{fig:10-f2} shows some typical results for the simulations.  The left plot gives the steady-state OPO output power of the $P_{a,\rm out}$, in units of photons per round-trip.  This is proportional to the photon number $N_{a}$.  If the cavity round-trip loss is $O(1)$, the photon number will be different at the beginning and end of the crystal: $N_a\bigr|_{z=L} = G_0 N_a\bigr|_{z=0}$.  The output power, neglecting cavity losses other than the out-coupler and $\chi^{(2)}$ gain medium, is given by $P_{a,\rm out} = (G_0 - 1)N_a\bigr|_{z=0}$.

The figure shows a clear set of resonances called {\it detuning peaks}.  At each detuning peak, the round-trip phase $\phi_0$ is either 0 or $\pi$, since both phases can be amplified by the crystal.  There is an optimal length detuning denoted $\ell = 0$ for which the threshold is the lowest, which is understandable because a nonzero $\ell$ creates a repetition-rate mismatch between the pump and signal, increasing the required pump power.  Adding a nonzero offset to the relation $\phi_0 = \pi\ell + \mbox{const}$ shifts the detuning peaks, but not the envelope; since the envelope is much broader than any peak, this does not have a significant effect on Fig.~\ref{fig:10-f2}.  There is an asymmetry in the plot, where $\ell > 0$ peaks have higher power if the pump is strong enough; this is a result of walkoff and pump depletion that will be explained using sech-pulse theory in Section \ref{sec:10-simulton}.

Another common figure will be the ``resonance diagrams'' on the top-right plot.  These are plots of the power spectrum $P(\lambda) = |a(\lambda)|^2$ as a function of $\lambda$ and cavity round-trip phase $\phi_0$.  They show how the steady-state spectrum varies within a detuning peak.  As the pump power increases above threshold, the resonance diagrams become increasingly structured.  This structure will be explained later in Sec.~\ref{sec:10-boxpulse} in terms of box-shaped pulses that tend to form well above threshold.

Simulations are performed for many values of $\ell$ in parallel and sweeping the ``normalized pump'' $p = b/b_0$, the ratio of peak pump amplitude to the CW threshold (Table~\ref{tab:10-t1}).  The stored output is a 3-dimensional array $a(k,p,\ell)$.  A typical run with 256 parallel simulations of 20000 round-trips each takes 15 hours with an Nvidia Tesla M2070 GPU.  Integrating $|a|^2$ over $k$ gives the power plot in Fig.~\ref{fig:10-f2}.  The resonance diagrams are $p$-slices of $|a|^2$.  Each $\phi$-slice of a resonance diagram is a spectrum.  The lower-right figure shows the simulated power spectrum for a 4-cm PPLN OPO with a a fiber to compensate the PPLN crystal GVD.  Experimental data are in agreement with this result \cite{Marandi2015}.

\section{Linear Eigenmode Theory}
\label{sec:10-linear}

In actively AM-mode-locked lasers, the pulse shape is set by a competition between two forces: a resonant cavity modulation confines the pulse in time, while the finite bandwidth of the gain medium confines it in frequency \cite{Kuizenga1970, Siegman1970, Haus2000}.  These effects give rise to a linear master equation for pulse evolution, which can be solved as an eigenvalue problem, the dominant eigenmode (typically a Gaussian) becoming the lasing mode.

The same story holds for SPOPOs.  In this case, the finite pump length confines the signal in time, while dispersion in the cavity and gain medium confines it in frequency \cite{Becker1974, Khaydarov1994, Khaydarov1995}.  Patera {\it et al.}\ followed a similar procedure for the SPOPO below threshold, linearizing the equations of motion and diagonalizing them to obtain squeezing ``supermodes'' \cite{DeValcarcel2006, Patera2010}.  However, their analysis was restricted to the low-gain, high-finesse case, which is not applicable here.

This section derives an eigenmode expansion that extends the work of Patera et al.\ to the high-gain regime with walkoff, where waveguide-based SPOPOs typically operate.  We do so using a split-step procedure -- a single round trip $a(t;n)\rightarrow a(t;n+1)$ is divided up as follows:

\begin{enumerate}
    \item Continuous-wave step: Solve equations with dispersion terms, but constant pump $b(t) = b_{\rm max}$.  Result: $\tilde{a}(\delta\omega) \rightarrow \Delta(\delta\omega) \tilde{a}(\delta\omega)$ (Sec.~\ref{sec:10-cw})
    \item Dispersionless step: Solve with pulsed pump $b(t) - b_{\rm max}$ (peak value subtracted), and no dispersion terms.  Result: $a(t) \rightarrow \Gamma(t)a(t)$ (Sec.~\ref{sec:10-dispersionless})
\end{enumerate}

This is analogous to the split-step Fourier method used for the nonlinear Schr{\"o}dinger equation \cite{AgrawalBook}.  The key assumption that the pulse shape does not change much during a single step (``gain without distortion ansatz'') is equally necessary here.  Here the ``step'' corresponds to a single pass through the entire waveguide; nevertheless this assumption tends to be true unless the pump is far above threshold.

Combining the two steps, the pulse satisfies the following round-trip equation:
\beq
	a(t;n+1) = \Gamma(t)\Delta(i\tfrac{\d}{\d t}) a(t;n)
\eeq
$\Gamma\Delta$ is related to a Hermitian matrix by transformation, so it is diagonalizable and the $k^{\rm th}$ eigenmode is found by solving the corresponding eigenvalue equation:
\beq
	\Gamma(t)\Delta(i\tfrac{\d}{\d t}) a_k(t) = \lambda_k a_k(t) \label{eq:10-eig}
\eeq

We define a {\it gain-clipping function} $G(t) \equiv \log\Gamma(t)$ and a {\it dispersion loss function} $D(\delta\omega) \equiv \log(\Delta(\delta\omega)/\Delta_{\rm max})$, where $\Delta_{\rm max} = \mbox{max}_{\delta\omega}\Delta(\delta\omega)$.  Both of these functions are negative.  Near threshold, $G(t), D(\delta\omega) \ll 1$ (so that $\Gamma \approx 1+G$, $\Delta \approx \Delta_{\rm max}(1+D)$) and we can obtain a master equation analogous to \cite{Haus2000}:
\beq
	a(t;n+1) = \Delta_{\rm max}\left[1 + G(t) + D(i\tfrac{\d}{\d t})\right] a(t;n) \label{eq:10-nt-rt}
\eeq
Again, one can convert (\ref{eq:10-nt-rt}) into an eigenvalue equation to extract the eigenmodes:
\beq
	\left[g_{\rm cw} + G(t) + D(i\tfrac{\d}{\d t})\right] a_k(t) = g_k a_k(t) \label{eq:10-nt-eig}
\eeq
Here $g_{\rm cw} = \log \Delta_{\rm max}$ is the CW gain and $g_k = \log \lambda_k$ is the eigenmode gain.  Because of the negativity of $G$ and $D$, $g_k \leq g_{\rm cw}$ for all eigenmodes.

\subsection{Continuous Wave Step}
\label{sec:10-cw}

To obtain the CW round-trip gain $\Delta(\delta\omega)$, consider the case of a signal $a_\sig$ at frequency $\omega + \delta\omega$ and idler $a_\idl$ at $\omega - \delta\omega$.  From these we define $a_+ = (a_\sig + a_\idl^*)/2$, $a_- = (a_\sig - a_\idl^*)/2$ (``real'' and ``imaginary'' parts of the field) and use (\ref{eq:10-at}-\ref{eq:10-bt}), excluding pump depletion, to get:
\begin{eqnarray}
    \frac{\d a_\pm}{\d z} & = & \left(-\tfrac{1}{2}\alpha_a \pm \epsilon\,b\right) a_\pm \mp \left(\tfrac{1}{2}\beta_2\delta\omega^2\right) a_\mp \label{eq:10-cw}
\end{eqnarray}

Unless the pump loss $\alpha_b L$ is large, the pump remains relatively constant during the propagation; we can replace it by its average value $b \rightarrow \bar{b} \approx b_{\rm in} e^{-\alpha_b L/4}$.  Equation (\ref{eq:10-cw}) can then be solved by matrix exponentiation.  After exiting the gain medium, the field passes through the dispersion element and is then re-inserted.  There will be additional loss due to out-coupling, giving a transmission factor of $G_0^{-1/2}$, and possibly additional delay and phase due to the cavity detuning.  Thus, the reinserted field is related to the exiting field by: $a_\sig \rightarrow G_0^{-1/2} e^{i(\phi + \psi)}a_\sig$, $a_\idl \rightarrow G_0^{-1/2} e^{i(\phi - \psi)}a_\idl$, where $\phi \equiv \phi_0 + \tfrac{1}{2}\phi_2\delta\omega^2$ is the symmetric phase shift, and $\psi \equiv \pi\ell$ as the asymmetric phase.  The overall round-trip propagation of $a_\pm$ is:
\begin{align}
	&\begin{bmatrix} a_+ \\ a_- \end{bmatrix} \rightarrow 
		G_0^{-1/2} e^{-\alpha_a L/2} e^{i\psi} \nonumber \\
		& \quad \times
		\underbrace{\begin{bmatrix} \cos\phi & -\sin\phi \\ \sin\phi & \cos\phi \end{bmatrix}}_{R(\phi)}
		\underbrace{\exp\left(\begin{bmatrix} \epsilon\,\bar{b} & -\tfrac{1}{2}\beta_2\delta\omega^2 \\ \tfrac{1}{2}\beta_2\delta\omega^2 & -\epsilon\,\bar{b} \end{bmatrix} L\right)}_{M}
		\begin{bmatrix} a_+ \\ a_- \end{bmatrix}
\end{align}

This equation has two eigenvalues: $\lambda_\pm$.  The round-trip gain is the larger of the two.  Note that $\det R(\phi) = \det M = 1$ (since $\det M = e^{\rm tr(\log M)}$), so the product of the eigenvalues must equal $G_0^{-1} e^{2i\psi} e^{-\alpha_a L}$, whose magnitude is less than one.  Thus, at most one of the modes experiences gain.  We now assume that the frequency components of the pulse $a(t)$ live primarily in the growing eigenmode, so that we can substitute $\Delta(\delta\omega) \approx \lambda_+(\delta\omega)$.  This eigenvalue is:
\beq
	\Delta(\delta\omega) \approx \lambda_+ = \mbox{sign}(T) G_0^{-1/2} e^{-\alpha_a L/2} e^{i\psi} \left[|Z| + \sqrt{Z^2 - 1}\right] \label{eq:10-Delta}
\eeq
where $Z \equiv \tfrac{1}{2}\mbox{Tr}[R(\phi)M]$.

\begin{figure}[tbp]
\begin{center}
\includegraphics[width=1.00\columnwidth]{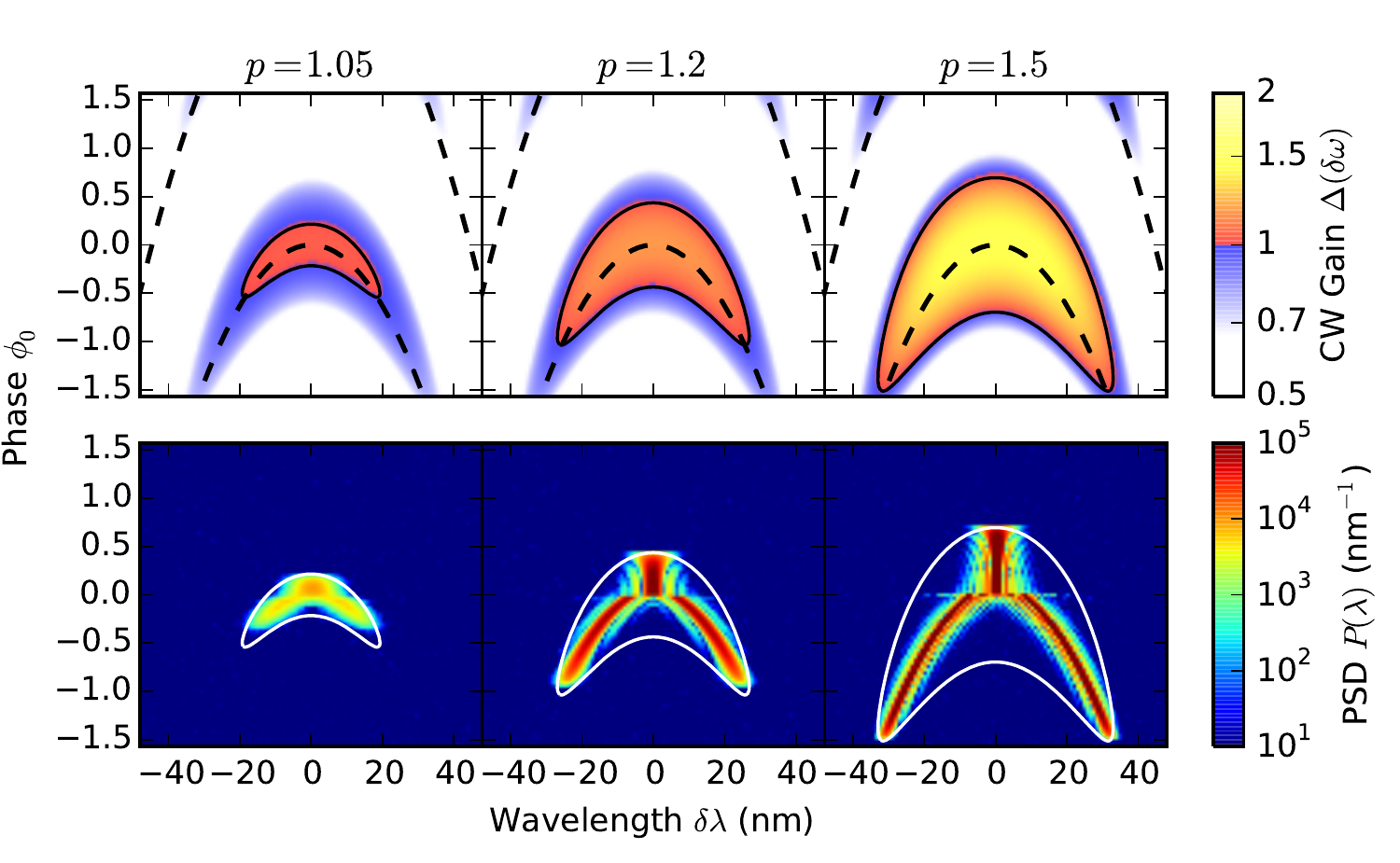}
\caption{Top: CW gain $|\Delta(\delta\omega)|$ as a function of $\delta\lambda = (-\lambda^2/2\pi c) \delta\omega$.  Bottom: plot of power spectral density $P(\lambda)$ (in photons/nm), from simulation.  White contour gives the threshold condition $|\Delta| = 1$.  Dashed line is Eq.~(\ref{eq:10-phimatch}).  PPLN OPO with $L = 4$ cm, free-space.}
\label{fig:10-f3a}
\end{center}
\end{figure}

\begin{figure}[tbp]
\begin{center}
\includegraphics[width=1.00\columnwidth]{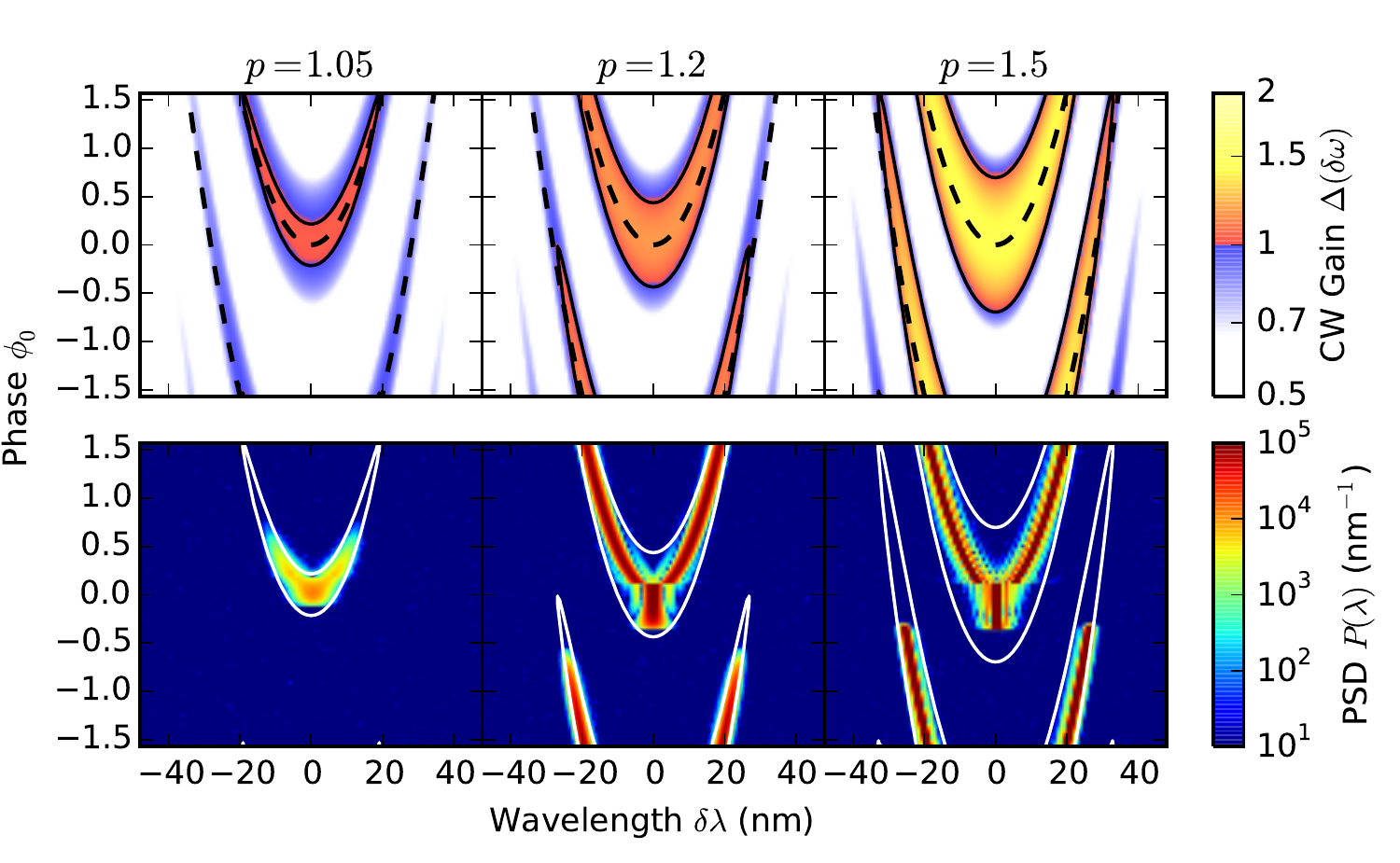}
\caption{PPLN OPO, 1-m SMF-28e fiber ($\beta_2 = -1.58 \times 10^{-26}$ s$^2$/m, $\beta_3 = 1.10 \times 10^{-40}$ s$^3$/m).  This fiber over-compensates the PPLN GVD by a factor $|\phi_2 / \beta_2 L| = 3.5$}
\label{fig:10-f3b}
\end{center}
\end{figure}

The pump can be written in terms of its normalized amplitude $p$, where $p = 1$ is the OPO threshold for a CW pump with the same peak intensity as $b(t)$.  Since the threshold depends on $\phi_0$, for specificity we take the lowest threshold, when $\phi_0 = 0$, $\delta\omega = 0$:
\beq
	\bar{b} = p\,\bar{b}_0,\ \ \ \bar{b}_{0} = \frac{1}{2L\epsilon} \log(G_0 e^{\alpha_a L})
\eeq
At $p$ times above threshold, the maximum gain is at $\phi_0 = 0$, $\delta\omega = 0$, where dispersion effects disappear:
\beq
	\mbox{max}_{\phi_0, \delta\omega} {\Delta(\delta\omega, \phi_0)} = (G_0 e^{\alpha_a L})^{p-1} \label{eq:10-maxgain}
\eeq

\begin{figure}[tbp]
\begin{center}
\includegraphics[width=1.00\columnwidth]{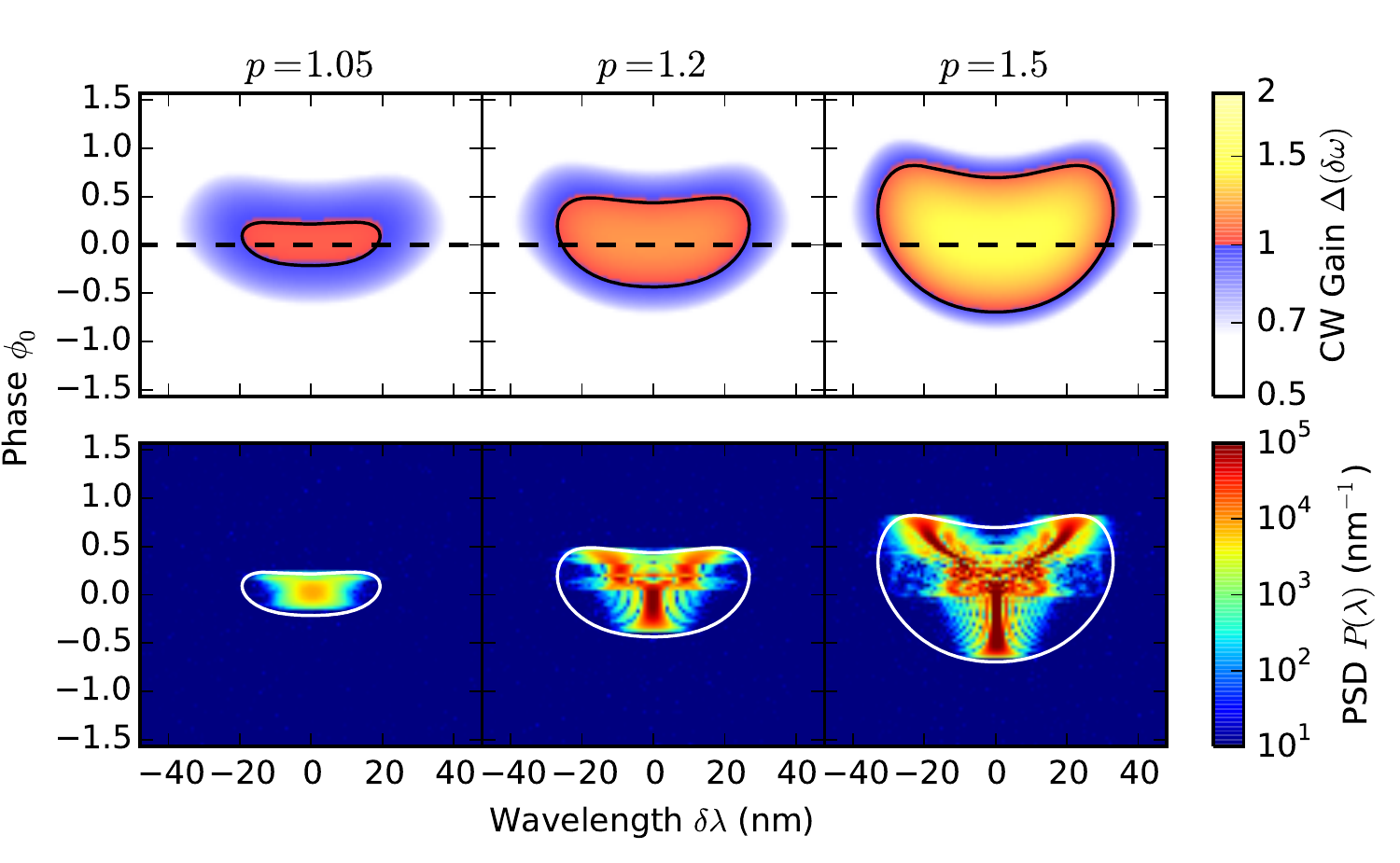}
\caption{PPLN OPO, GVD-compensating fiber ($\phi_2 = -\beta_2 L = -4.49 \times 10^{-27}$ s$^2$, $\phi_3 = 5.14 \times 10^{-41}$ s$^3$).}
\label{fig:10-f3c}
\end{center}
\end{figure}

Figures \ref{fig:10-f3a}-\ref{fig:10-f3c} compare the CW gain from Eq.~(\ref{eq:10-Delta}) to numerical spectra.  The power spectrum $P(\delta\omega)$ of the OPO signal is confined to the frequency-gain window $|\Delta(\delta\omega)| > 1$, as expected, centered on the resonance condition
\beq
	\phi_0 + \tfrac{1}{2}\underbrace{(\phi_2 + \beta_2 L)}_{\phi'_2} \delta\omega^2 = n\pi \label{eq:10-phimatch}
\eeq
which essentially says that the line-center phase shift $\phi_0$ must be compensated by the total (waveguide plus fiber) dispersion.  The shape of the spectrum depends on independent factors, which we will revisit in Sec.~\ref{sec:10-boxpulse}.

\subsubsection{Approximate Forms}

Equation (\ref{eq:10-Delta}) gives an accurate model of the CW round-trip gain, but it is cumbersome so it would be helpful to have an approximate form that is easier to work with analytically.  

Naturally, one expects the gain to be maximized when the fiber dispersion compensates the waveguide dispersion, that is: $\phi_0 + \tfrac{1}{2}\phi'_2\delta\omega^2 = n\pi$ (with $\phi'_2 = \phi_2 + \beta_2 L$).  There are two possible limits:

\begin{enumerate}
	\item $\phi_0\phi'_2 \geq 0$.  This is the degenerate limit, because no value of $\delta\omega$ can satisfy the phase relation.  We assume that $a(t)$ is real when it exits the crystal.  This is not exact (Eq.~(\ref{eq:10-cw}) assumes $a(t)$ can have arbitrary phase), but is approximately true because the amplification is phase-sensitive.
	
	Next, we treat the dispersion as a lumped element.  Thus, $a(t)$ entering the cavity has a phase $\phi = \phi_0 + \tfrac{1}{2}\phi'_2\delta\omega^2$.  Since we are only keeping track of the real part of the field as per the first assumption, this amounts to a round-trip gain of:
	\beq
		\Delta(\delta\omega) \approx \Delta_{\rm max} \cos\left(\phi_0 + \tfrac{1}{2}\phi'_2\delta\omega^2\right) \label{eq:10-Delta-app}
	\eeq
	The cosine term can be expanded, giving an approximation for $D(\delta\omega) = \log(\Delta(\delta\omega)/\Delta_{\rm max})$
	\beq
		\quad\quad\quad\ \  D(\delta\omega) \approx -\frac{\phi'_2\tan\phi_0}{2} \delta\omega^2 - \frac{(\phi'_2\sec\phi_0)^2}{8} \delta\omega^4 \label{eq:10-dw-deg}
	\eeq
	
	\item $\phi_0\phi'_2 < 0$.  This is the nondegenerate limit.  We make the same assumptions as before, but this time there exists a $\delta\omega_0 \equiv \sqrt{-2\phi_0/\phi'_2}$ that satisfies the phase relation.  At this frequency, $\Delta(\delta\omega)$ is (approximately) maximized.  Expanding the formula (\ref{eq:10-Delta-app}) about that point, we obtain:
	\beq
		D(\delta\omega) \approx -|\phi_0\phi'_2| (\delta\omega - \delta\omega_0)^2 \label{eq:10-dw-nd}
	\eeq
	Section~\ref{sec:10-emshapes} makes use of Eqs.~(\ref{eq:10-Delta-app}-\ref{eq:10-dw-nd}) to obtain an analytic form for the pulse shape.	
\end{enumerate}

\subsection{Dispersionless Step}
\label{sec:10-dispersionless}

\begin{figure}[tbp]
\begin{center}
\includegraphics[width=1.0\columnwidth]{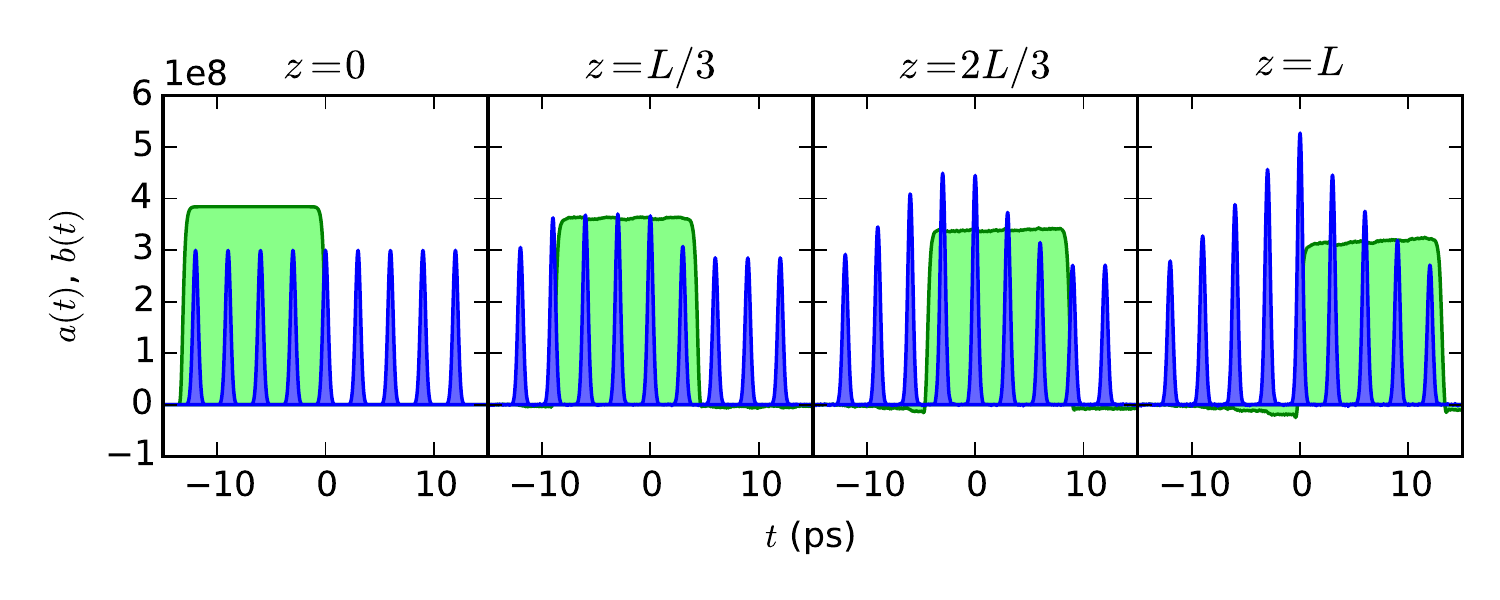}
\caption{Illustration of gain clipping.  A train of femtosecond pulses (blue) is amplified by a picosecond pump pulse (green).}
\label{fig:10-f4}
\end{center}
\end{figure}

The dispersionless step treats Eqs.~(\ref{eq:10-at}-\ref{eq:10-bt}) without the dispersion terms with the residual pump $b_{\rm in}(t) - b_{\rm max}$ (since $b(t) = b_{\rm max}$ was used in the continuous-wave pump, and we need to avoid double-counting the gain).  Since this section is about linear effects, we ignore pump depletion (but see Sec.~\ref{sec:10-nonlinear}), so the pump integrates to $(b_{\rm in}(t - uz) - b_{\rm max})e^{-\alpha_b z/2}$ ($u = v_b^{-1}-v_a^{-1}$ is the temporal walkoff) and Eq.~(\ref{eq:10-at}) becomes:
\beq
    \frac{\partial a(z,t)}{\partial z} =  - \frac{1}{2}\alpha_a a(z,t) + \epsilon\,a(z,t)^* (b_{\rm in}(t - uz) - b_{\rm max}) e^{-\alpha_b z/2} \label{eq:10-da-gc}
\eeq

We assume that $a(z,t)$ is close to real, because the imaginary component experiences loss when propagating through the waveguide.  This is only approximate when there is dispersion ($\beta_2 \neq 0, \phi_2 \neq 0$) or detuning ($\phi_0 \neq 0$).  Integrating (\ref{eq:10-da-gc}) we obtain the input-output map:
\beq
	a(t) \rightarrow \underbrace{\exp\Bigl(\int_0^L{\!\epsilon(b_{\rm in}(t - uz) \!-\! b_{\rm max}) e^{-\alpha_b z/2} \d z}\Bigr)}_{\Gamma(t)} a(t) \label{eq:10-gammat}
\eeq
The {\it gain-clipping function}, defined after Eq.~(\ref{eq:10-eig}) as $G(t) = \log \Gamma(t)$, is:
\beq
	G(t) = \int_0^L{\epsilon(b_{\rm in}(t - uz) - b_{\rm max}) e^{-\alpha_b z/2} \d z} \label{eq:10-gcf}
\eeq
This function is always negative, so the dispersionless step always gives rise to loss. We call this effect ``gain-clipping'' because it results in a temporal localization of gain, and confines the pulse in time.

The concept is illustrated in Figure \ref{fig:10-f4}.  As a signal pulse propagates through the waveguide, it walks through the pump.  The pulse gain depends on the amount of pump that it passes through, which in turn depends on the pulse's position.  Thus $G(t)$ takes the form of an integral.  For box pulses whose duration matches the walkoff time in the crystal ($T_p = L u$), it is given by:
\beq
	G(t) \approx -\frac{\epsilon\,b_{\rm max}}{u}|t| = -p \frac{\log(G_0 e^{\alpha_a L})}{2T_p} |t| \label{eq:10-gain-gc0}
\eeq
%The total gain in the split-step approximation
The total gain in the split-step approach is $\Gamma(t)\Delta(\delta\omega)$.  Assuming a box pump and negligible dispersion, we can replace $\Delta(\delta\omega) = \Delta_{\rm max}$ with (\ref{eq:10-maxgain}) and thus the gain is
\beq
	\Delta_{\rm max}\Gamma(t) = \exp\left[\frac{\log(G_0 e^{\alpha_a L})}{2} \left((p-1) - p\frac{|t|}{T_p}\right)\right] \label{eq:10-gain-gc}
\eeq

\begin{figure}[tbp]
\begin{center}
\includegraphics[width=1.00\columnwidth]{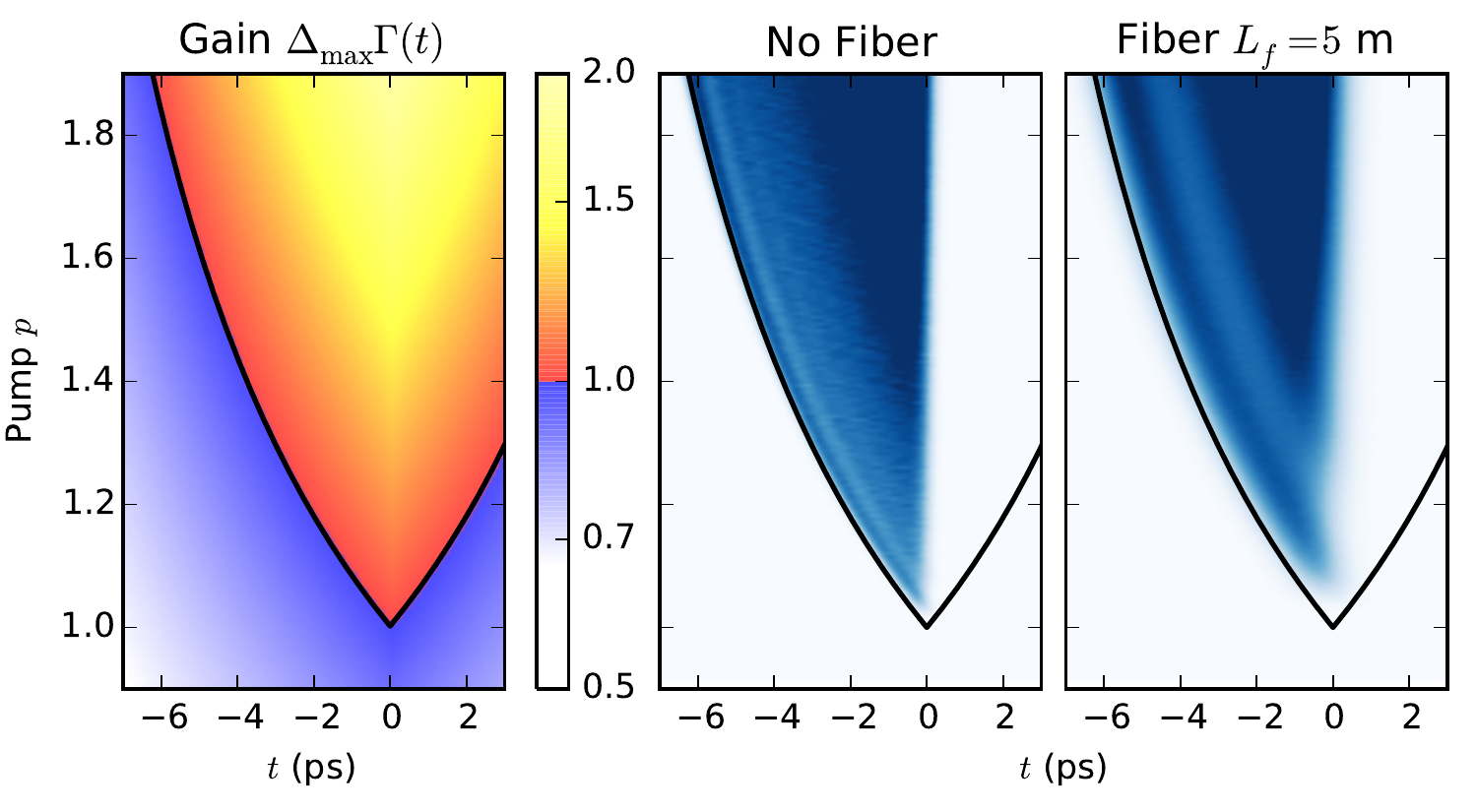}
\caption{Left: Dispersionless round-trip gain $\Delta_{\rm max} \Gamma(t)$ as a function of pump power and time, given by Eq.~(\ref{eq:10-gain-gc}).  Right: Pulse power $|a(t)|^2$ for PPLN waveguide, fiber lengths $L_f = 0$ m and 5 m (which overcompensates the GVD by a factor of 17.6).}
\label{fig:10-f5}
\end{center}
\end{figure}

As Figure \ref{fig:10-f5} shows, the pulse is confined to the positive-gain region ($\Delta_{\rm max} \Gamma(t) > 1$).  The signal pulses become longer as the pump power is increased, since the gain window becomes larger the larger $p$.  But only the left half of the gain window is filled.  This behavior is explored in more detail in Sec.~\ref{sec:10-boxpulse}, but in short is a result of walkoff and pump depletion: the right-side region only reaches the pump after it is depleted by the left side, and is no longer sufficient for amplification.  By this reasoning, the pulse width is derived from (\ref{eq:10-gain-gc}) to be half the gain-window width:
\beq
	T_s = \frac{p-1}{p} T_p
\eeq

This result is consistent with the simulations.  The agreement is strongest when the cavity dispersion is weakest.  As we add dispersion to the cavity, we filter out the high-frequency modes and force $a(t)$ to take a smoother waveform (Fig.~\ref{fig:10-f5}, right plot).  To model the case with dispersion we need both $\Gamma(t)$ and $\Delta(\delta\omega)$ -- this is done in the following section.

\subsection{Shapes of Eigenmodes}
\label{sec:10-emshapes}

Now that we have the gain-clipping and dispersion terms, Eqs.~(\ref{eq:10-Delta}, \ref{eq:10-gammat}), we are ready to find the eigenmodes.  There are two ways to do this: using Eq.~(\ref{eq:10-eig}) gives $a_k(t)$ exactly, along with the round-trip gain $g_k \equiv \log\lambda_k$; however, this approach must be done numerically.  Alternatively, one can take the near-threshold approximation Eq.~(\ref{eq:10-nt-eig}), and using analytic approximations for $G(t)$, $D(i\tfrac{\d}{\d t})$, obtain analytic expressions for the eigenmodes.  The analytic method is presented first, and compared to Eq.~(\ref{eq:10-eig}) and simulations in the following subsection.

\subsubsection{Analytic Form, Degenerate Case ($\phi_0 \phi'_2 \geq 0$)}

As the resonance diagrams in Figs.~\ref{fig:10-f3a}-\ref{fig:10-f3c} make clear, there are two regimes of interest: degenerate and non-degenerate.  The OPO is degenerate when $\phi'_2\phi_0 > 0$, where $\phi'_2 = \phi_2 + \beta_2 L$ (Eq.~(\ref{eq:10-phimatch})).  In this case, using Eq.~(\ref{eq:10-nt-eig}) and substituting (\ref{eq:10-dw-deg}) and (\ref{eq:10-gain-gc0}) for the $D(i\tfrac{\d}{\d t})$ and $G(t)$ respectively, we find near threshold ($p \approx 1$) that $a(t)$ gets mapped after one round trip to
\beq
	\Bigl[g_{\rm cw} \underbrace{-\ \tfrac{\log(G_0 e^{\alpha_a L})}{2T_p} |t|}_{G(t)} + \underbrace{\vphantom{\tfrac{\log(G_0 e^{\alpha_a L})}{2T_p} |t|}\tfrac{\phi'_2\tan\phi_0}{2}\tfrac{\d^2}{\d t^2} - \tfrac{(\phi'_2\sec\phi_0)^2}{8}\tfrac{\d^4}{\d t^4}}_{D(i\tfrac{\d}{\d t})}\Bigr] a(t) \label{eq:10-deg-eig}
\eeq
and thus $[g_{\rm cw} + G(t) + D(i\tfrac{\d}{\d t})] a_k = g_k a_k$ is the eigenvalue equation.  (Eq.~(\ref{eq:10-deg-eig}) was obtained for a box-pulse pump matched to the crystal length; for other pump shapes $G(t)$ changes, see Eq.~(\ref{eq:10-gcf}))

The general case is not solvable analytically, but usually one of the time-derivative terms is much larger than the other, leading to one of two limits:

\begin{enumerate}

\item $\phi_0 \sim O(1)$.  Since $g_k$ is small near threshold, both $G(t)$ and $D(i\tfrac{\d}{\d t})$ must be small, and are typically of the same order.  But if $(\phi'_2\tfrac{\d^2}{\d t^2})a(t) \sim O(g_k) \ll 1$, then $(\phi'_2\tfrac{\d^2}{\d t^2})^2 a(t) \sim O(g_k^2) \ll (\phi'_2\tfrac{\d^2}{\d t^2})a(t)$ and so the fourth-derivative term can be neglected.  In this case (\ref{eq:10-deg-eig}) gives Airy's equation, with the solutions: %The result is Airy's equation and can be solved to give:
\bea
	a_k(t) & \!\!=\!\! & \mbox{sign}(t)^k \mbox{Ai}\left[\left(\frac{T_p \phi'_2\tan\phi_0}{\log(G_0 e^{\alpha_a L})}\right)^{-1/3} \!|t| - \xi_k \right] \label{eq:10-eig-airy} \\
	g_k & \!\!=\!\! & g_{\rm cw} - \frac{1}{2}\left(\frac{\phi'_2\tan\phi_0}{T_p^2}\log(G_0 e^{\alpha_a L})^2\right)^{1/3} \xi_k \label{eq:10-eig-airy2}
\eea
where $\{-\xi_k\}$ is the set of all roots and extrema of the Airy function $\mbox{Ai}(\tau)$ (Table \ref{tab:10-t2}).

\item $\phi_0 \approx 0$.  In this case the second-derivative term is discarded because it goes as $\tan\phi_0$.  The result is a fourth-order analog of Airy's equation: $\d^4y/\d x^4 + xy = 0$, which has two linearly independent solutions that satisfy the boundary conditions at $|t| \rightarrow \infty$: $R_1(\zeta), R_2(\zeta)$ (see Eq.~(\ref{eq:10-rf1}-\ref{eq:10-rf2})). The solution is given by the linear combination
\bea
	a_k(t) & \!\!=\!\! & \mbox{sign}(\zeta)^k \bigl[c_{1,k} R_1(|\zeta| - \zeta_k) + c_{2,k} R_2(|\zeta| - \zeta_k)\bigr]
	 \nonumber \\
	& & \zeta \equiv \left(\tfrac{T_p (\phi'_2)^2}{4\log(G_0 e^{\alpha_a L})}\right)^{-1/5}t \label{eq:10-phizero-ak}  \\
	g_k & \!\!=\!\! & g_{\rm cw} - \frac{1}{2}\left(\frac{\log(G_0 e^{\alpha_a L})^4 (\phi'_2)^2}{4T_p^4}\right)^{1/5}\zeta_k \label{eq:10-eig-hg2}
\eea
that satisfies the differentiability conditions at $t = 0$.  This condition constrains $\zeta_k$ (and thus $g_k$), since these conditions can be reduced to finding a matrix null-space:

\begin{table}[tbp]
\begin{center}
\begin{tabular}{c|cccccccc}
\hline\hline
$k$       & 0    & 1    & 2    & 3    & 4    & 5    & 6    & 7    \\ \hline
$\xi_k$   & 1.02 & 2.34 & 3.25 & 4.09 & 4.82 & 5.52 & 6.16 & 6.79 \\ 
$\zeta_k$ & 0.97 & 2.36 & 3.56 & 4.66 & 5.71 & 6.70 & 7.66 & 8.59 \\ \hline \hline
\end{tabular}
\caption{$\xi_k$ and $\zeta_k$ used in Eqs.~(\ref{eq:10-eig-airy}-\ref{eq:10-eig-hg2})}
\label{tab:10-t2}
\end{center}
\end{table}

\end{enumerate}

\beq
	\underbrace{\begin{bmatrix} R_1'(-\zeta_k) & R_2'(-\zeta_k) \\ R_1'''(-\zeta_k) & R_2'''(-\zeta_k) \end{bmatrix}
	\!\!\begin{bmatrix} c_{1,k} \\ c_{2,k} \end{bmatrix} \!=\! 0}_{k\,=\,0,2,\ldots\ \text{(even solutions)}},\ \
	\underbrace{\begin{bmatrix} R_1(-\zeta_k) & R_2(-\zeta_k) \\ R_1''(-\zeta_k) & R_2''(-\zeta_k) \end{bmatrix}
	\!\!\begin{bmatrix} c_{1,k} \\ c_{2,k} \end{bmatrix} \!=\! 0}_{k\,=\,1,3,\ldots\ \text{(odd solutions)}} 
\eeq
The roots $\zeta_k$ are listed in Table \ref{tab:10-t2}.  For reference, $R_1(\zeta)$ and $R_2(\zeta)$ can be expressed in terms of hypergeometric functions:
\begin{align}
	&\!\!\!R_1(\zeta) = 
			{}_0F_3\!\bigl(;\tfrac{2}{5}, \tfrac{3}{5},\tfrac{4}{5};\tfrac{-\zeta^5}{625}\bigr) 
		- \tfrac{2\pi}{5^{1/20}\phi^{3/2}\Gamma(\tfrac{1}{5})\Gamma(\tfrac{3}{5})}\; 
			\!{}_0F_3\!\bigl(;\tfrac{3}{5},\tfrac{4}{5},\tfrac{6}{5}; \tfrac{-\zeta^5}{625}\bigr) \zeta \nonumber \\
		& \!\!-\! \tfrac{5^{3/20}\pi}{\phi^{3/2}\Gamma(\tfrac{1}{5})\Gamma(\tfrac{2}{5})}\;
			\!{}_0F_3\!\bigl(;\tfrac{4}{5},\tfrac{6}{5},\tfrac{7}{5}; \tfrac{-\zeta^5}{625}\bigr) \zeta^2
		\!+\! \tfrac{5^{3/5}\Gamma(\tfrac{4}{5})}{6\Gamma(\tfrac{1}{5})}\;
			\!{}_0F_3\!\bigl(;\tfrac{6}{5},\tfrac{7}{5},\tfrac{8}{5}; \tfrac{-\zeta^5}{625}\bigr) \zeta^3 \nonumber \\
	\label{eq:10-rf1} \\
	&\!\!\!R_2(\zeta) =
			-{}_0F_3\!\bigl(;\tfrac{3}{5},\tfrac{4}{5},\tfrac{6}{5}; \tfrac{-\zeta^5}{625}\bigr) \zeta
		+ \tfrac{5^{1/5} \phi\,\Gamma(\tfrac{3}{5})}{2\Gamma(\tfrac{2}{5})}\;
			\!{}_0F_3\!\bigl(;\tfrac{4}{5},\tfrac{6}{5},\tfrac{7}{5}; \tfrac{-\zeta^5}{625}\bigr) \zeta^2 \nonumber \\
		& - \tfrac{5^{13/20}\phi^{1/2} \Gamma(\tfrac{3}{5})\Gamma(\tfrac{4}{5})}{12\pi}\;
			\!{}_0F_3\!\bigl(;\tfrac{6}{5},\tfrac{7}{5},\tfrac{8}{5}; \tfrac{-\zeta^5}{625}\bigr) \zeta^3 \label{eq:10-rf2}
\end{align}

\comment{
\bea
	\!\!R_1(\zeta) & \!\!=\!\! & 
			{}_0F_3\left(;\tfrac{2}{5}, \tfrac{3}{5},\tfrac{4}{5};\tfrac{-\zeta^5}{625}\right) 
		- \tfrac{2\pi}{5^{1/20}\phi^{3/2}\Gamma(\tfrac{1}{5})\Gamma(\tfrac{3}{5})}\; 
			{}_0F_3\left(;\tfrac{3}{5},\tfrac{4}{5},\tfrac{6}{5}; \tfrac{-\zeta^5}{625}\right) \zeta \nonumber \\
		& & - \tfrac{5^{3/20}\pi}{\phi^{3/2}\Gamma(\tfrac{1}{5})\Gamma(\tfrac{2}{5})}\;
			{}_0F_3\left(;\tfrac{4}{5},\tfrac{6}{5},\tfrac{7}{5}; \tfrac{-\zeta^5}{625}\right) \zeta^2
		+ \tfrac{5^{3/5}\Gamma(\tfrac{4}{5})}{6\Gamma(\tfrac{1}{5})}\;
			{}_0F_3\left(;\tfrac{6}{5},\tfrac{7}{5},\tfrac{8}{5}; \tfrac{-\zeta^5}{625}\right) \zeta^3 \label{eq:10-rf1} \\
	\!\!R_2(\zeta) & \!\!=\!\! & 
			-{}_0F_3\left(;\tfrac{3}{5},\tfrac{4}{5},\tfrac{6}{5}; \tfrac{-\zeta^5}{625}\right) \zeta
		+ \tfrac{5^{1/5} \phi\,\Gamma(\tfrac{3}{5})}{2\Gamma(\tfrac{2}{5})}\;
			{}_0F_3\left(;\tfrac{4}{5},\tfrac{6}{5},\tfrac{7}{5}; \tfrac{-\zeta^5}{625}\right) \zeta^2 \nonumber \\
		& & - \tfrac{5^{13/20}\phi^{1/2} \Gamma(\tfrac{3}{5})\Gamma(\tfrac{4}{5})}{12\pi}\;
			{}_0F_3\left(;\tfrac{6}{5},\tfrac{7}{5},\tfrac{8}{5}; \tfrac{-\zeta^5}{625}\right) \zeta^3 \label{eq:10-rf2}
\eea
}

\subsubsection{Analytic Form, Non-degenerate Case ($\phi_0 \phi'_2 < 0$)}

In the nondegenerate case, most of the frequency content is contained around $\delta\omega_0 = \sqrt{-2\phi_0/\phi'_2}$, which satisfies the phase condition $\phi_0 + \tfrac{1}{2}\phi'_2 \delta\omega_0^2 = 0$.  We thus make the substitution:
\beq
	a(t) = \mbox{Re}\left[\bar{a}(t) e^{-i\,\delta\omega_0 t}\right] \label{eq:10-abar}
\eeq
The eigenvalue equation (\ref{eq:10-nt-eig}) can be solved with the help of (\ref{eq:10-dw-nd}) and (\ref{eq:10-gain-gc0}); neglecting higher-order derivative terms we obtain:
\beq
	\Bigl[g_{\rm cw} \underbrace{-\ \frac{\log(G_0 e^{\alpha_a L})}{2T_p} |t|}_{G(t)} + \underbrace{|\phi'_2\phi_0|\frac{\partial^2}{\partial t^2}}_{D(i\tfrac{\d}{\d t})}\Bigr] \bar{a}_k(t) = g_k \bar{a}_k(t) \label{eq:10-nd-eig}
\eeq
Note that Eq.~(\ref{eq:10-nd-eig}) is the same as (\ref{eq:10-deg-eig}) if we remove the fourth-order derivative and replace $\tfrac{1}{2}\phi'_2\tan\phi_0 \rightarrow |\phi'_2\phi_0|$.  Thus, the solutions are Airy functions:
\bea
	\!\bar{a}_k(t) & \!=\! & \mbox{sign}(t)^k \mbox{Ai}\left[\left(\frac{2T_p |\phi'_2\phi_0|}{\log(G_0 e^{\alpha_a L})}\right)^{-1/3} \!|t| - \xi_k \right] \label{eq:10-nd-airy} \\
	\!g_k & \!=\! & g_{\rm cw} - \frac{1}{2}\left(\frac{2|\phi'_2\phi_0|}{T_p^2}\log(G_0 e^{\alpha_a L})^2\right)^{1/3} \xi_k \label{eq:10-nd-airy2}
\eea

\subsubsection{Full Form}

\begin{figure}[t]
\begin{center}
\includegraphics[width=1.00\columnwidth]{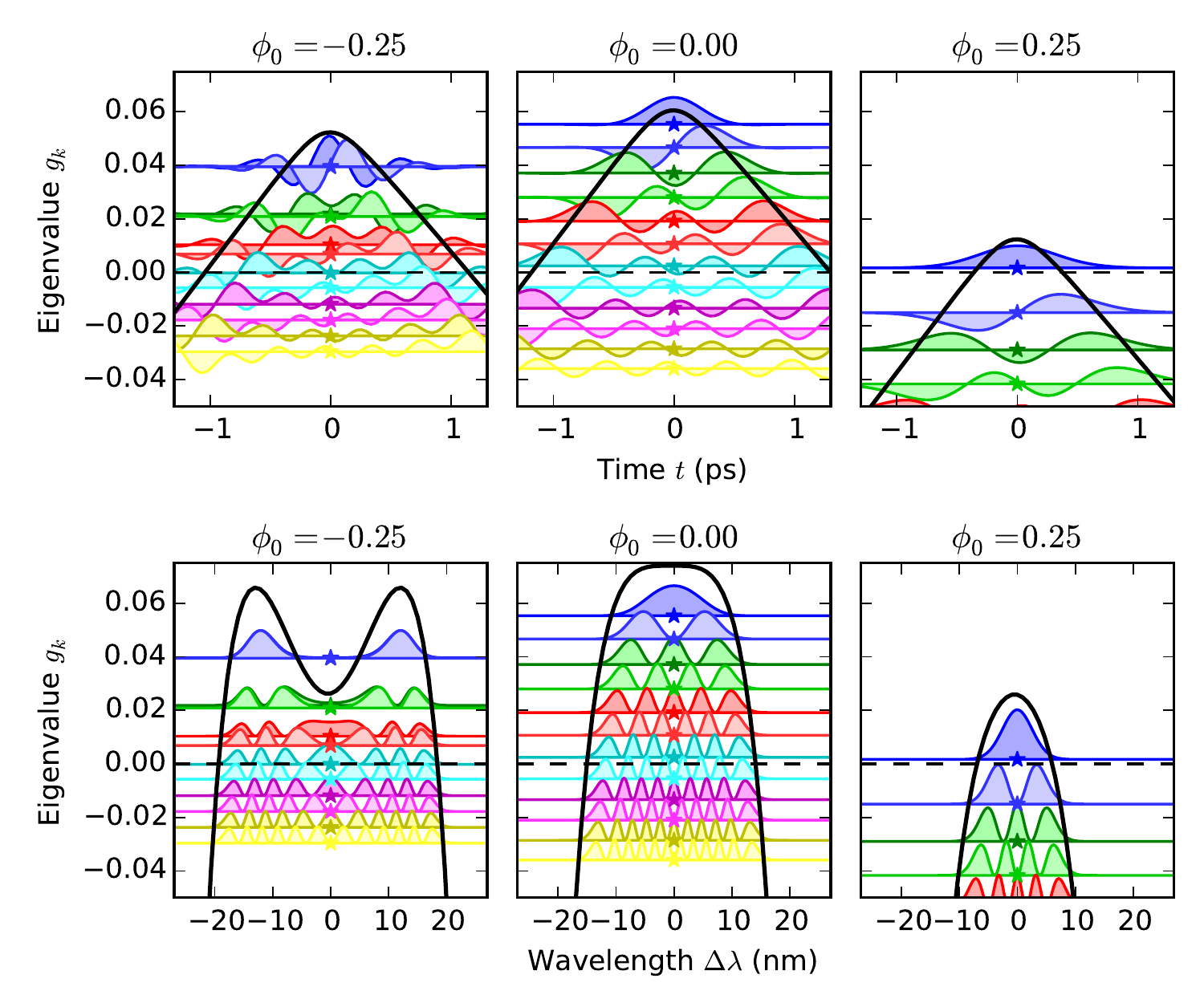}
\caption{Top: shapes of eigenmodes $a_k(t)$ as a function of $\phi_0$, PPLN OPO with $p = 1.1$ and no fiber.  Dark line is the dispersionless gain $\log(\Delta_{\rm max}) + G(t)$.  Bottom: Power spectra of eigenmodes $|a_k(\omega)|^2$, dark line is the CW gain $\log(\Delta_{\rm max}) + D(\delta\omega)$.}
\label{fig:10-f7c}
\end{center}
\end{figure}

\begin{figure}[t]
\begin{center}
\includegraphics[width=1.00\columnwidth]{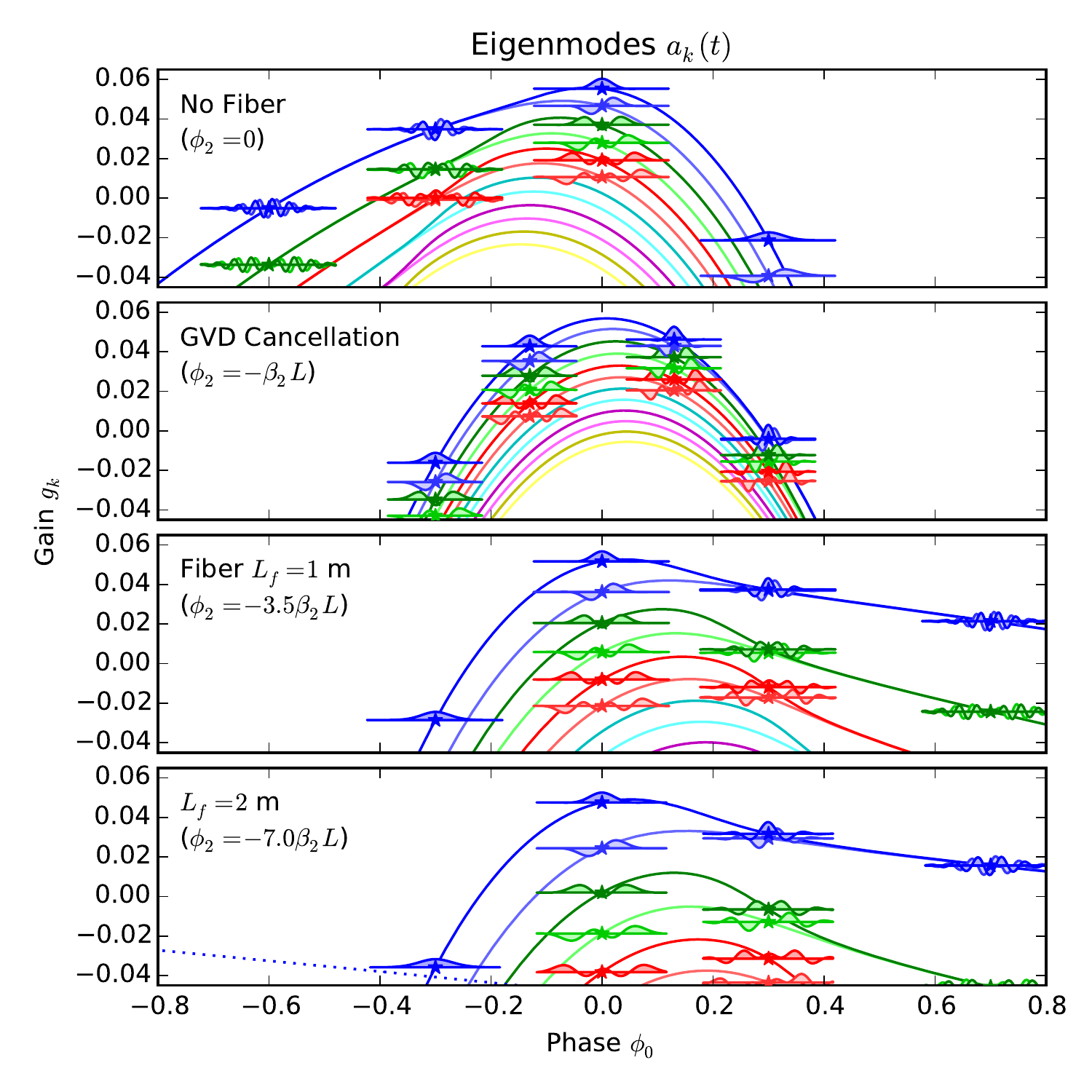}
\caption{Plots of eigenmodes $a_k(t)$ and eigenvalues $g_k$ at pump $p = 1.1$ as a function of cavity phase $\phi_0$ and fiber length $L$.  Pulse widths not to scale between graphs.}
\label{fig:10-f6}
\end{center}
\end{figure}

One can solve the eigenmode equation exactly without resorting to approximations, diagonalizing (\ref{eq:10-eig}) numerically using (\ref{eq:10-Delta}) and (\ref{eq:10-gammat}) for $\Delta(i\tfrac{\d}{\d t})$ and $\Gamma(t)$, respectively.  This approach is necessary in the GVD-compensated case, where the lumped-element approximations (\ref{eq:10-dw-deg}-\ref{eq:10-dw-nd}) break down.  Numerically, it is much easier to diagonalize $\Gamma(t)^{1/2} \Delta(i\tfrac{\d}{\d t}) \Gamma(t)^{1/2}$, which is Hermitian and whose eigenvectors are related to those of $\Gamma(t)\Delta(t)$ by a (nearly constant) function of $t$.

Figure \ref{fig:10-f7c} shows the temporal and frequency structure of the eigenmodes $a_k(t)$.  The system studied here is the PPLN-waveguide OPO without any fiber.  Like particles in a potential well, each eigenmode wavefunction $a_k(t)$ is largely confined to the region $\log(\Delta_{\rm max} \Gamma(t)) > g_k$, since $\Gamma(t) = e^{G(t)}$ plays the role of the potential here.

The power spectra in Fig.~\ref{fig:10-f7c} show that the OPO smoothly transitions from degenerate to nondegenerate operation as the phase is scanned from positive to negative, consistent with the analysis in the previous sections.  This transition happens because the CW gain function $\Delta(\delta\omega)$ plays the role of a potential here.  This function is quadratic for $\phi_0 > 0$ but transitions to a double-well structure for $\phi_0 < 0$, leading to nondegenerate operation in that regime.

Fiber dispersion is accounted for in Figure \ref{fig:10-f6}.  Here the eigenvalues $g_k$ are plotted against $\phi_0$ for a range of fiber lengths.  As the fiber becomes longer, the spacing between eigenvalues increases, largely consistent with the scaling laws in Eqs.~(\ref{eq:10-eig-airy2}, \ref{eq:10-phizero-ak}, \ref{eq:10-nd-airy}).  As the phase passes through zero, the eigenvalues ``pair up'' into degenerate doublets.  It may be confusing notationally, but having degenerate eigenmodes corresponds to {\it non}-degenerate oscillation.  Nondegenerate OPOs will always have degenerate pairs of eigenmodes, each pair corresponding to the real and imaginary components of signal and idler, both which can be amplified.  In a degenerate OPO, only the real quadrature can be amplified, so the eigenmodes do not form degenerate pairs.

\begin{figure}[t]
\begin{center}
\includegraphics[width=1.00\columnwidth]{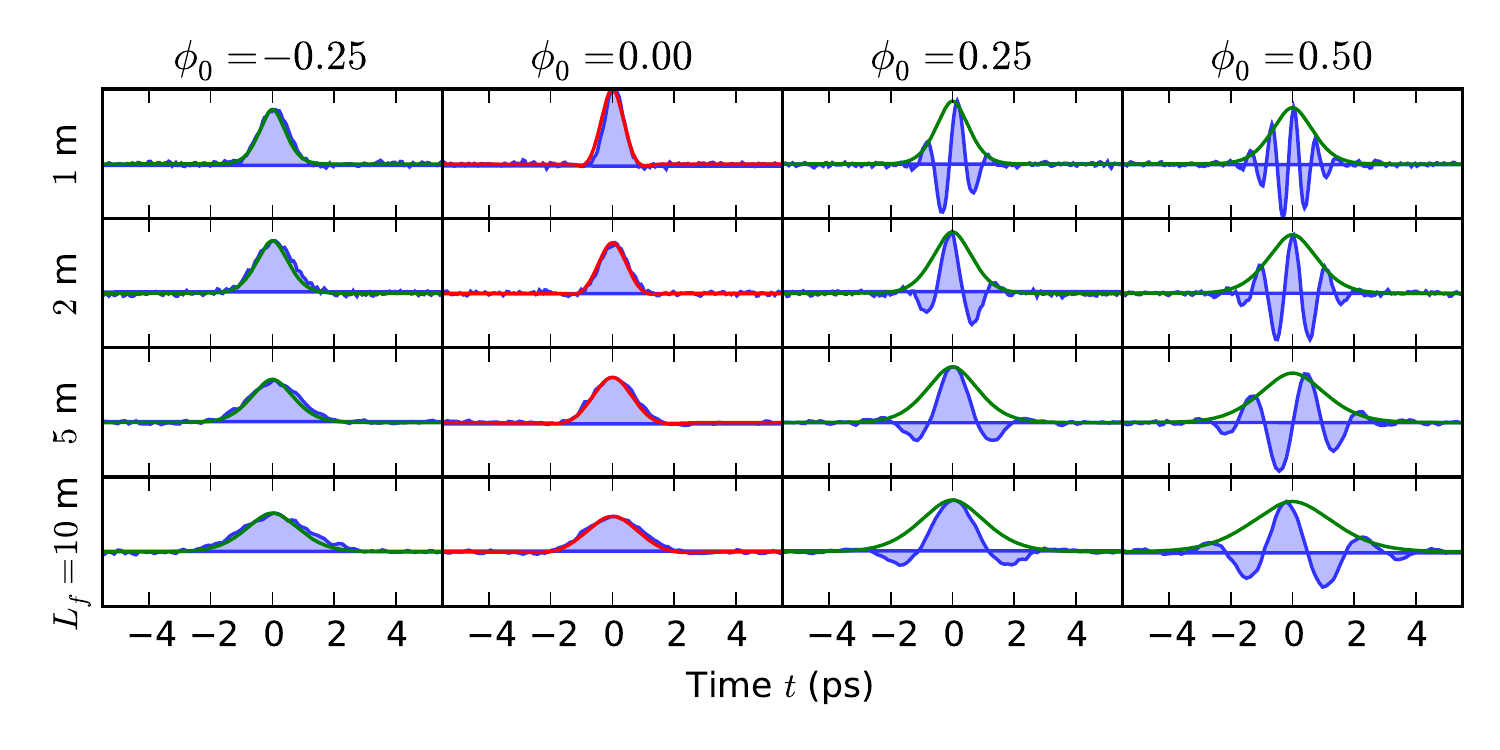}
\caption{OPO steady-state pulse shape just above threshold.  Blue (filled): numerical result.  Green (solid, $\phi_0 \neq 0$): Airy-function solution, (\ref{eq:10-eig-airy}) for degenerate case and (\ref{eq:10-nd-airy}) for nondegenerate case.  Envelope $\bar{a}_k(t)$ is plotted for nondegenerate case.  Red (solid, $\phi_0 = 0$): hypergeometric result (\ref{eq:10-phizero-ak}).}
\label{fig:10-f8}
\end{center}
\end{figure}

Figure \ref{fig:10-f8} compares the pulse shapes from Eqs.~(\ref{eq:10-eig-airy2}, \ref{eq:10-phizero-ak}, \ref{eq:10-nd-airy}) against simulation data.  The simulation data are taken very close to threshold, so that nonlinear effects do not distort the pulse shape.

In addition to the obvious agreement between theory and simulation, Fig.~\ref{fig:10-f8} shows several important trends in the behavior of pulsed OPOs.  First, the pulses become longer the more fiber is inserted into the OPO ($L = 1$ m already over-compensates the PPLN dispersion).  In addition, the larger one makes $\phi_0$ in the nondegenerate region, the larger the signal-idler splitting, consistent with the signal-idler splitting $\delta\omega = \sqrt{-2\phi_0/\phi'_2}$ (Eq.~(\ref{eq:10-phimatch})).

\subsection{Threshold}

Threshold is both straightforward to measure and easy to derive from the linearized model.  It is the pump power needed to make the principal eigenmode have the highest gain: $g_0 = 0$.  Since the eigenmode gain depends on $\phi_0$, threshold depends on $\phi_0$ as well, giving rise to the detuning peaks in Fig.~(\ref{fig:10-f2}).  For a CW pump at $\phi_0 = 0$, the threshold is clearly $p = 1$.

We can compute thresholds near the center of a detuning peak by inverting the eigenmode gain expression.  Recall from (\ref{eq:10-eig-airy2}, \ref{eq:10-eig-hg2}, \ref{eq:10-nd-airy2}) that the eigenmode gain takes the form:
\beq
	g_k = g_{\rm cw} + g'_k \label{eq:10-gk-zero}
\eeq
where $g'_k$ depends on the differential equation being solved.  Near the center of the detuning peak, the CW gain goes as $\Delta \approx (G_0 e^{\alpha_a L})^{p-1}$ (Eq.~(\ref{eq:10-maxgain})), so we can write $g_{\rm cw}(p) \approx g_{\rm cw}(p=1) + \tfrac{p-1}{2}\log(G_0 e^{\alpha_a L})$.  Setting the gain (\ref{eq:10-gk-zero}) to zero, we obtain an approximate formula for the threshold:
\beq
	p_{\rm th} = 1+\frac{-g_0(p=1)}{\tfrac{1}{2}\log(G_0 e^{\alpha_a L})}
\eeq
This relation is valid for $|g_0| \ll 1$.  In the same way, we can compute the thresholds for the higher eigenmodes.

\begin{figure}[tbp]
\begin{center}
\includegraphics[width=1.00\columnwidth]{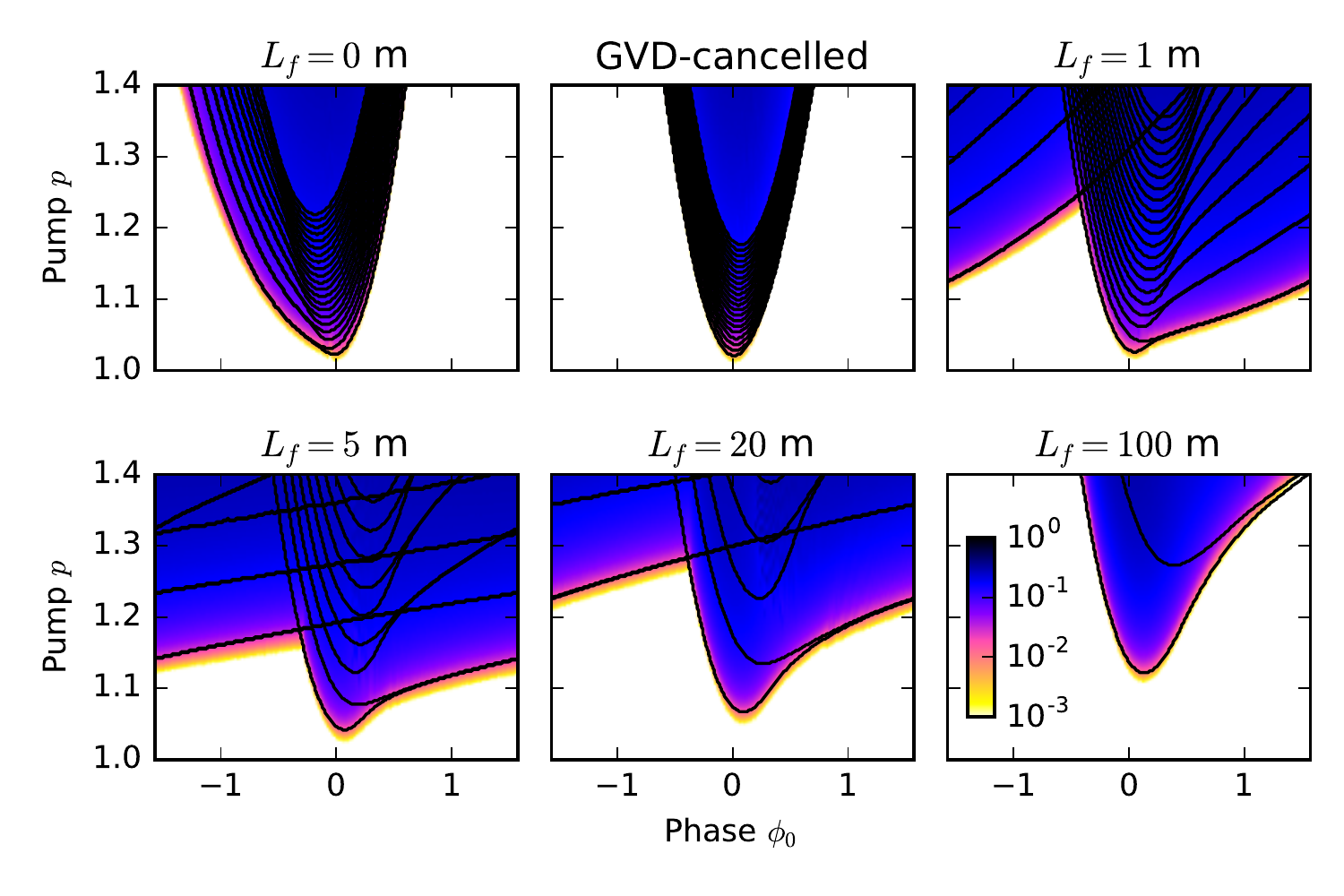}
\caption{Plot of OPO efficiency $\eta = P_{a,\rm out}/P_{b,\rm in}$ as a function of $p$ and $\phi_0$, with cavity dispersion provided by various lengths of fiber $L_f$; ``GVD-cancelled'' refers to a fiber that compensates the dispersion of the $\chi^{(2)}$ medium.  Contours are thresholds for the first 20 eigenmodes $a_k(t)$.}
\label{fig:10-f9}
\end{center}
\end{figure}

By definition, the OPO turns on when the pump power exceeds threshold.  In the simulation results of Fig.~\ref{fig:10-f9}, the OPO efficiency $\eta = P_{a,\rm out}/P_{b,\rm in}$ is plotted against cavity phase and pump power.  In simulations, the OPO turns on right at the point where the highest eigenmode goes above threshold ($g_0 = 0$).  Thus, the eigenmode model should give accurate predictions of pulsed OPO thresholds.

Note that the structure of these thresholds matches that of the eigenmodes.  Consistent with Fig.~\ref{fig:10-f6}, the eigenmodes ``pair up'' in the nondegenerate regime $\phi_0\phi'_2 < 0$.  Also, as the fiber length is increased, the spacing between thresholds increases.

Figure \ref{fig:10-f9} is useful because it tells us when a pulsed OPO is in single-mode operation.  If the pump is below the threshold for the first excited mode $a_1(t)$, then the device behaves like a single-mode OPO.  But once it passes that threshold, multiple modes can oscillate in principle, and the dynamics may become more complex.  Multi-mode effects, coupled with nonlinearity, can give rise to oscillation (Sec.~\ref{sec:10-twomode}), instabilities (Sec.~\ref{sec:10-simstab}), centroid drift (Sec.~\ref{sec:10-ansatz}), and the formation of flat-top pulses (Sec.~\ref{sec:10-boxpulse}).  More complex behavior is possible with multimode OPO networks; recent experiments have hinted towards a multimode description \cite{Takata2016}, and the topic is being actively investigated.

\section{Nonlinear Corrections to Eigenmode Theory}
\label{sec:10-nonlinear}

For an OPO above threshold, we must add nonlinearity to the model since it prevents signals from diverging to infinity.  It also makes the otherwise-independent eigenmodes interact.  The resulting pulse shape will depend on OPO parameters like $p$, $\phi_0$.

This section treats nonlinearity as a perturbation to the eigenmode dynamics.  This will only be valid reasonably close to threshold.  Moreover, it is necessary to truncate the nonlinear model by keeping only a finite number of eigenmodes in the basis.  The required number of eigenmodes grows as the pump power increases and more modes go above threshold (Fig.~\ref{fig:10-f9}).  The method described here has $O(N^4)$ complexity, where $N$ is the number of modes, so if too many modes are included it becomes impractical.  However, we will show in this section that a reasonable number ($N \lesssim 20$) gives good agreement with numerical data.  Thus, the nonlinear eigenmode theory is a good alternative ``reduced model'' that captures the dynamics of the full simulations, but takes $10^2$--$10^3$ times less computation time.

In addition to nonlinearity, cavity detuning will be treated in this section.  To treat these two effects, first we introduce the equations of motion and project them onto the eigenmode basis (Sec.~\ref{sec:10-nl-eom}).  Next we discuss the results of an analytic ``two-mode'' model (Sec.~\ref{sec:10-twomode}) which provides insight into pulse stability and dynamics, and finally compare the nonlinear eigenmode model with full simulations (Sec.~\ref{sec:10-nl-num}).

\subsection{Equations of Motion}
\label{sec:10-nl-eom}

The normal modes derived in Section \ref{sec:10-linear} allow us to describe the field of the OPO pulse in terms of a few mode amplitudes rather than hundreds of Fourier components.  This greatly reduces the complexity of the problem, at the cost of having to compute the modes in the first place and being restricted to a subspace spanned by the dominant modes.  Supposing that $a(t;n)$ is the pulse at the $n^{\rm th}$ round trip.  This can be written in terms of the normal modes $a_k(t)$ and their amplitudes $c_k(n)$:
\beq
	a(t;n) = \sum_k a_k(t) c_k(n)
\eeq
In the absence of pump depletion or any other effects, the equation of motion is:
\beq
	c_k(n+1) = e^{g_k} c_k(n) \label{eq:10-dc-disc}
\eeq
In the near-threshold case where $g_k \ll 1$, this can be converted to a differential equation:
\beq
	\frac{\d c_k}{\d n} = g_k c_k \label{eq:10-dc-cont}
\eeq
Pump depletion and cavity length detuning (repetition-rate mismatch) give corrections to the linear model, as described in the sections below.

\subsubsection{Detuning}

When the cavity is detuned by a length $\ell$, the signal picks up a round-trip phase $\pi\ell$ and its envelope shifts by $(\lambda/2c)\ell$:
\beq
	a(t) \rightarrow a(t - \tfrac{\lambda}{2c}\ell) e^{i \pi \ell}
\eeq
The phase shift was accounted for when the normal modes were chosen.  In the normal-mode picture, the envelope shift is accounted for using the map
\beq
	c_k \rightarrow S_{kl}(\ell) c_l,\ \ \ S_{kl}(\tau) = \int{a_k(t) a_l(t-\tfrac{\lambda}{2c}\ell)\d t} \label{eq:10-rrrm0}
\eeq
Combining both (\ref{eq:10-dc-disc}) and (\ref{eq:10-rrrm0}), one arrives at the relation $c_k(n+1) =  \sum_l S_{kl} e^{g_l} c_l(n)$.  If the field changes slowly between round trips, e.g. $g_k, S_{k\neq l} \ll 1$, then one has:
\beq
	\underbrace{\frac{\d}{\d n}\begin{bmatrix} c_0 \\ c_1 \\ \vdots \\ c_m \end{bmatrix}}_{\d c/\d n} = 
	\underbrace{\begin{bmatrix} g_0 & \ell J_{01} & \cdots & \ell J_{0m} \\
		-\ell J_{01} & g_1 & \cdots & \ell J_{1m} \\
		\vdots & \vdots & \ddots & \vdots \\
		-\ell J_{0m} & -\ell J_{1m} & \cdots & g_m \end{bmatrix}}_{\ell J + G} 
	\underbrace{\begin{bmatrix} c_0 \\ c_1 \\ \vdots \\ c_m \end{bmatrix}}_{c} \label{eq:10-det-dc}
\eeq
where the coupling matrix $J$ is:
\beq
	J_{kl} = \left.\frac{\d S_{kl}}{\d\ell}\right|_{\ell=0} = -\frac{\lambda}{2c}\int{a_k(t) \frac{\d a_l(t)}{\d t} \d t}
\eeq

Integration by parts shows that $J_{kl}$ is antisymmetric and only mixes modes of opposite parity.  The linear dynamics are set by the matrix $G + J$.  This mixes modes of positive and negative eigenvalue.  If the mixing is strong enough, all of the eigenvalues will be negative and the oscillation is suppressed.  Thus the oscillation threshold will increase with increasing $|\ell|$.

\subsubsection{Pump Depletion}

To calculate the effect of pump depletion, go back to Eqs.~(\ref{eq:10-at}-\ref{eq:10-bt}).  During the dispersionless step in Sec.~\ref{sec:10-dispersionless}, we solved these equations in the absence of GVD.  The pump equation can be integrated using the method of characteristics to give:
\beq
    b(z, t)\!=\!b_{\rm in}(t - u z)e^{-\alpha_b z/2} - \frac{\epsilon}{2}\!\int_0^z{\!\!e^{\alpha_b (z'-z)/2} a\!\left(z', t\!+\!u(z'\!-\!z)\right)^2 \!\d z'}
\eeq

We now invoke the ``gain-without-distortion ansatz'' used to derive the linear eigenmode theory.  In this case it takes the form: $a(z', t) \approx G_{z\rightarrow z'} a(z, t)$.  For small $|z'-z|$, say of order one walkoff length, we can expand $G_{z\rightarrow z'}$ in terms of the $z$ coordinate $a(z', t) \approx e^{g(z)(z'-z)/2} a(z, t)$.  When this is so, we can account for the $z'$ dependence in the integral on the right with a factor of $e^{g(z)(z'-z)/2}$,  change the integration variable to $t' = t+u(z'-z)$ and (in the limit that the walkoff length $L u$ is much longer than the signal) set the left bound to $-\infty$, and obtain:
\beq
b(z, t) = b_{\rm in}(t - u z)e^{-\alpha_b z/2} - \frac{\epsilon}{2u} \int_{-\infty}^{t}{e^{(g(z)+\alpha_b/2)(t'-t)/u} a(z, t')^2 \d t'}
\eeq
Substituting this into the differential equation for $a$, we can eliminate the pump and obtain an equation of motion that depends only on the signal:
\begin{align}
    & \!\!\!\!\frac{\partial a(z,t)}{\partial z} = -\frac{1}{2} \alpha_a a(z, t) + \epsilon\,a^*(z,t) b_{\rm in}(t - u z)e^{-\alpha_b z/2} \nonumber \\
    & \qquad \underbrace{-\frac{\epsilon^2}{2u} a^*(z,t)\!\int_{-\infty}^{t}{\!\!e^{(g(z)+\alpha_b/2)(t'-t)/u} a(z, t')^2 \d t'}}_{\partial a/\partial z\bigr|_{\rm NL}} \label{eq:10-intdiff}
\end{align}

Now one can apply the gain without distortion approximation so that $a(z,t)$ can be related to its initial condition, expressing the right-hand side of (\ref{eq:10-intdiff}) in terms of $a_{in}(t)$.  

For a constant pump, $g$ is constant in $z$, but in general it will go as $g = 2G_{0\rightarrow z}^{-1} \d G_{0\rightarrow z}/\d z$.  Although $G_{0\rightarrow z}$ depends on $t$, the dependence is weak in the region where the pulse forms (at least for the waveguide OPOs), so it can be taken to be constant in $t$.  Taking $a_{\rm in}$ to be real, we can integrate through (\ref{eq:10-intdiff}) to obtain the perturbation on $a_{\rm out}$:
\begin{align}
	& a_{\rm out}(t)\bigr|_{\rm NL} = -\frac{\epsilon^2}{2u} G_{0\rightarrow L} \int_0^L {\Bigl[G_{0\rightarrow z}^2 a_{\rm in}(t)} \nonumber \\
	& \qquad \times \int_{-\infty}^t{e^{(g(z)+\alpha_b/2)(t'-t)/u} a_{\rm in}(t')^2 \d t'}\Bigr]\d t \label{eq:10-depl-3d}
\end{align}

At threshold, PPLN gain matches cavity loss, so the loss near threshold is approximately $1/G_{0\rightarrow L}$.  This fact combined with (\ref{eq:10-depl-3d}) gives a round-trip equation for $a(t)$.  In terms of the coefficients $c_k$, this may be written as:
\beq
	\Delta c_k\bigr|_{\rm NL} = -2\beta\sum_{lmn} {\Psi_{klmn} c_l c_m c_n} \label{eq:10-3oc}
\eeq
where the $\beta$ (pump back-conversion term) and $\Psi_{klmn}$ are:
\begin{align}
	& \beta = \frac{\epsilon^2}{4u} \int_0^L {G_{0\rightarrow z}^2 \d z} \nonumber \\
	& \Psi_{klmn} = \frac{1}{\int_0^L {G_{0\rightarrow z}^2 \d z}} \int_0^L{\Bigl[G_{0\rightarrow z}^2 \int_{-\infty}^\infty{a_k(t)a_l(t)}} \nonumber \\
	& \ \ \times\int_{-\infty}^t{e^{(g(z)+\alpha_b/2)(t'-t)/u} a_m(t') a_n(t')\d t'}\,\d t\Bigr]\d z \label{eq:10-psi0000}
\end{align}

If the gain is constant ($G_{0\rightarrow z} = e^{gz/2}$, $g(z) = g$ constant) and small per walkoff length ($gt/u \ll 1$) then one can simplify this further.  These assumptions generally hold for waveguide OPOs pumped with flat-top pulses.  Using $G_{0\rightarrow L} \approx G_0^{1/2}$ near threshold, one can substitute $g \rightarrow \tfrac{1}{2L}\log(G_0)$; one can then evaluate the integrals in (\ref{eq:10-psi0000}), and applying the formulas in Table \ref{tab:10-t1}, express the remaining constants in terms of the threshold gain $G_0$ and photon number $N_{b,0}$:
\bea
	\beta & \!=\! & \frac{e^{\alpha_b L/2}(G_0-1) \log(G_0 e^{\alpha_a L})^2}{16N_{b,0} \log G_0} \nonumber \\
	\Psi_{klmn} & \!=\! & \int_{-\infty}^\infty{\!a_k(t)a_l(t) \int_{-\infty}^t{\!a_m(t') a_n(t')\d t'}\,\d t} \label{eq:10-betapsi}
\eea

Equation (\ref{eq:10-betapsi}) divides the physics into two terms: $\beta$ is a property of the pump and the waveguide, while $\Psi_{klmn}$ is a geometric factor that depends only on the shape of the normal modes $a_k(t)$.  $\Psi_{klmn}$ also satisfies a few important identities.  Integration by parts gives:
\beq
	\Psi_{klmn} = \delta_{kl}\delta_{mn} - \Psi_{mnkl} \label{eq:10-psi-id1}
\eeq
Typically, the fields $a_k$ have inversion symmetry.  Let's suppose that the $a_k$ are numbered so that the odd-indexed ones are odd and the even-indexed ones are even: $a_k(-t) = (-1)^k a_k(t)$.  Then one finds that exactly half of the $\Psi_{klmn}$ are either zero or a half:
\beq
	\Psi_{klmn} = \frac{1}{2}\delta_{kl}\delta_{mn}\ \ \ (\mbox{if}\ k+l+m+n\ \mbox{even}) \label{eq:10-psi-id2}
\eeq
Combining Equations (\ref{eq:10-det-dc}, \ref{eq:10-3oc}), one has all the physics needed to simulate the OPO near threshold.  Writing these for convenience in continuous-time, the equations of motion are:
\beq
	\frac{\d c_k}{\d n} = g_k c_k + \sum_l J_{kl} c_l - 2\beta \sum_{lmn} \Psi_{jklm} c_l c_m c_n \label{eq:10-dck-qd}
\eeq

In the single-mode limit, this resembles the classic result for a single-mode singly-resonant OPO, with $\Psi$ playing the role of a pump depletion term \cite{Kinsler1991}.  The single-mode theory was extended for high-finesse resonators \cite{DeValcarcel2006, Patera2010}, and the form resembles (\ref{eq:10-dck-qd}).  Note, however, that $c_k$ is constrained to be a real number here, so (\ref{eq:10-dck-qd}) will not capture the squeezing dynamics of the OPO.  A more careful treatment of the eigenmodes, which accounts for both the real and imaginary parts of the field, will be needed to model squeezing.

\subsection{Two-Mode Model}
\label{sec:10-twomode}

Consider a two-mode model.  This model is simple enough that it can be solved analytically, shedding important insight into the bifurcations and stability of the pulsed OPO.  

The time-delay matrix $J_{kl}$ only has two nonzero elements: $J_{10} = -J_{01} \equiv J$.  Most of the values of $\Psi_{klmn}$ are set by identities (\ref{eq:10-psi-id1}-\ref{eq:10-psi-id2}), giving:
\begin{align}
	& \Psi_{0000} = \Psi_{0011} = \Psi_{1100} = \Psi_{1111} = \tfrac{1}{2} \nonumber \\
	& \Psi_{0001} = \Psi_{0010} = -\Psi_{0100} = -\Psi_{1000} \nonumber \\
	& \Psi_{0111} = \Psi_{1011} = -\Psi_{1101} = -\Psi_{1110} \nonumber \\
	& \Psi_{0101} = \Psi_{0110} = \Psi_{1001} = \Psi_{1010} = 0
\end{align}
Putting this all together, we have an equation that depends on 6 parameters $(g_0, g_1, J, \beta, \Psi_{0001}, \Psi_{0111})$:
\bea
	\!\!\dot{c}_0 & \!\!=\!\! & g_0 c_0 \!-\! J\, c_1 \!+\! \beta\left[-(c_0^2 + c_1^2) c_0 \!-\! 2(\Psi_{0001} c_0^2 \!+\! \Psi_{0111} c_1^2) c_1\right] \label{eq:10-dc0}\ \ \ \ \ \ \\
	\!\!\dot{c}_1 & \!\!=\!\! & g_1 c_1 \!+\! J\, c_0 \!+\! \beta\left[-(c_0^2 + c_1^2) c_1 \!+\! 2(\Psi_{0001} c_0^2 \!+\! \Psi_{0111} c_1^2) c_0\right] \label{eq:10-dc1}\ \ \ \ \ \ 
\eea

Since $g_0 > g_1$ are the largest eigenvalues $g_0 \leq 0$ means no signal.  Assuming $g_0$ positive, one can reduce (\ref{eq:10-dc0}-\ref{eq:10-dc1}) by scaling time by $g_0^{-1}$ and the fields by $\sqrt{\beta/g_0}$:
\bea
	\!\!\frac{\d\bar{c}_0}{\d\bar{n}} & \!\!=\!\! & \bar{c}_0 - \bar{J}\, \bar{c}_1 - \left[(\bar{c}_0^2 + \bar{c}_1^2) \bar{c}_0 + 2(\Psi_{0001} \bar{c}_0^2 + \Psi_{0111} \bar{c}_1^2) \bar{c}_1\right]\ \ \ \ \ \  \label{eq:10-dc0-2} \\
	\!\!\frac{\d\bar{c}_1}{\d\bar{n}} & \!\!=\!\! & \bar{g} \bar{c}_1 + \bar{J}\, \bar{c}_0 - \left[(\bar{c}_0^2 + \bar{c}_1^2) \bar{c}_1 - 2(\Psi_{0001} \bar{c}_0^2 + \Psi_{0111} \bar{c}_1^2) \bar{c}_0\right]\ \ \ \ \ \  \label{eq:10-dc1-2}
\eea
Now we only have four parameters ($\bar{J}=J/g_0, \bar{g}=g_1/g_0, \Psi_{0001}, \Psi_{0111}$).  Since the model is two-dimensional, textbook dynamical-systems theory is very useful here \cite{StrogatzBook}.  In particular, we can draw a phase-space diagram and plot the critical points, limit cycles and separatrices.  This can be done by brute force using numerical solvers, but system (\ref{eq:10-dc0-2}-\ref{eq:10-dc1-2}) is simple enough that it has an analytic solution.  Making the substitution $c_1 = c_0\xi$, one can combine the two equations to remove $c_0$, leaving a fourth-order polynomial in $\xi$
\begin{align}
	& (1 + \xi^2) \left[\bar{J}\xi^2 + (\bar{g}-1)\xi + \bar{J}\right] \nonumber \\
	& \quad + 2(1+\bar{g}\xi^2) \left[\Psi_{0001} + \Psi_{0111}\xi^2 \right] = 0
\end{align}
Once this is found, one can plug the result into (\ref{eq:10-dc0-2}) to get $c_0$:
\beq
	c_0^2 = \frac{1 - \bar{J}\xi}{(1 + \xi^2) + 2(\Psi_{0001} + \Psi_{0111}\xi^2)\xi}
\eeq
For given eigenmodes, $\Psi_{0001}$ and $\Psi_{0111}$ are fixed.  As long as the general shape of the eigenmodes remains the same, they will not vary by much.  Thus, the reduced system (\ref{eq:10-dc0-2}-\ref{eq:10-dc1-2}) only has two parameters.  For typical Hermite-Gauss or sech-like eigenmodes, one has $\Psi_{0001} \approx -0.26,\ \Psi_{0111} \approx 0.11$.  

\begin{figure}[tbp]
\begin{center}
\includegraphics[width=1.0\columnwidth]{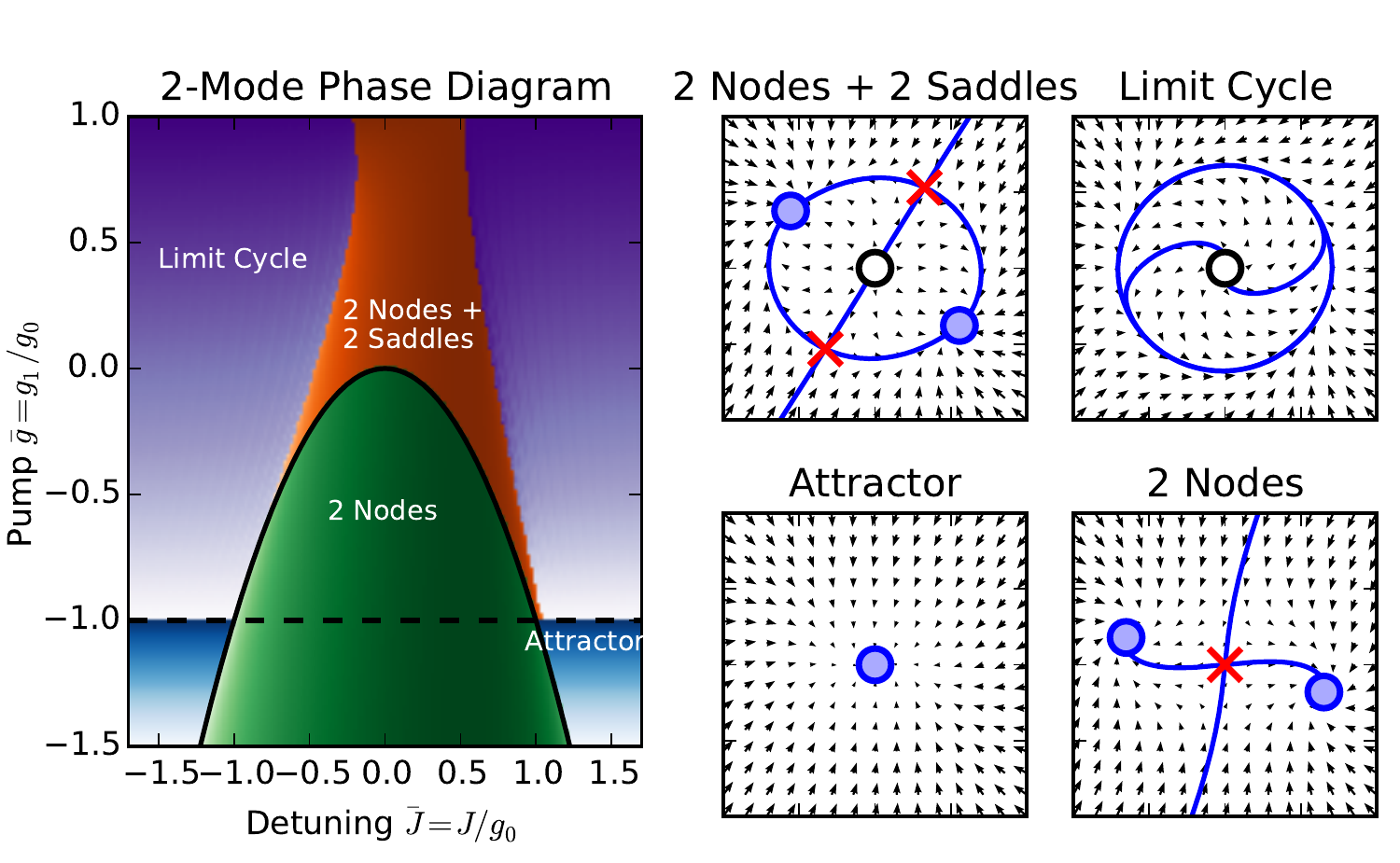}
\caption{Left: phase diagram of the two-mode model in terms of normalized parameters $\bar{J} = J/g_0$, and $\bar{g} = g_1/g_0$.  Right: typical phase-space plots corresponding to the four regions in the diagram.}
\label{fig:10-f10}
\end{center}
\end{figure}

Four types of behavior are possible, as illustrated in Figure \ref{fig:10-f10}.  They are:

\begin{enumerate}
	\item Single attractor.  This occurs if $g_0, g_1 < 0$ or if $g_0 + g_1 < 0$ and $J^2 > -g_0g_1$.  It corresponds to the OPO below threshold.
	\item 2 nodes.  As the pump power is increased, the attractor undergoes a pitchfork bifurcation, creating a saddle point at the origin and two neighboring attractors.  In the limit $g_1/g_0 \rightarrow -\infty$, this reduces to the case of a single-mode OPO above threshold, since the second mode decays too quickly to participate in the dynamics.  In this regime, the OPO behaves qualitatively like the single-mode model.
	\item 2 nodes + 2 saddles.  If the pump increases further, $g_1$ becomes positive and the saddle point at zero splits into two saddles and an unstable node.
	\item Limit cycle.  In the previous picture, nonzero delay causes the attractors and saddle points to move towards each other.  If $|J|$ is large enough, these fixed points annihilate in a saddle-node bifurcation, giving rise to a limit cycle.  Alternatively, one could start in the single-attractor region with sufficiently large $T$, and increasing $g_1$ will lead to the limit-cycle region by way of a Hopf bifurcation.
\end{enumerate}

\begin{figure}[tbp]
\begin{center}
\includegraphics[width=1.0\columnwidth]{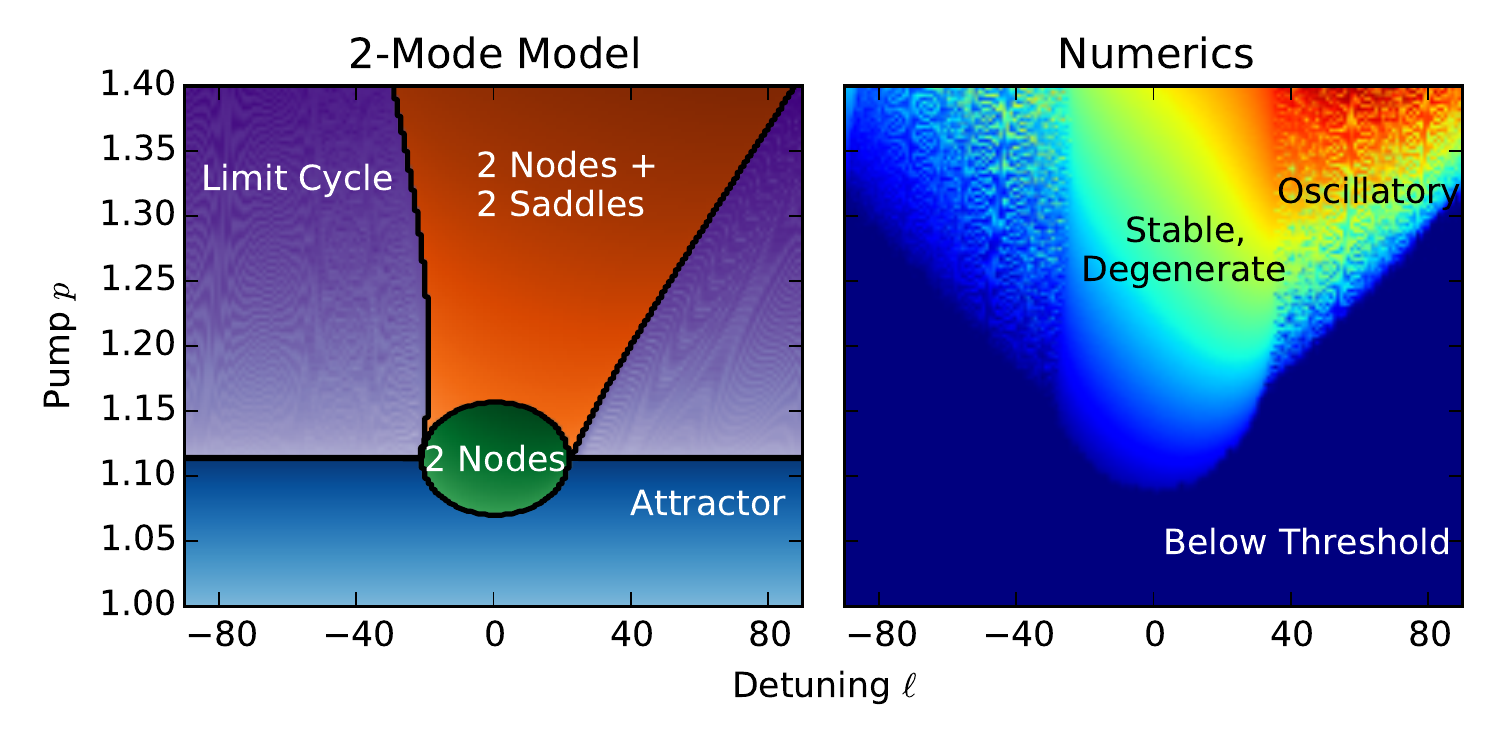}
\caption{Left: 2-mode model phase plot for PPLN OPO with 20-m fiber, $\phi_0 = 0$.  Right: photon number plot for numerical simulation.}
\label{fig:10-f22}
\end{center}
\end{figure}

The OPO pump and detuning are related to the two-mode parameters $\bar{J}, \bar{g}$, so the phase diagram in Fig.~\ref{fig:10-f10} can be mapped onto $(\ell, p)$.  Figure \ref{fig:10-f22} shows the phase diagram as a function of $(\ell, p)$ for an OPO with 20-m of fiber (at the centers of the detuning peaks, $\phi_0 = 0$).  The right plot gives the photon number from a simulation where the pump is swept from $p = 1.0$ to $p = 1.4$.

Qualitatively, many of the features from the numerical plot agree with the two-mode model.  Near $\ell = 0$, the threshold is lowest, increasing quadratically with $\ell$.  The two-mode model does not predict the threshold correctly for larger $\ell$, since higher-order modes start mixing with $a_0(t), a_1(t)$, raising the threshold still further.

The two-mode model gives a region of stability at low $\ell$, surrounded by a limit-cycle region with no stable fixed points.  The width of this region roughly matches the simulations, although it deviates for large $p$ where higher-order modes become important.  The only way to make the model more accurate is to add more modes as will be discussed in the next section.

While a two-mode model with real coefficients cannot support period-doubling or chaos \cite{StrogatzBook} (such phenomena are, however, possible in pump-resonant OPOs with detuning \cite{Lugiato1988}), it is likely that with three or more modes, or with multiple interacting pulses, one could realize these and more complex dynamics \cite{Takata2016}.

In the frequency domain, the limit cycle in Figs.~\ref{fig:10-f10}-\ref{fig:10-f22} is associated with the coexistence of two separate ``signal'' and ``idler'' frequency combs with carrier-envelope offsets that differ by $2\Omega$, where $\Omega$ is the limit-cycle frequency (this is the above-threshold analog of \cite{Jiang2012}).  If the envelopes of these combs overlap, they will beat against each other, leading to an RF photocurrent signal at frequency $\Omega$.  This effect has been reported in the literature \cite{Wolf2013}.

\subsection{Comparison to Numerics}
\label{sec:10-nl-num}

\begin{figure}[tbp]
\begin{center}
\includegraphics[width=1.00\columnwidth]{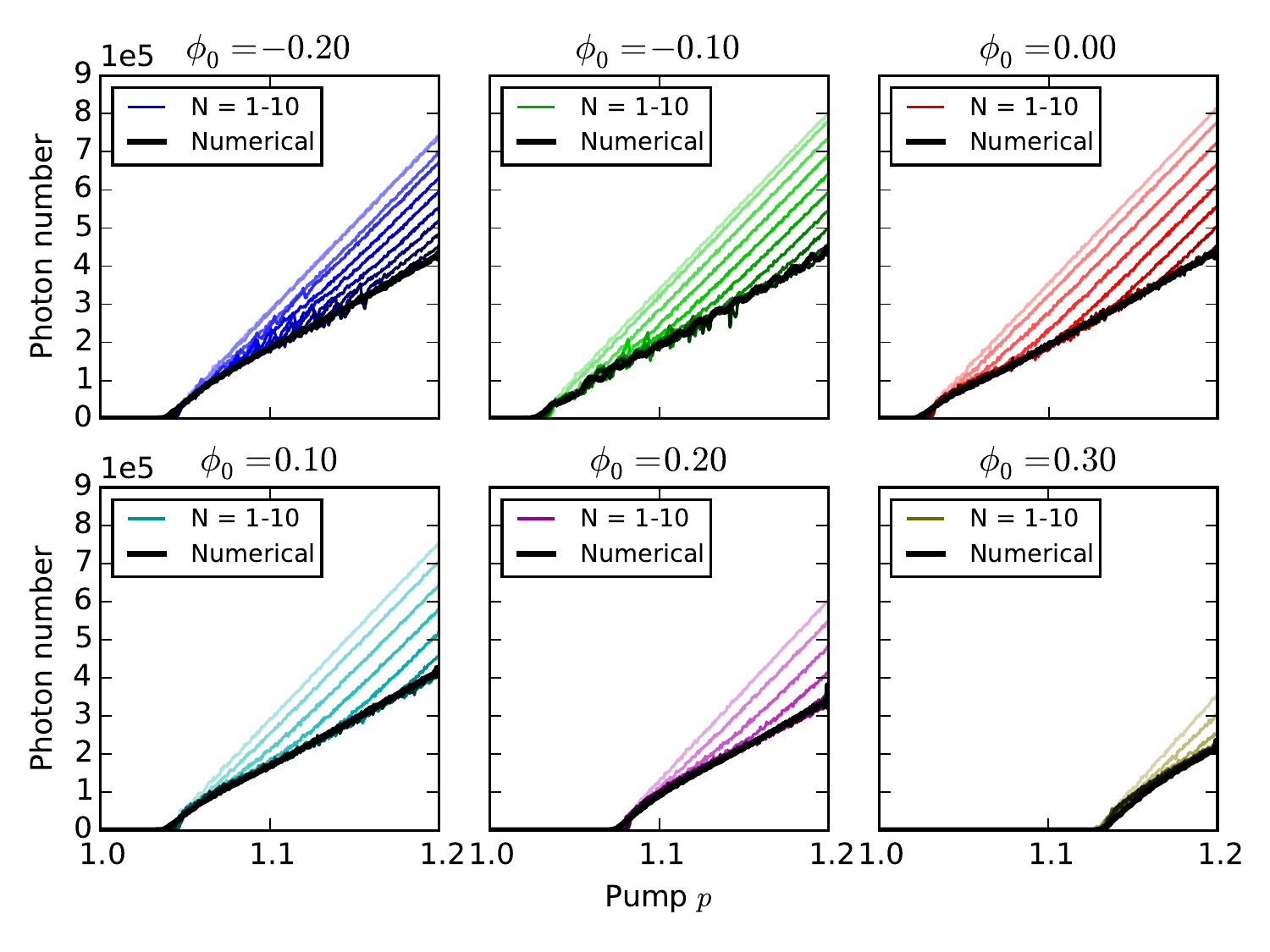}
\caption{Photon number as a function of pump amplitude.  Darker colored lines are eigenmode models with increasing $N$.  Black line is the numerical result.}
\label{fig:10-f11}
\end{center}
\end{figure}

\begin{figure}[t]
\begin{center}
\includegraphics[width=1.0\columnwidth]{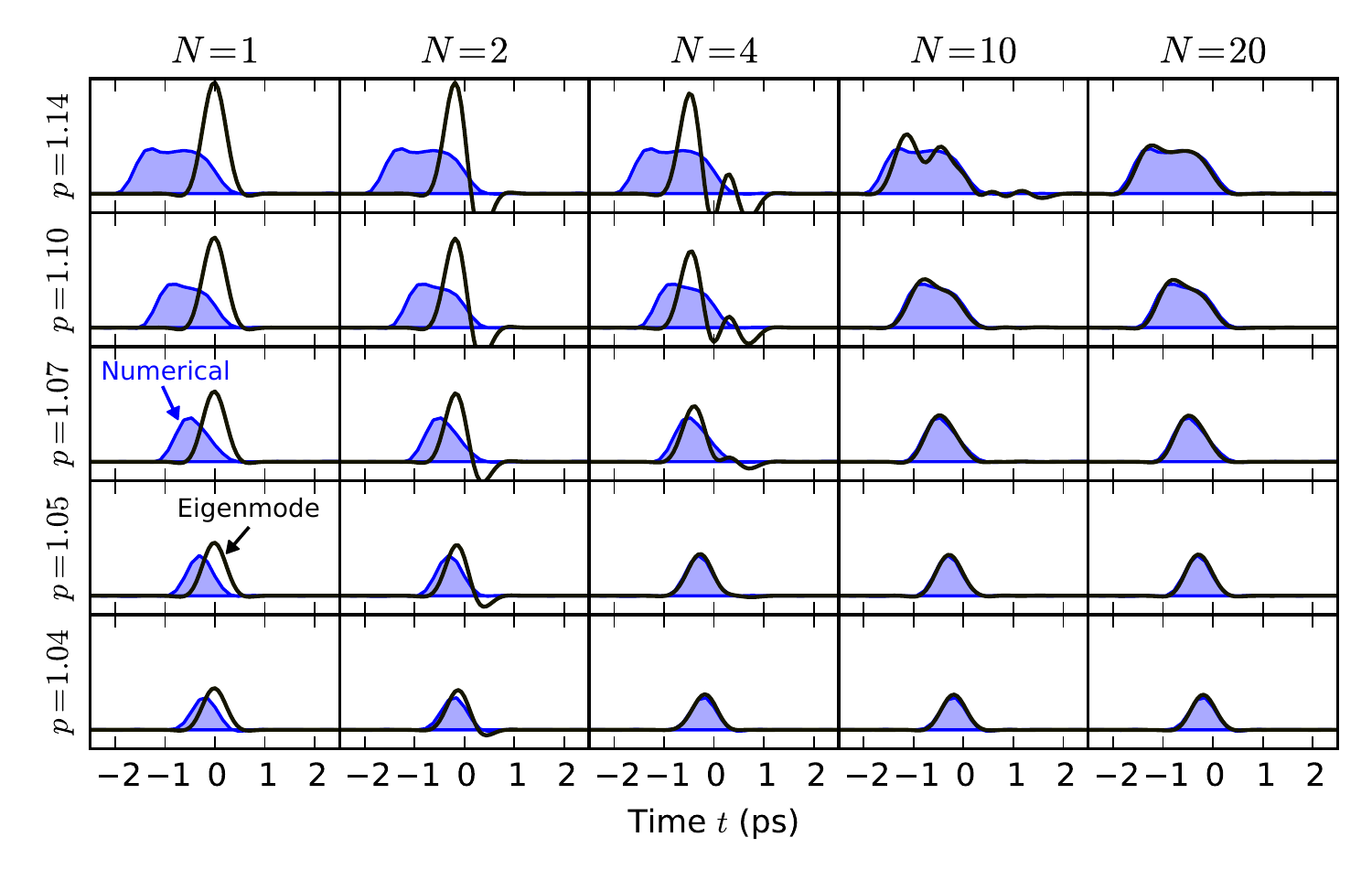}
\caption{Signal pulse shape, computed numerically (filled blue) and with the eigenmode theory (black line).}
\label{fig:10-f12}
\end{center}
\end{figure}

As the number of modes $N$ is increased, the eigenmode model becomes more accurate.  However, the accuracy depends on how far one is from threshold.  The further above threshold, the more modes get excited and the larger $N$ must be to accurately model the OPO.

Figure \ref{fig:10-f11} gives the signal photon number (upon entering the crystal) as a function of pump amplitude.  The colored lines denote results from the eigenmode models, with darker lines for larger values of $N$.  For $N \gtrsim 10$, these lines match the numerical result.

\begin{figure}[t]
\begin{center}
\includegraphics[width=1.00\columnwidth]{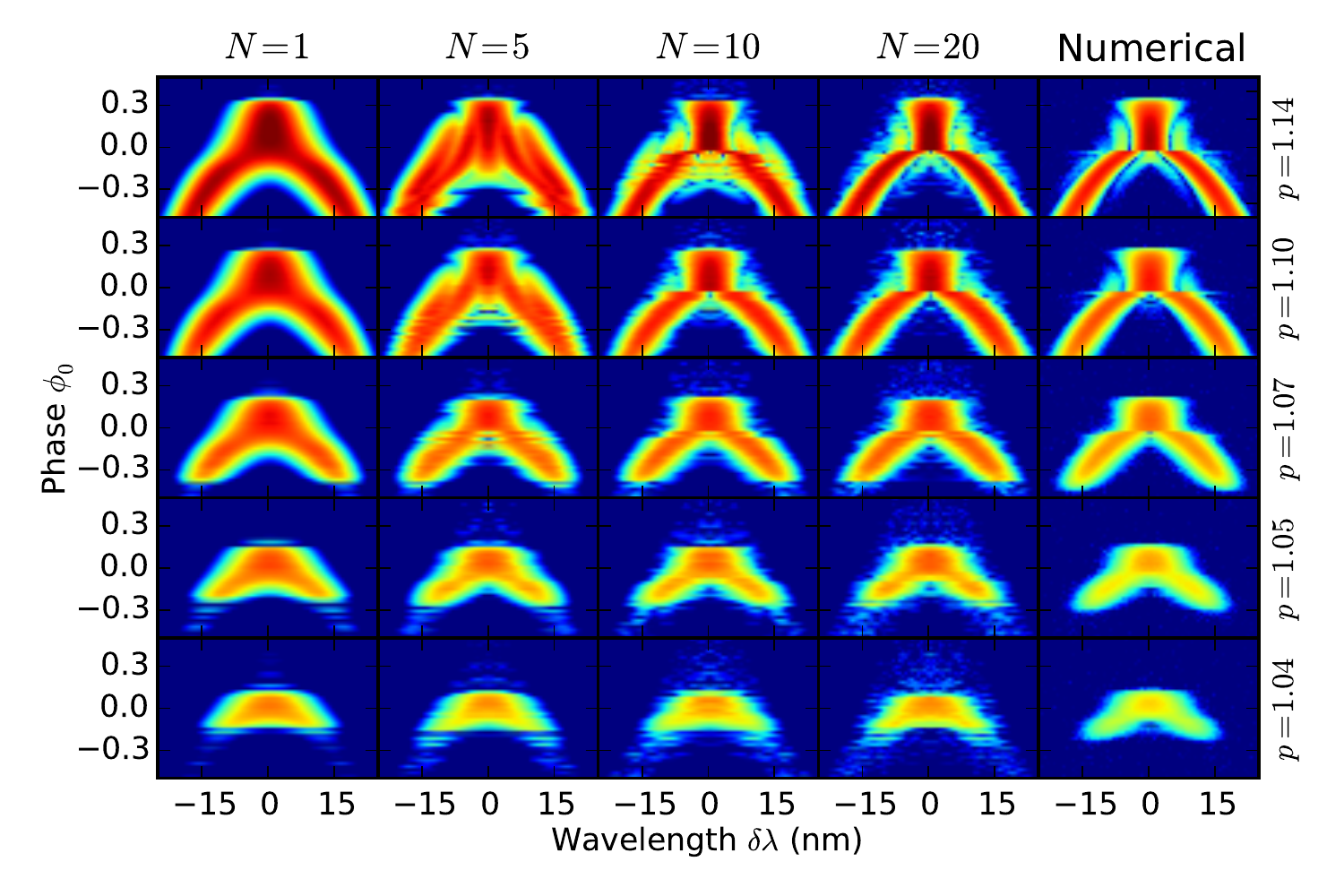}
\caption{Resonance diagrams, computed numerically (right column) and with eigenmode theories of increasing $N$ (left columns).}
\label{fig:10-f13}
\end{center}
\end{figure}

Likewise, the eigenmode model does a good job predicting the steady-state signal pulse shape, provided that enough modes are used.  Figure \ref{fig:10-f12} compares the actual pulse shapes with the eigenmode model.  A linearized treatment would predict a signal centered at the maximum of the gain-clipping function (black curve, left column), but a combination of pump depletion and walkoff push it to the left.  This ``simulton acceleration'' term (see Sec.~\ref{sec:10-simulton}) can be treated to first order in an $N=2$ model, which predicts the centroid drift up to about $p = 1.06$.  Beyond that point, the pulse becomes increasingly elongated and more and more modes must be included to describe it.

This effect can also be seen in the resonance diagrams in Fig.~\ref{fig:10-f13}.  As in Fig.~\ref{fig:10-f2}, these give the power spectrum as a function of cavity phase.  All such diagrams show the same general shape, but as the power is increased, the numerical plot acquires finer structure.  This structure is only reproduced if enough modes are kept in the eigenmode expansion, and with insufficient modes, agreement is quite poor.

\section{Sech-pulse Ansatz}
\label{sec:10-simulton}

A common way to model pulse propagation is to assume that the pulse maintains a given shape, and obtain equations of motion for its parameters using manifold projection or Lagrangian techniques \cite{AgrawalBook}.  The eigenmode model of Sec.~\ref{sec:10-nonlinear} is an example of {\it linear} projection, where $a(t)$ is projected onto a linear subspace spanned by the $a_k(t)$.  Unfortunately, this model required many modes in order to reproduce the full OPO dynamics.

This section studies the pulsed OPO using {\it nonlinear} manifold projection onto the space of sech-like pulses

\beq
	a(z,t) = \frac{A(z)}{\sqrt{2\tau}} \mbox{sech}\bigl((t - T(z))/\tau(z)\bigr) \label{eq:10-simulton}
\eeq

The sech pulse is a natural choice because of its relation to the $\chi^{(2)}$ simulton, a bright signal soliton which co-propagates with a dark pump soliton \cite{Akhmanov1968, Trillo1996}.  In fact, if we take, Eq.~(\ref{eq:10-intdiff}) and assume in the weak gain per walkoff length
\bea
    \frac{\partial a(z,t)}{\partial z} & = & -\frac{1}{2} \alpha_a a(z, t) + \epsilon\,a^*(z,t) b_{\rm in}(t - u z)e^{-\alpha_b z/2} \nonumber \\
    & & - \frac{\epsilon^2}{2u} a^*(z,t)\int_{-\infty}^{t}{a(z, t')^2 \d t'} \label{eq:10-intdiff-2}
\eea
then for a flat-top pump, the sech pulse maintains its shape as it propagates through the waveguide.  This observation suggests that, absent other effects, sech pulses should naturally form in PPLN-waveguide SPOPOs, particularly when a flat-top pump is used.  This view is corroborated by the eigenmode model, which gives a nearly sech-shaped pulse in the degenerate regime (Fig.~\ref{fig:10-f14} below) as well as the sech-shaped spectra in experimental data (Fig.~\ref{fig:10-f2}, see also Refs.~\cite{Marandi2016, Jankowski2016})

In this section, we begin with the sech-pulse ansatz (\ref{eq:10-simulton}) and obtain equations of motion for the parameters $A, T, \tau$ (Sec.~\ref{sec:10-ansatz}) and perturbation terms due to gain clipping and dispersion (Sec.~\ref{sec:10-simpert}).  The near-threshold limit is discussed (\ref{sec:10-simgc}) and the sech waveform is compared to first-order eigenmode.  Finally, we compare predictions of the sech-pulse theory to numerical simulations (Sec.~\ref{sec:10-simnum}).

\subsection{Ansatz and Equations of Motion}
\label{sec:10-ansatz}

Assume the simulton-like sech solution (\ref{eq:10-simulton}).  This confines the field $a(t)$ to a 3-dimensional manifold in the state space.  This solution has three free parameters: amplitude $A$ (normalized so that $|A|^2$ is the photon number), centroid $T$ and width $\tau$.  We obtain the reduced model by projecting equation of motion (\ref{eq:10-intdiff-2}) onto the manifold (\ref{eq:10-simulton}).  Projection requires an inner-product, so we use $\ip{f}{g} = \int{f(t)g(t)\d t}$.  Each of the three variables $\xi \in \{A, T, \tau\}$ evolves according to the projection rules:
\beq
	\frac{\d\xi}{\d z} = \frac{\int{\partial_\xi a\,\partial_z a\,\d t}}{\int{\partial_\xi a\,\partial_\xi a\,\d t}} \label{eq:10-manproj}
\eeq
where $\partial_\xi a$, computed from the ansatz (\ref{eq:10-simulton}), is the tangent vector along $\xi$, and $\partial_z a$ is computed from (\ref{eq:10-intdiff-2}) \cite{Mabuchi2008b, VanHandel2005}.  The equations for $A, T, \tau$ are:

\begin{eqnarray}
    \!\!\frac{\d A}{\d z} & \!\!=\!\! & \left[\int{\gamma(t, z) \frac{\mbox{sech}^2(\tfrac{t-T}{\tau})}{2\tau}\d t} - \frac{\epsilon^2}{4u} A^2\right] A \label{eq:10-sim-da} \\
    \!\!\frac{\d T}{\d z} & \!\!=\!\! & - \tau \frac{\epsilon^2}{4u} A^2 + \frac{3}{2} \int{\gamma(t, z) \mbox{sech}^2(\tfrac{t-T}{\tau})\mbox{tanh}(\tfrac{t-T}{\tau}) \d t} \label{eq:10-sim-dt} \\
    \!\!\frac{\d\tau}{\d z} & \!\!=\!\! & \frac{18}{3+\pi^2} \int{\gamma(t, z) \left[\tfrac{t-T}{\tau} \tanh(\tfrac{t-T}{\tau}) - \tfrac{1}{2}\right] \mbox{sech}^2(\tfrac{t-T}{\tau}) \d t} \nonumber \\ \label{eq:10-sim-dtau}
\end{eqnarray}
where $\gamma(t, z) = \epsilon\,b_{\rm in}(t-uz)e^{-\alpha_b z/2} - \alpha_a/2$.

Three effects come into play here: gain, gain-clipping, and pump depletion.  As in Sec.~\ref{sec:10-linear}, we separate the continuous-wave dynamics from gain-clipping: first we solve the equations of motion assuming a constant-pump gain $\gamma(t, z) \rightarrow \epsilon \bar{b} - \alpha_a/2$, then treat deviations using perturbation theory.  We also add dispersion terms as a perturbations.  The solution will take the form:
\beq
	A = A_0 + \delta A,\ \ \ 
	T = T_0 + \delta T,\ \ \ 
	\tau = \tau_0 + \delta\tau
\eeq
where $A_0, T_0, \tau_0$ satisfy the continuous-wave, lossless equations and $\delta A, \delta T, \delta\tau$ are the gain-clipping and dispersion perturbation terms.

Taking Eqs.~(\ref{eq:10-sim-da}-\ref{eq:10-sim-dtau}) and assuming a constant pump $b(t, z) \rightarrow \bar{b}$, one obtains $\d\tau_0/\d z = 0$ and the equations for $A_0, T_0$:
\beq
    \!\frac{\d A_0}{\d z} = \Bigl[\epsilon\,(\bar{b}_{\rm in} - \tfrac{1}{2}\alpha_a) - \frac{\epsilon^2}{4u} A_0^2\Bigr] A_0,\ \ 
    \frac{\d T_0}{\d z} = -\tau \frac{\epsilon^2}{4u} A_0^2 \label{eq:10-ideal}
\eeq

If the pump field is nearly constant (as is the case with flat pulses or sufficiently long Gaussian pulses) and the waveguide is nearly lossless, $A_0, T_0, \tau_0$ will be a good approximation to the pulse parameters.  The constant pump $\bar{b}$ is chosen to be close to the average value for a CW field of the same peak intensity as $b_{\rm in}(t)$:
\beq
	\bar{b} = \frac{1}{L} \int{b_{\rm max} e^{-\alpha_b z/2} dz} \approx b_{\rm max} e^{-\alpha_b L/4}
\eeq
Solving Eq.~(\ref{eq:10-ideal}) one finds:
\beq
	A_0(z) = \sqrt{\frac{2u(2\epsilon\,\bar{b} - \alpha_a) e^{(2\epsilon\,\bar{b} - \alpha_a)z}}{2u(2\epsilon\,\bar{b} - \alpha_a) + (e^{(2\epsilon\,\bar{b} - \alpha_a)z} - 1) \epsilon^2 A_0(0)^2}}\,A_0(0) \label{eq:10-inout1}
\eeq
At threshold $p = 1$, the constants $\bar{b}$, $u$, $\epsilon$ can be expressed in terms of two experimental parameters: pump intensity $N_{b,0} \approx e^{\alpha_b L/2} \bar{b}_0^2$ and waveguide gain $G_0 = e^{(2\epsilon\bar{b}_0-\alpha_a)L}$ at threshold (Table~\ref{tab:10-t1}).  The pump amplitude is proportional to $p$, so $\bar{b} = p \bar{b}_0$.  Making the substitutions $N_b = p^2\,N_{b,0}$, $G = G_0^{p} e^{(p-1)\alpha_a L}$, we rewrite Eq.~(\ref{eq:10-inout1}) as:
\beq
	A_0(z) = \left[\frac{G^{z/L}}{1 + \left(G^{z/L} - 1\right) \frac{\log(G e^{\alpha_a L})^2}{\log(G)} \frac{A_0(0)^2}{8 N_{b} e^{-\alpha_b L/2}}}\right]^{1/2} A_0(0) \label{eq:10-aout-1}
\eeq
Combining the first two equations in (\ref{eq:10-ideal}), we can obtain the centroid shift $T$ in terms of the amplitude:
\beq
	T_0(z) = T_0(0) - \tau\,\log\left(\frac{G^{z/2L}}{A(z)/A(0)}\right)% = T_0(0) - \frac{\tau}{2} \log\left[1 + \bigl(G^{z/L} - 1\bigr) \frac{\log(G e^{\alpha_a L})^2}{\log(G)} \frac{A_0(0)^2}{8 N_{b} e^{-\alpha_b L/2}}\right] 
	\label{eq:10-tout-1}
\eeq

Eqs.~(\ref{eq:10-aout-1}-\ref{eq:10-tout-1}) govern the pulse evolution in the presence of a CW pump.  The width $\tau$ does not change.  Note that the pump depletion shifts the centroid of the pulse in addition to reducing its gain.  This {\it simulton acceleration} is caused by pump-signal walkoff: as the pulse walks through the pump, the leading side experiences gain from the undepleted pump while the gain on the trailing side is depleted, shifting the centroid forward.  

\subsection{Perturbations}
\label{sec:10-simpert}

\subsubsection{Gain-Clipping Terms}

Gain clipping gives rise to perturbations in $A$, $T$ and $\tau$.  To find these, we first rewrite (\ref{eq:10-sim-da}-\ref{eq:10-sim-dtau}) as:
\begin{eqnarray}
    \frac{\d(\delta A/A_0)}{\d z} & = & -\frac{\epsilon^2}{2u} A_0^2 (\delta A/A_0) + g(T,\tau,z) \label{eq:10-pert1} \\
    \frac{\d (\delta T)}{\d z} & = & \frac{3\tau_0^2}{2} \frac{\partial g(T_0,\tau_0,z)}{\partial T_0} \label{eq:10-pert2} \\
    \frac{\d (\delta\tau)}{\d z} & = & \frac{18\tau_0^2}{3+\pi^2} \frac{\partial g(T_0,\tau_0,z)}{\partial\tau_0} \label{eq:10-pert3}
\end{eqnarray}
where $g(T,\tau,z)$ is the differential gain-clipping function of the sech-pulse, defined by:
\beq
	g(T, \tau, z) = \int{\epsilon(b_{\rm in}(t-uz)e^{-\alpha_b z/2} - \bar{b}) \frac{\mbox{sech}^2((t-T)/\tau)}{2\tau}\d t}
\eeq
Up to a constant, this is the convolution of the pump $b_{\rm in}(t-uz)$ and sech intensity $(2\tau)^{-1}\mbox{sech}^2((t-T)/\tau)$.

Equations (\ref{eq:10-pert1}-\ref{eq:10-pert3}) can be integrated to give:
\begin{eqnarray}
	\!\!\!\!\!\!\delta A(z) & = & A_0(z) \int_0^z{g(T_0,\tau_0,z') \frac{(A_0(z)/A_0(0))^2}{G^{z/L}}\d z} %\nonumber \\
	%& = &  A_0(z) \int_0^z{g(T_0,\tau_0,z') \frac{1 + \bigl(G^{z'/L} - 1\bigr) \frac{\log(G e^{\alpha_a L})^2}{\log(G)} \frac{A_0(0)^2}{8 N_{b} e^{-\alpha_b L/2}}}{1 + \bigl(G^{z/L} - 1\bigr) \frac{\log(G e^{\alpha_a L})^2}{\log(G)} \frac{A_0(0)^2}{8 N_{b} e^{-\alpha_b L/2}}}\d z'} 
	\label{eq:10-p-da} \\
	\!\!\!\!\!\!\delta T(z) & = & \frac{3\tau_0^2}{2} \int_0^z{\frac{\partial g(T_0,\tau_0,z')}{\partial T_0}\d z'} \label{eq:10-p-dt} \\
	\!\!\!\!\!\!\delta \tau(z) & = & \frac{18\tau_0^2}{3+\pi^2} \int_0^z{\frac{\partial g(T_0,\tau_0,z')}{\partial\tau_0}\d z'} \label{eq:10-p-dtau}
\end{eqnarray}
	
Equation (\ref{eq:10-p-da}) gives the gain-clipping correction to the linear gain.  Although the full form is complicated, it simplifies in the near-threshold regime, where the fraction on the right side of the integral can be ignored.  Equations (\ref{eq:10-p-dt}-\ref{eq:10-p-dtau}) can be simplified if we assume that $T$ and $\tau$ change slowly enough in a single round-trip that we can replace them inside the integral by their initial values.  The input-output relations become:
\begin{eqnarray}
	\delta A(z) & = & A_0(z) G(T_0, \tau_0) \label{eq:10-p2-da} \\
	\delta T(z) & = & \frac{3\tau_0^2}{2} \frac{\partial G(T_0, \tau_0)}{\partial T_0} \label{eq:10-p2-dt} \\
	\delta \tau(z) & = & \frac{18\tau_0^2}{3+\pi^2} \frac{\partial G(T_0, \tau_0)}{\partial\tau_0} \label{eq:10-p2-dtau}
\end{eqnarray}
where
\beq
	G(T, \tau) = \int_0^L {g(T-uz,\tau,z)\d z} \label{eq:10-gttau}
\eeq
is the integrated sech-pulse gain-clipping function.  Up to a constant factor and offset, it is equal to the convolution of the the gain-clipping function $G(t)$ from (\ref{eq:10-gcf}) and the sech waveform.

Combining Eqs.~(\ref{eq:10-aout-1}-\ref{eq:10-tout-1}, \ref{eq:10-p2-da}-\ref{eq:10-p2-dtau}), one obtains the full PPLN input-output relations accounting for both gain-clipping and pump depletion:
\begin{eqnarray}
	\!\!A_{\rm out} & \!=\! & \left[\frac{G}{1 + \left(G - 1\right) \frac{\log(G e^{\alpha_a L})^2}{\log(G)} \frac{A_{\rm in}^2}{8 N_{b} e^{-\alpha_b L/2}}}\right]^{1/2} \nonumber \\
	& & \times \bigl(1 + G(T_{\rm in}, \tau_{\rm in})\bigr)A_{\rm in} \label{eq:10-inout-a} \\
	\!\!T_{\rm out} & \!=\! & T_{\rm in} \!-\! \tau \log\left(\frac{G^{1/2}}{A_{\rm out}/A_{\rm in}}\right) + \frac{3\tau_{\rm in}^2}{2} \frac{\partial G(T_{\rm in},\tau_{\rm in})}{\partial T_{\rm in}}\ \  \label{eq:10-inout-t} \\
	\!\!\tau_{\rm out} & \!=\! & \tau_{\rm in} + \frac{18\tau_{\rm in}^2}{3+\pi^2} \frac{\partial G(T_{\rm in},\tau_{\rm in})}{\partial\tau_{\rm in}} \label{eq:10-inout-tau}
\end{eqnarray}

\subsubsection{Dispersion and Detuning}

Following Sec.~\ref{sec:10-linear}, we employ the lumped-element model to treat dispersion, since the pulse shape changes only slightly between round trips and dispersion is a linear effect that does not depend on the pulse amplitude.  Restricting ourselves to the degenerate regime $\phi_0\phi'_2 > 0$ where we expect to see simulton-like solutions and following (\ref{eq:10-deg-eig}), we have:
\beq
	\Delta a(t)\bigr|_{\rm dispersion} = \frac{\phi'_2\tan\phi_0}{2} \frac{\d^2 a(t)}{\d t^2} - \frac{(\phi'_2\sec\phi_0)^2}{8} \frac{\d^4 a(t)}{\d t^4} \label{eq:10-disp}
\eeq
where $\phi_0$ is the round-trip phase and $\phi'_2$ is the total (PPLN plus fiber) dispersion.  We enforce the simulton-like form (\ref{eq:10-simulton}) by projecting (\ref{eq:10-disp}) onto the 3-dimensional sech-pulse manifold.  As before, each of the three variables $A, T, \tau$ changes according to Eq.~(\ref{eq:10-manproj}).  Performing the necessary integrals, one finds:
\begin{eqnarray}
	\Delta A & = & \left[-\frac{1}{3} \frac{\phi'_2\tan\phi_0}{2} \tau^{-2} - \frac{7}{15} \frac{(\phi'_2\sec\phi_0)^2}{8} \tau^{-4} \right] A \\
	\Delta \tau & = & \frac{12}{3+\pi^2} \frac{\phi'_2\tan\phi_0}{2} \tau^{-1} + \frac{168}{5(3+\pi^2)}\frac{(\phi'_2\sec\phi_0)^2}{8} \tau^{-3} \nonumber \\
\end{eqnarray}
Higher-order effects such as third-order dispersion and $\chi^{(3)}$ are not included here, but could also be treated with this perturbation theory.  GVD gives no centroid shift.  However, there is a nonzero $\Delta T$ due to cavity detuning: $\Delta T = (\lambda/2c)\ell$.  Combining these with Eqs.~(\ref{eq:10-inout-a}-\ref{eq:10-inout-tau}) and adding a loss factor $G_0^{-1}$, one obtains round-trip propagation equations for $A, T, \tau$ in the OPO:

%Cavity detuning is easy to incorporate.  Let's suppose that the cavity detuning length is $\ell\lambda/2c$.  This shifts the pulse to the right for positive $\ell$, by an amount:
%\beq
%	\Delta T = \frac{\lambda}{2c}\ell
%\eeq
%Combining these with Eqs.~(\ref{eq:10-inout-a}-\ref{eq:10-inout-tau}) and adding a loss $G_0^{-1}$, one obtains round-trip propagation equations for $A, T, \tau$ in the OPO:

\begin{widetext}

\begin{eqnarray}
	A & \rightarrow & \left[1 + G(T, \tau) - \frac{1}{3} \frac{\phi'_2\tan\phi_0}{2} \tau^{-2} - \frac{7}{15} \frac{(\phi'_2\sec\phi_0)^2}{8}\tau^{-4} \right] \left[\frac{G/G_0}{1 + \left(G - 1\right) \frac{\log(G e^{\alpha_a L})^2}{\log(G)} \frac{A^2}{8 N_{b} e^{-\alpha_b L/2}}}\right]^{1/2} A \label{eq:10-prop-a} \\
	T & \rightarrow & T + \frac{\lambda}{2c}\ell - \frac{\tau}{2} \log\left[1 + (G - 1) \frac{\log(G e^{\alpha_a L})^2}{\log(G)} \frac{A^2}{8 N_{b} e^{-\alpha_b L/2}}\right] + \frac{3\tau^2}{2} \frac{\partial G(T,\tau)}{\partial T} \label{eq:10-prop-t} \\
	\tau & \rightarrow & \tau + \frac{18\tau^2}{3+\pi^2} \frac{\partial G(T,\tau)}{\partial\tau_{\rm in}} + \frac{12}{3+\pi^2} \frac{\phi'_2\tan\phi_0}{2} \tau^{-1} + \frac{168}{5(3+\pi^2)} \frac{(\phi'_2\sec\phi_0)^2}{8} \tau^{-3} \label{eq:10-prop-tau}
\end{eqnarray}

\end{widetext}

\subsection{Near-Threshold Limit}
\label{sec:10-simgc}

Near threshold, the sech-pulse model should match the eigenmode model derived in Sec.~\ref{sec:10-linear}.  In that limit, we can truncate all of the nonlinear gain terms in (\ref{eq:10-prop-a}-\ref{eq:10-prop-tau}) at third order and replace $G \rightarrow G_0$, the at-threshold gain.  In addition, supposing a flat-top pump pulse, the gain-clipping function becomes $G(t) = -\tfrac{1}{2}|t/T_p| \log(G_0 e^{\alpha_a L})$.  Using Eq.~\ref{eq:10-gttau}, $G(T, \tau)$ becomes:
\beq
	G(T, \tau) = -\frac{\tau}{2T_p} \log(G_0 e^{\alpha_a L})\log\bigl[2\cosh(T/\tau)\bigr]
\eeq
This function is maximized for $T = 0$, i.e.\ for a signal pulse located at the trailing edge of the pump (Fig.~\ref{fig:10-f4}).  Since $A, T, \tau$ change slowly on each round trip, we can convert (\ref{eq:10-prop-a}-\ref{eq:10-prop-tau}) to a differential equation analogous to (\ref{eq:10-dc-cont}); performing the near-threshold substitutions, we obtain:

\begin{eqnarray}
	\!\!\!\!\!\frac{\d A}{\d n} & \!\!=\!\! & \left[\frac{p-1}{2}\log(G_0 e^{\alpha_a L}) \!-\! \frac{\log(G_0 e^{\alpha_a L})}{2T_p} \log\bigl[2\cosh(T/\tau)\bigr]\tau \right. \nonumber \\
	& & \ \left.- \frac{1}{3} \frac{\phi'_2\tan\phi_0}{2\tau^{2}} - \frac{7}{15} \frac{(\phi'_2\sec\phi_0)^2}{8\tau^{4}}\right] A - \beta A^3 \label{eq:10-prop-gc-a}\\
	\!\!\!\!\!\frac{\d T}{\d n} & \!\!=\!\! & \frac{\lambda}{2c}\ell - \tau\beta A^2 - \frac{3\tau^2}{4 T_p} \log(G_0 e^{\alpha_a L}) \tanh(T/\tau) \label{eq:10-prop-gc-t} \\
	\!\!\!\!\!\frac{\d\tau}{\d n} & \!\!=\!\! & \frac{18}{3+\pi^2} \frac{\log(G_0 e^{\alpha_a L})}{2T_p} \left[\log\bigl[2\cosh(\tfrac{T}{\tau})\bigr] \!-\! \tfrac{T}{\tau} \tanh(\tfrac{T}{\tau})\right] \tau^2 \nonumber \\
	& & \ + \frac{12}{3+\pi^2} \frac{\phi'_2\tan\phi_0}{2\tau} + \frac{168}{5(3+\pi^2)} \frac{(\phi'_2\sec\phi_0)^2}{8\tau^3} \label{eq:10-prop-gc-tau}
\end{eqnarray}

Most of these terms make intuitive sense.  For the $A$ equation, the $p-1$ term is the CW gain and the $O(\tau)$, $O(\tau^{-1})$ and $O^(\tau^{-3})$ terms account for gain clipping and dispersion, which reduce the overall gain of the signal.  An $O(A^3)$ term accounts for pump depletion in the near-threshold limit; $\beta$ is given by
\beq
    \beta = \frac{e^{\alpha_b L/2} (G_0 - 1) \log(G_0 e^{\alpha_a L})^2}{16 N_{b,0} \log G_0}
\eeq
which matches Eq.~(\ref{eq:10-betapsi}) from the eigenmode theory.

% Note -- I think this is irrelevant right now -- need to come back to it...
\comment{

As equation (\ref{eq:10-prop-gc-t}) shows, gain clipping and simulton acceleration both shift the center of the pulse.  In steady state, these effects cancel out, giving a steady-state centroid shift:

\beq
	T = -\tau \tanh^{-1}\left[\frac{4 T_p \beta A^2}{3\tau \log(G_0 e^{\alpha_a L})}\right]
\eeq

The sech pulse is only stable if the argument in the $\tanh^{-1}$ lies in $(-1, 1)$.  Otherwise it will run away and decay to zero.  Not included here is the effect of cavity detuning.  Detuning can either enhance or counteract the simulton acceleration.

}
% Note [above] -- I think this is irrelevant right now -- need to come back to it...

Equation (\ref{eq:10-prop-gc-tau}) lets us compute the pulse width.  The $O(\tau^2)$ gain-clipping term is compensated by the $O(\tau^{-1})$, $O(\tau^{-3})$ dispersion terms.  Working at $\phi_0 = 0$ and close enough to threshold that the simulton acceleration can be neglected ($T = 0$), one finds the steady-state pulse width:
\beq
	\tau_{\rm sech} = \left(\frac{7}{15} \frac{(\phi'_2)^2 T_p}{\log(G_0 e^{\alpha_a L})\log 2}\right)^{1/5} \label{eq:10-tau-zero}
\eeq
Gain-clipping theory says that signal pulses at $\phi_0 = 0$ are given by combinations of hypergeometric functions (Sec.~\ref{sec:10-emshapes}): $a(t) \sim f(t/\tau_L)$, where $\tau_L = \left((\phi'_2)^2 T_p/4\log(G_0 e^{\alpha_a L})\right)^{1/5}$.  Comparing to (\ref{eq:10-tau-zero}), we find $\tau_{\rm sech} = 1.21\tau_L$.

In the degenerate $\phi_0 \neq 0$ limit, the $\tau^{-1}$ term in (\ref{eq:10-prop-gc-tau}) dominates and the steady-state pulse width is:
\beq
	\tau_{\rm sech} = \left(\frac{2 T_p \phi'_2\tan\phi_0}{3\log(G_0 e^{\alpha_a L})\log 2}\right)^{1/3} \label{eq:10-tau-nz}
\eeq
This result should be compared to the eigenmode model, in which the pulse shape is given by an Airy function $\mbox{Ai}(t/\tau_{\rm Ai} + \mbox{const})$, with the time constant given by $\tau_{\rm Ai} = \bigl(T_p \phi'_2\tan\phi_0/\log(G_0 e^{\alpha_a L})\bigr)^{1/3}$.  We find that $\tau_{\rm sech} = 0.987 \tau_{\rm Ai}$.

\begin{figure}[b!]
\begin{center}
% From SimultonPerturbations-Fig4.pdf
\includegraphics[width=0.83\columnwidth]{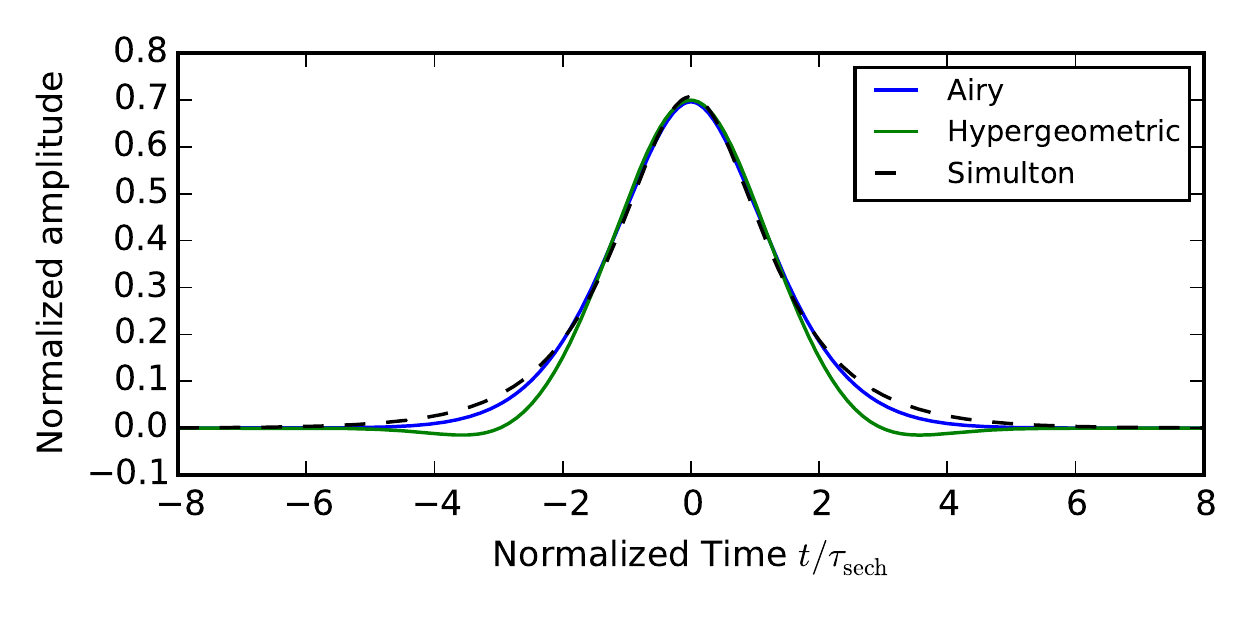}
\caption{Plot of the simulton solution $(2\tau)^{-1/2}\mbox{sech}(t/\tau_{\rm sech})$ against the Airy (Eq.~\ref{eq:10-eig-airy}) and hypergeometric (Eq.~\ref{eq:10-phizero-ak}) eigenfunctions.}
\label{fig:10-f14}
\end{center}
\end{figure}

Although the pulse widths $\tau_{\rm Ai}$, $\tau_{L}$ and $\tau_{\rm sech}$ differ, the respective functions have different shapes, so that the pulse waveforms predicted by eigenmode and simulton theory happen to lie right on top of each other, and their full-width half-maxima agree to a few percent (Fig.~\ref{fig:10-f14}).

\subsection{Comparison to Numerics}
\label{sec:10-simnum}

Numerical simulations for the waveguide OPO show that the simulton model is accurate when the OPO exhibits degenerate, singly-peaked behavior.  This happens in a limited range of circumstances:

\begin{enumerate}
	\item Power: The pulse is sech-shaped near threshold.  Far above threshold, pulses become box-shaped and are better described by the theory in Sec.~\ref{sec:10-boxpulse}.
	\item Phase: One must be near the center of a detuning peak ($\phi_0 \approx 0$) to use the simulton description.  Far from the center for $\phi_0\phi'_2 < 0$, the pulse that resonates starts to resemble a nondegenerate pulse, which is not described by a sech-pulse.
	\item Detuning: The cavity detuning $\ell$ cannot be too large; otherwise the sech-pulse goes unstable and the field amplitude starts to oscillate.
\end{enumerate}

\subsubsection{Steady-State Behavior}

The sech-pulse model does a good job predicting the pulse shape near threshold, provided that the oscillating mode is degenerate.  For $\phi_0 = 0$ or $\phi_0$ sufficiently large, Eqs.~(\ref{eq:10-tau-zero}) and (\ref{eq:10-tau-nz}) can be used to get the pulse width, respectively.  For general $\phi_0$, one must solve for the steady-state of (\ref{eq:10-prop-gc-tau}).  (Near threshold one can take $T \rightarrow 0$ in that equation, resulting in a $5^{\rm th}$-order polynomial in $\tau$.)

\begin{figure}[tbp]
\begin{center}
% From SimultonPerturbations-Fig3b.pdf
% (SimultonPerturbations2.ipynb)
\includegraphics[width=1.0\columnwidth]{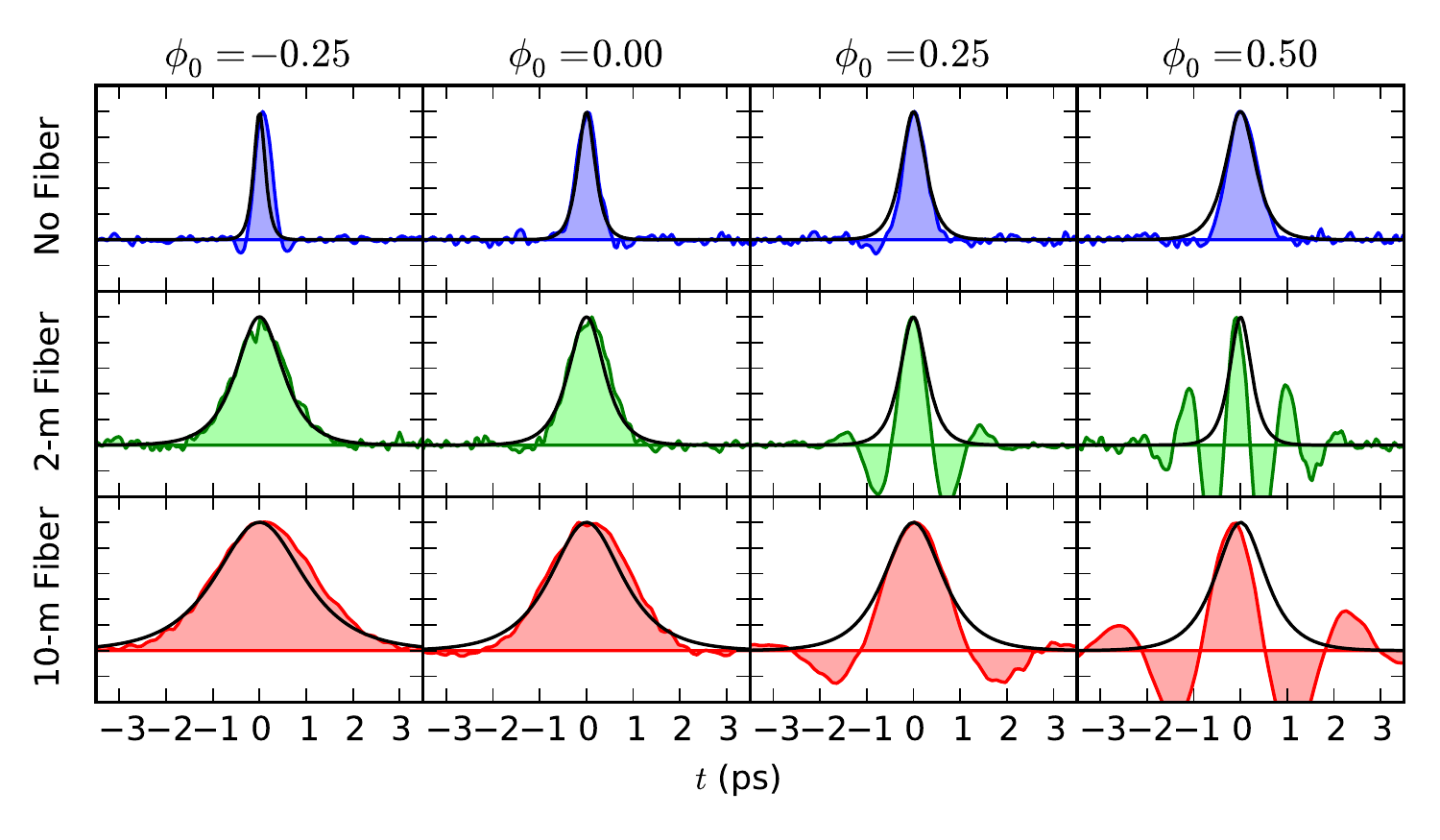}
\caption{Near-threshold pulse shape computed numerically (filled curve), compared to the steady-state sech solution (black line).}
\label{fig:10-f15}
\end{center}
\end{figure}

However, as Figure \ref{fig:10-f15} shows, one cannot use the sech-pulse model when the OPO oscillates nondegenerately.  Also, Eqs.~(\ref{eq:10-prop-gc-a}-\ref{eq:10-prop-gc-tau}) must be modified when dispersion compensation is used to set $\phi'_2 \rightarrow 0$, and higher-order dispersion must be taken into account.  Since dispersion is treated as a lumped element here, this causes the pulse width to shrink to zero (as in Sec~\ref{sec:10-emshapes}).  An OPO with dispersion compensation must be studied numerically or with the eigenmode model, or a more careful approach must be taken, avoiding lumping the dispersion into one element.  In the dispersion-engineered limit where both $\beta_2$ and $\phi_2$ are zero, one must go further and include higher-order dispersion terms.

At threshold, the pulse is centered at the point of maximum gain.  As the pump increases and the amplitude grows, the simulton acceleration causes its centroid to drift towards negative $T$.  In the absence of detuning $\ell = 0$, a steady state is reached in (\ref{eq:10-prop-gc-t}) when $\beta A^2 = (3\tau^2/4T_p) \log(G_0 e^{\alpha_a L}) \tanh(T/\tau)$.  One can replace $\beta A^2 \rightarrow \tfrac{1}{2}(p-1)\log(G_0 e^{\alpha_a L})$ by making the assumption that those two terms are dominant in the amplitude equation (\ref{eq:10-prop-gc-a}).  Assuming a small $T$ and expanding the hyperbolic tangent, we get:

\beq
	T = \frac{2(p-1) T_p}{3} \label{eq:10-delay}
\eeq

\begin{figure}[tbp]
\begin{center}
% From SimultonEquations-Fig2.pdf
\includegraphics[width=1.0\columnwidth]{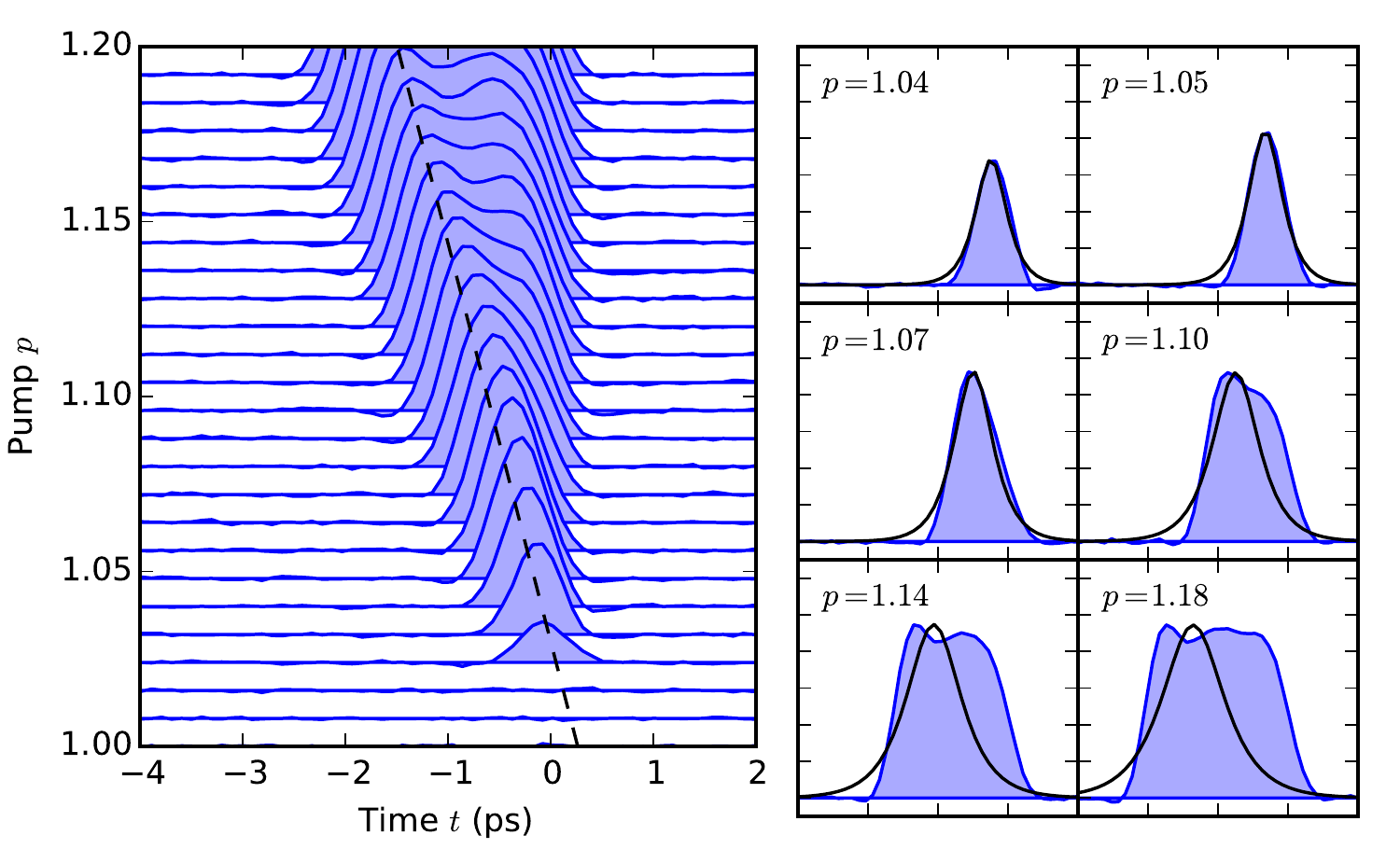}
\caption{Plot of the simulated pulse shape (filled), compared to the simulton steady-state of Eqs.~(\ref{eq:10-prop-a}-\ref{eq:10-prop-tau}) (black line).  Dashed line is the relation (\ref{eq:10-delay}).}
\label{fig:10-f16}
\end{center}
\end{figure}

To go beyond this approximation, one must simulate Eqs.~(\ref{eq:10-prop-a}-\ref{eq:10-prop-tau}) or (\ref{eq:10-prop-gc-a}-\ref{eq:10-prop-gc-tau}) numerically.  Figure (\ref{fig:10-f16}) compares numerical data against the simulton model for the free-space PPLN OPO.  The pulse shape matches the sech form well in the linear regime, and continues to match reasonably well as the pulse is displaced from the maximum-gain point.  However, at high pump powers its shape becomes deformed and it begins to resemble a flat-top pulse.

In Sec.~\ref{sec:10-nl-num}, we made a similar comparison with the eigenmode theory.  Figs.~\ref{fig:10-f16} and \ref{fig:10-f12} are computed for the same OPO system, allowing a direct comparison.  We see that for these OPO parameters, the simulton model is accurate up to about $p = 1.10$, does better than the $N=4$ eigenmode model, but not as good as $N=10$.

\begin{figure}[tbp]
\begin{center}
% From SimultonEquations-Fig1-phase.pdf
\includegraphics[width=1.00\columnwidth]{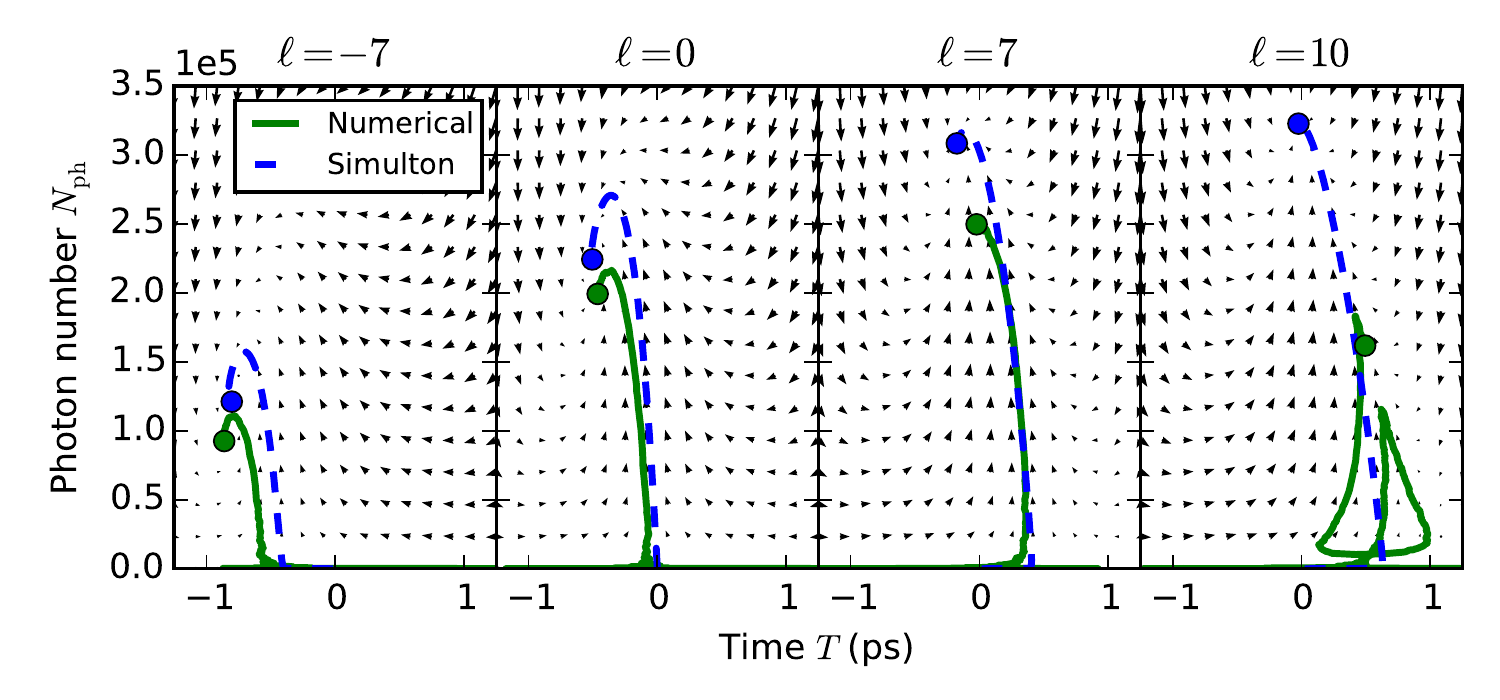}
\caption{Evolution of pulse photon number $N_{\rm ph}(t)$ and centroid $T(t)$ for sech-pulse model (dashed lines) and full numerics (solid).  Four different detuning values shown.  PPLN OPO, no fiber.}
\label{fig:10-f17-phase}
\end{center}
\end{figure}

\subsubsection{Transient Behavior}

We now consider the evolution of the signal using the Sech-pulse ansatz, assuming the pump is turned on abruptly.  In the absence of detuning, the pulse first grows at the maximum-gain point, as per the linear model.  Once pump depletion becomes significant, the pulse shifts forward, reaching an equilibrium when its amplitude saturates.  Both the simulation and simulton model agree here (Fig.~\ref{fig:10-f17-phase}, center-left plot).  This figure visualizes the dynamics with a phase space plot.  The full system is three-dimensional, but the pulse width can be assumed constant, giving a dynamical system with two variables.  This has one attractor, which is a spiral, explaining the initial overshoot in photon number.

This behavior changes with cavity detuning.  For negative detuning ($\ell = -7$, left plot), the pulse first grows at $T < 0$ and is shifted further by the simulton acceleration.  In this case, both detuning and simulton acceleration move the pulse in the same direction, away from the maximum-gain point, so its amplitude is reduced.

In contrast, for positive detuning (center-right plot), simulton acceleration opposes the detuning shift.  When the pulse is weak, the latter is dominant, so it grows at $T > 0$, but once pump depletion kicks in, it eventually drifts back to the maximum-gain point, where simulton acceleration and detuning cancel out.  Not surprisingly, photon number is larger than without detuning.

For a given pump power, the optimal detuning is the one that cancels the simulton acceleration, so that the pulse can be amplified at the maximum-gain point.  This happens when $T = 0$ is a steady state to (\ref{eq:10-prop-gc-t}).  Applying the same substitution to $\beta A^2$, we find:
\beq
	\frac{\ell_{\rm max}\lambda}{2c} = \frac{p-1}{2}\log(G_0 e^{\alpha_a L})\tau \label{eq:10-ellmax}
\eeq
where $\tau$ is computed from (\ref{eq:10-prop-gc-tau}), which becomes independent of the other variables when $T = 0$.  This depends on the pump power; the larger $p-1$, the larger $\ell$ should be to form the optimal signal pulse.  Overshooting gives rise to weaker signal pulses, and can also cause instabilities that suppress the amplitude and are not captured by the simulton model (Fig.~\ref{fig:10-f17-phase}, right plot).

\subsubsection{Detuning and Stability}
\label{sec:10-simstab}

We can see from Figure \ref{fig:10-f17-phase} that the detuning has a substantial effect on the energy of the pulse that forms.  If $\ell$ is not too large, the numerical result matches the simulton description.

\begin{figure}[t]
\begin{center}
\includegraphics[width=1.0\columnwidth]{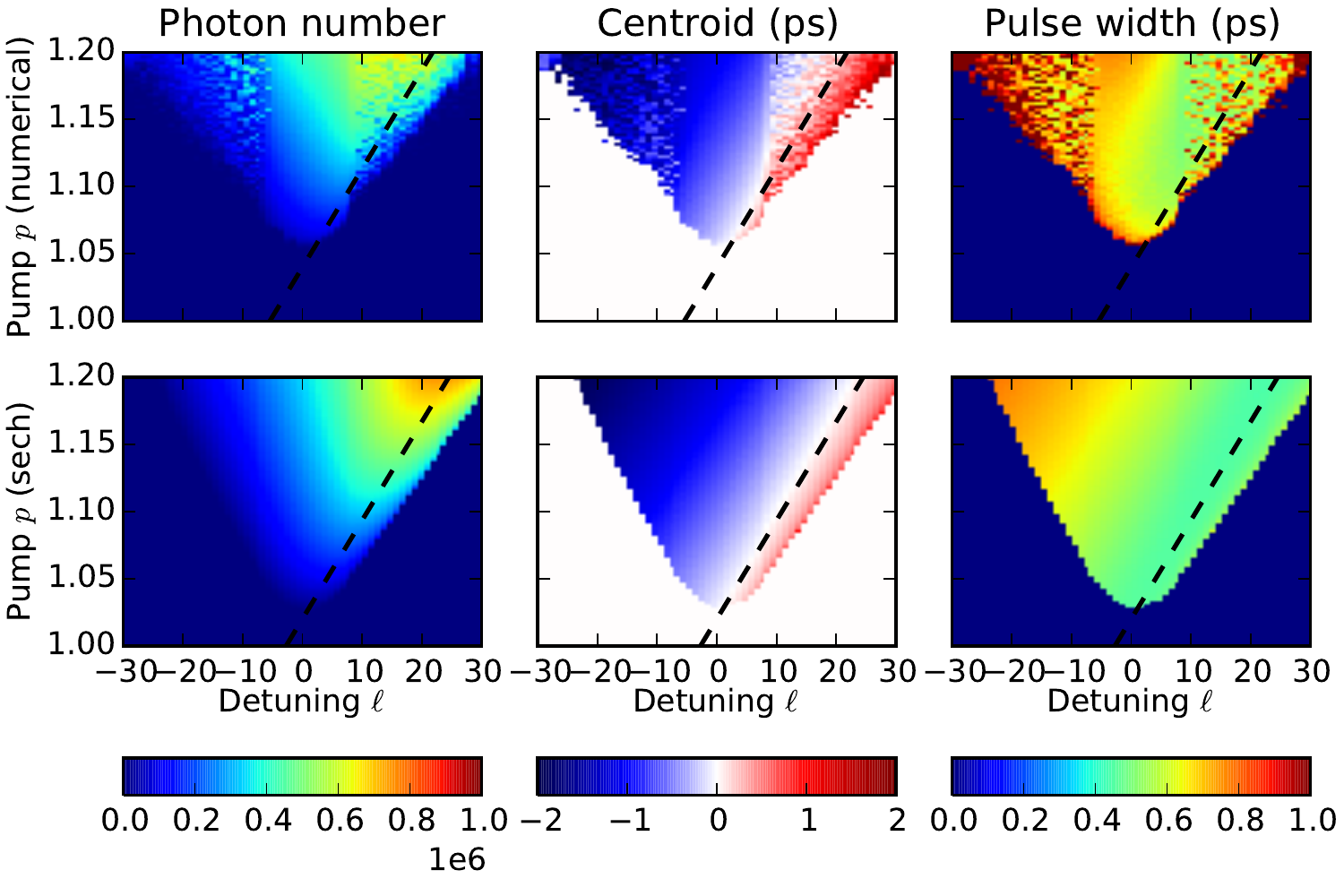}
\caption{Plots of the steady-state photon number (left), centroid (center) and pulse width (right) as a function of detuning $\ell$ and pump $p$.  Numerical simulations (top) are compared with the sech-pulse model (bottom).  Dashed line is Eq.~(\ref{eq:10-ellmax}).  Simulations are for PPLN OPO, 4-m fiber.}
\label{fig:10-f18}
\end{center}
\end{figure}

A more complete way to capture this behavior is to look at the pulse properties as a function of both pump $p$ and detuning $\ell$, as shown in Fig.~\ref{fig:10-f18}.

In the upper plots in Fig.~\ref{fig:10-f18}, several features stand out.  The threshold varies close to linearly with detuning, consistent with the simulton theory (lower plots).  The simulton theory also predicts that when Eq.~(\ref{eq:10-ellmax}) is satisfied, the pulse amplitude is maximized and the pulse width is shortest and the centroid lies at $T = 0$, the trailing edge of the pump.  This is roughly consistent with the data, although there is an overall offset in the thresholds.  The pulse width and photon number also roughly match.  

However, these plots show that the simulton picture is only valid for a limited range of $\ell$.  If $\ell$ is too large, additional effects destabilize the sech-pulse.  Thus, the pulse amplitude $\langle a(t) \rangle / \langle |a(t)| \rangle$, which is constant in the simulton picture, oscillates.  These amplitude fluctuations cause smaller oscillations in the photon number, centroid and pump width.

\section{Box Pulse Theory}
\label{sec:10-boxpulse}

Well above threshold, both the eigenmode and simulton theories fail.  An eigenmode expansion becomes impractical because too many modes must be used and the computation time scales as $O(N^4)$.  Simulton theory fails because in this regime the pulses are no longer sech-shaped.  We need a new theory that predicts the pulse shapes in this regime.

Simulations show that pulses get longer the further one goes above threshold (Figs.~\ref{fig:10-f5}, \ref{fig:10-f12}, \ref{fig:10-f13}, \ref{fig:10-f16}).  This is a result of the pulse filling the leading side of the positive-gain region $\Delta_{\rm max}\Gamma(t) > 0$ (Sec.~\ref{sec:10-dispersionless}).  Long pulses mean narrow spectra and weak dispersion effects.  The result is a competition between gain and pump depletion, with dispersion playing only a secondary role.

In this section, we ignore dispersion and derive an analytic formula for the pulse shape that is reasonably accurate in this regime.  Dispersion will be treated later, but its main effect will be to add a modulation on the pulse shape when $\phi_0\phi'_2 < 0$, giving rise to a nondegenerate box-like pulse.

\subsection{Degenerate Case $\phi_0 = 0$}
\label{sec:clipping}

First, let's treat the center of the detuning peak $\phi_0 = 0$.  Later on we will treat the general case, but the results are simplest for $\phi_0 = 0$.  Recalling (\ref{eq:10-intdiff}), we drop dispersion terms and invoke the gain-without-distortion ansatz to obtain:
\begin{align}
    & \frac{\partial a(z,t)}{\partial z} = -\frac{1}{2} \alpha_a a(z, t) + \epsilon\,a^*(z,t) b_{\rm in}(t - u z)e^{-\alpha_b z/2} \nonumber \\
    & \quad - \frac{\epsilon^2}{2u} a^*(z,t)\int_{-\infty}^{t}{e^{(g+\alpha_b/2)(t'-t)/u} a(z, t')^2 \d t'} \label{eq:10-intdiff3}
\end{align}
Here $g = \tfrac{1}{L}\log(G_0)$ is the gain per unit length at steady state.  Now make the substitution
\beq
	a(z, t) = e^{gz/2} \bar{a}(z, t)
\eeq
where $\bar{a}(z, t)$ is real and slowly-varying in $z$.  This is valid for flat-top pump pulses, where the gain is roughly constant because the pulse amplitude is constant.  We choose $g$ so that $e^{gL/2}$ is the cavity loss, since in steady state, gain equals loss and thus the single-pass gain should be $e^{gL/2}$.  Deviations will be handled by perturbation theory on $\bar{a}$.  Equation (\ref{eq:10-intdiff3}) becomes:
\begin{align}
    &\frac{\partial\bar{a}(z,t)}{\partial z} = \left[\epsilon\,(b_{\rm in}(t-uz)e^{-\alpha_b z/2} - \bar{b}_0)\right]\bar{a}(z,t) \nonumber \\
    &\qquad - \frac{\epsilon^2 e^{gz}}{2u}\bar{a}(z,t)\int_{-\infty}^t{e^{(g+\alpha_b/2)(t'-t)/u} \bar{a}(z,t')^2\d t} \label{eq:10-nodist}
\end{align}

To obtain the output field, one must integrate (\ref{eq:10-nodist}) from $z=0$ to $L$.  Gain without distortion means that the integrand is close to constant over that interval, so we can approximate the integral with one Picard step, setting $z=0$ everywhere in the integrand.  The evolution over one round-trip is:
\begin{align}
	& \Delta a(t) = a(t)\biggl[\overbrace{\int_0^L{\epsilon(b_{\rm in}(t-uz) e^{-\alpha_b z/2} - \bar{b}_0)\d z}}^{F(t)} \nonumber \\
	& \qquad - \frac{\epsilon^2(e^{gL}-1)}{2gu} \int_{-\infty}^t {e^{(g+\alpha_b/2)(t'-t)/u} a(t')^2 \d t'}\biggr] \label{eq:10-gclip-inout}
\end{align}

\begin{figure}[tbp]
\begin{center}
% From SimultonTheory-Rig19.pdf
\includegraphics[width=1.0\columnwidth]{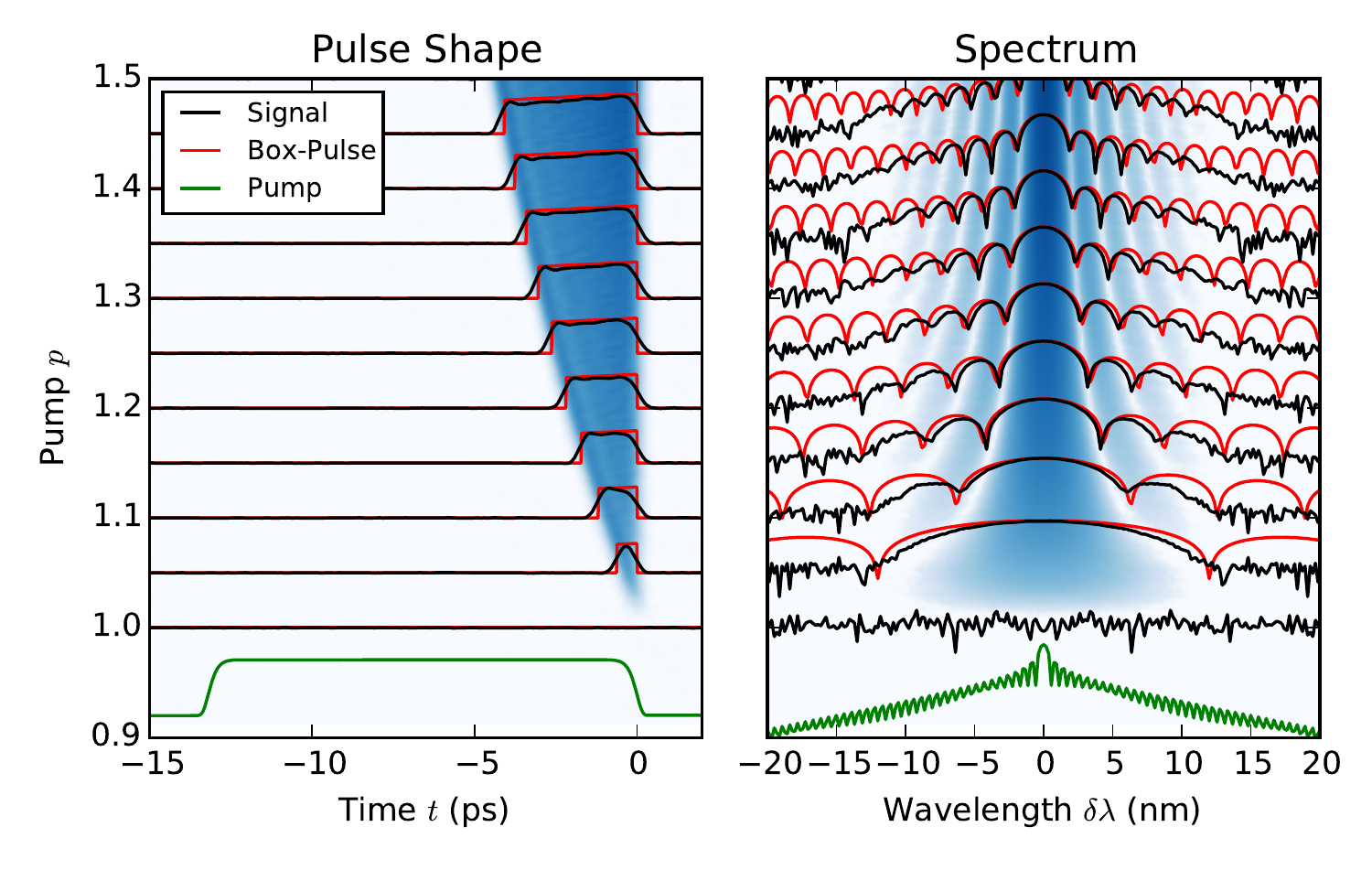}
\caption{Left: plot of the pulse shape for the signal and its box-pulse approximation via Eq.~(\ref{eq:10-box}), for $\phi_0 = 0$, $p \in [1.0, 1.5]$.  Pump is shown as the bottom trace, in green.  Right: power spectrum for the same data, on a log scale.}
\label{fig:10-f19}
\end{center}
\end{figure}

There are two linear terms in (\ref{eq:10-gclip-inout}).  The first is the gain-clipping term, where $F(t)$ is related to $G(t)$ by:

\bea
	F(t) & = & G(t) + \int_0^L {(b_{\rm max} e^{-\alpha_b z/2} - \bar{b}_0)\d z} \nonumber \\
	& = & G(t) + \frac{p-1}{2} \log(G_0 e^{\alpha_a L}) \nonumber \\
	& = & G(t) + \log \bigl[\Delta_{\rm max}(\phi_0 = 0)\bigr] \label{eq:10-boxft}
\eea

In steady state, $a(t)$ stays constant between round trips, so the right-hand side of (\ref{eq:10-gclip-inout}) must equal zero.  There are two ways this can happen:

\begin{enumerate}
\item $F(t) < 0$ or $F(t)$ decreasing.  Since the second integral is always positive and increasing, it is impossible to set the term in square brackets in (\ref{eq:10-gclip-inout}) to zero.  The only way to satisfy the steady-state condition is to set $a(t) = 0$.
\item $F(t) > 0$ and increasing.  In this case, $a(t) \neq 0$ and the terms in the square brackets must cancel out.  Combining (\ref{eq:10-gclip-inout}) with its time derivative (both which must equal zero), we find:
\beq
	a(t)^2 = \frac{2gu}{\epsilon^2(e^{gL}-1)} \left[F'(t) - \frac{g+\alpha_b/2}{u} F(t)\right] \label{eq:10-asq-box}
\eeq
\end{enumerate}

For a flat-top pump pulse, the analytic formula for $G(t)$ (Eq.~\ref{eq:10-gain-gc0}) will suffice; from this we can calculate $F(t) = \tfrac{1}{2}\log(G_0 e^{\alpha_a L}) \bigl[(p-1) - p|t|/T_p\bigr]$.  Using (\ref{eq:10-asq-box}) and substituting $g, \epsilon, u, b_0$ for $G_0, T_p, N_{b,0}$ (Table~\ref{tab:10-t1}) we find the solution
\begin{align}
	& a(t)^2 = \frac{4N_{b,0} e^{-\alpha_b L/2}\log(G_0)}{T_p (G_0 - 1) \log(G_0 e^{\alpha_a L})} \nonumber \\
	& \qquad \times \left[p + \left(\log G_0 + \tfrac{1}{2}\alpha_b L\right) \left((p-1) - p\frac{|t|}{T_p}\right)\right] \label{eq:10-box}
\end{align}
for $-T_p(1-p^{-1}) < t < 0$ (and $a(t) = 0$ otherwise).  This can be integrated to give the total photon number:
\begin{align}
	& N_a = \frac{4N_{b,0} e^{-\alpha_b L/2}\log(G_0)}{T_p (G_0 - 1) \log(G_0 e^{\alpha_a L})} \nonumber \\
	& \qquad \times \left[(p-1) + \frac{(p-1)^2}{2p} \left(\log G_0 + \tfrac{1}{2}\alpha_b L\right)\right]
\end{align}

Figure \ref{fig:10-f19} compares the waveform (\ref{eq:10-box}) and its Fourier transform to full simulations.  The amplitude and the general shape are modeled well by the theory, although it says nothing about the shape of the edges.  As the pulse gets longer with increasing pump power, the spectrum narrows, a fact confirmed in experiments and consistent with previous work \cite{Becker1974}.

\begin{figure}[tbp]
\begin{center}
\includegraphics[width=1.0\columnwidth]{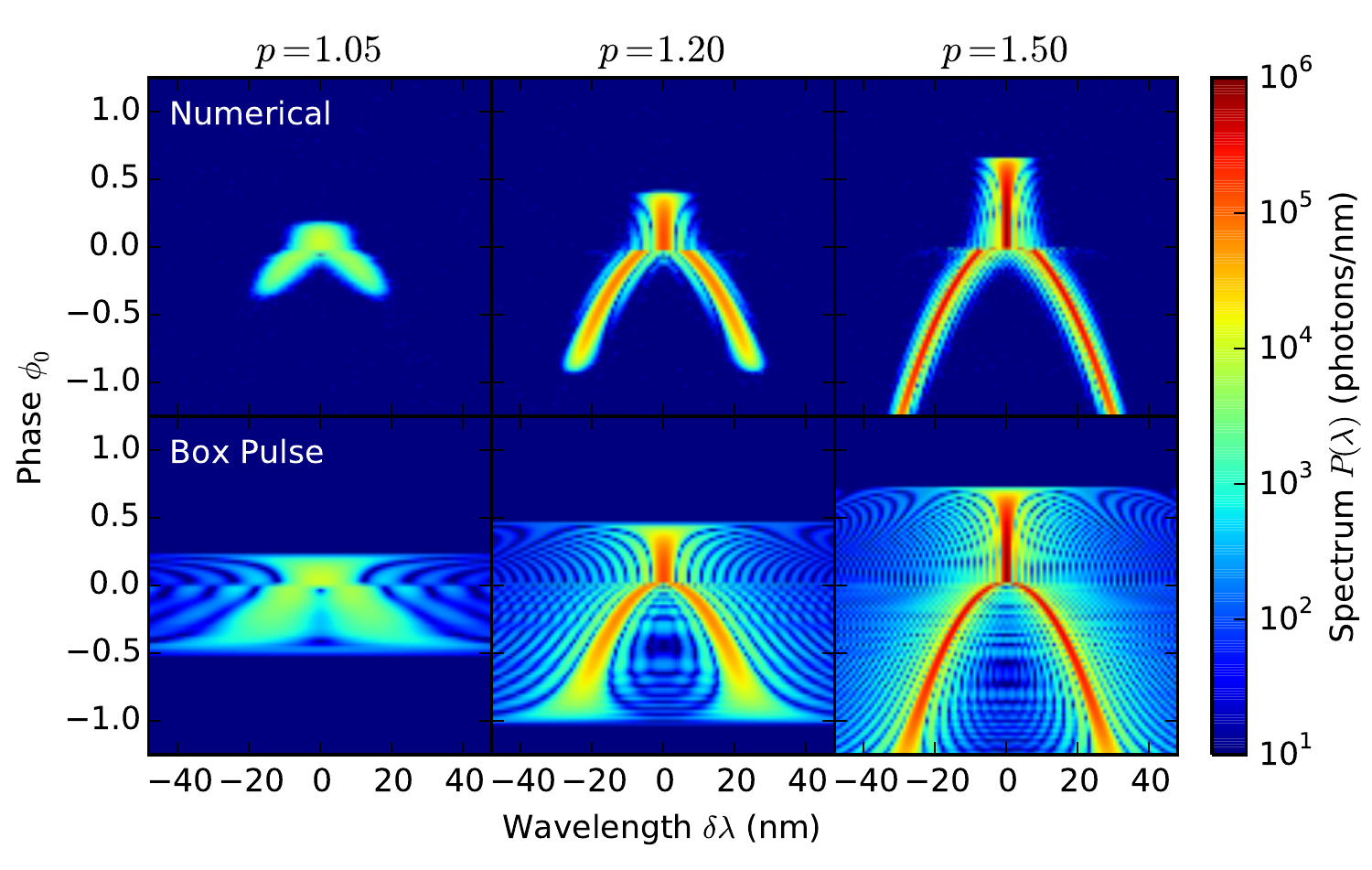}
\caption{Resonance diagrams for the box pulse model (Eq.~\ref{eq:10-box2}) compared to numerical result.}
\label{fig:10-f23}
\end{center}
\end{figure}

\subsection{Dispersion}

Gain clipping sets the overall pulse shape, while dispersion evens out the edges and sets the signal-idler splitting.  If $\phi_0 \phi'_2 > 0$, the OPO is degenerate so there is no signal-idler splitting; however, nonzero $\phi_0$ reduces the overall gain, which reduces the signal power.  The most straightforward way to do this is to say that Eq.~(\ref{eq:10-boxft}) should be modified to read
\beq
	F(t) = G(t) + \log\bigl[\Delta_{\rm max}(\phi_0)\bigr]
\eeq
and the rest of the results carry over unchanged.  Eq.~(\ref{eq:10-box}) becomes:
\begin{align}
	& a(t)^2 = \frac{4N_{b,0} e^{-\alpha_b L/2}\log(G_0)}{T_p (G_0 - 1) \log(G_0 e^{\alpha_a L})} \biggl[p + \left(\log G_0 + \tfrac{1}{2}\alpha_b L\right) \nonumber \\
	& \qquad\qquad \times \left(\frac{2\log(\Delta_{\rm max}(p,\phi_0))}{\log(G_0 e^{\alpha_a L})} - p\frac{|t|}{T_p}\right)\biggr] \label{eq:10-box2}
\end{align}

For $\phi_0 \phi'_2 < 0$, the pulse is box-shaped but nondegenerate: $a(t) = \mbox{Re}\bigl[\bar{a}(t) e^{-i\,\delta\omega_0 t}\bigr]$, (see Eq.~\ref{eq:10-abar}), and $\bar{a}(t)$ takes the same form as (\ref{eq:10-box2}) but with a $\sqrt{2}$ factor to preserve the overall energy.  

A good way to visualize (\ref{eq:10-box2}) is to plot resonance diagrams for the box-pulse model and compare them to the numerics, as in Fig.~\ref{fig:10-f23}.  The general structure of the resonance plots are the same, but the features on the tails differ, consistent with the smoothing in Fig.~\ref{fig:10-f19}.  However, these tails are suppressed by several orders of magnitude and only show up on the plot because of the log scale.

\begin{figure}[tbp]
\begin{center}
\includegraphics[width=1.0\columnwidth]{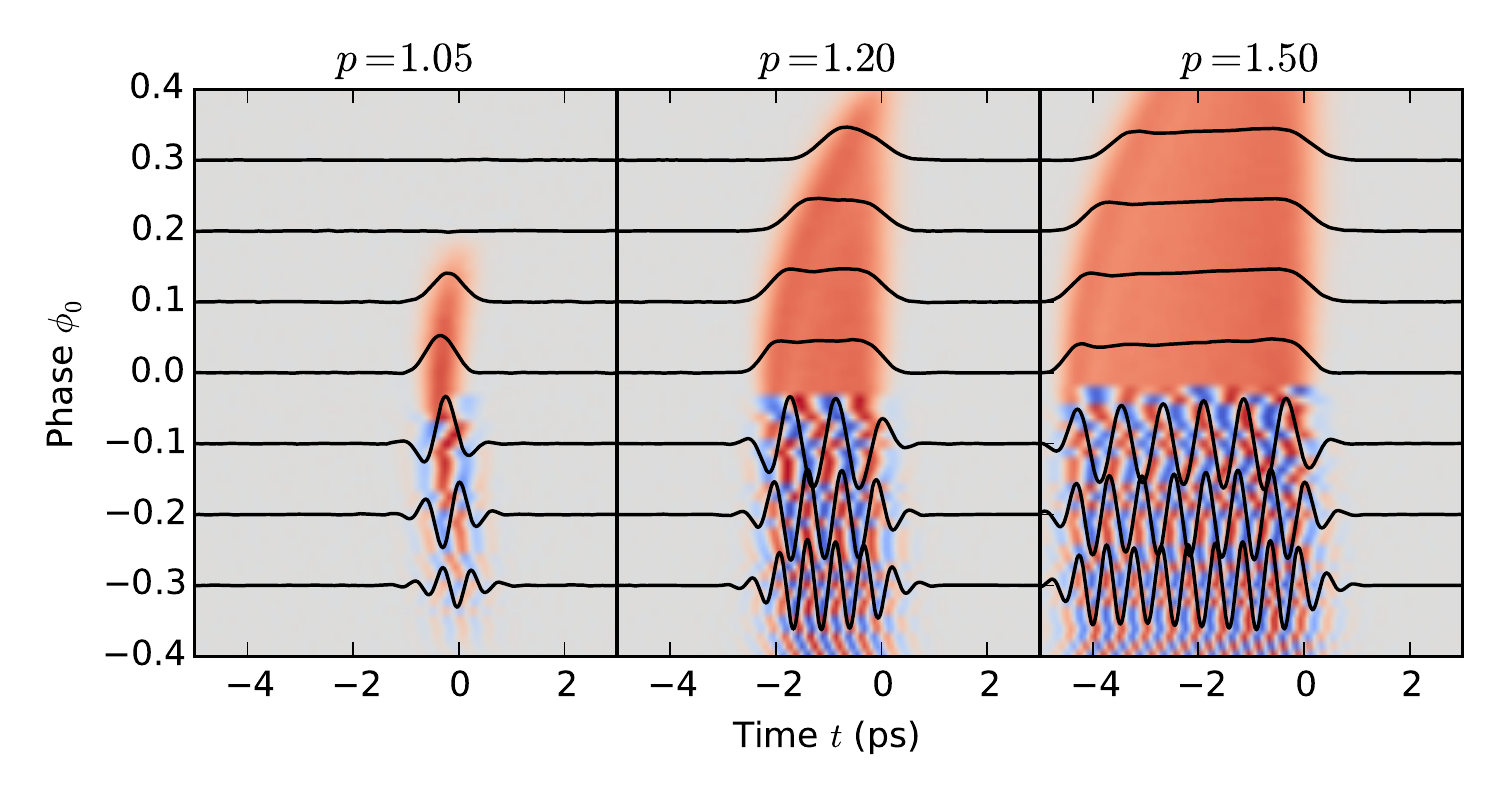}
\caption{Time-domain pulse shapes as function of phase and power, numerical.}
\label{fig:10-f21}
\end{center}
\end{figure}

The Fourier transform of this waveform is given in Fig.~\ref{fig:10-f21}.  The OPO is nondegenerate for $\phi_0 < 0$, but nondegeneracy does not affect the overall shape of the pulse.  Aside from a sinusoidal modulation, the pulse remains box-shaped.

\section{Conclusion}

This paper has introduced three reduced models that aid the understanding, simulation, and design of synchronously pumped OPOs.  These models are based on mathematical approximations and physical intuition, and show good agreement with numerical simulations for predicting steady-state pulse shapes, transient behavior and stability.  Because the models run several orders of magnitude faster than numerical simulations, they will be a useful tool for simulating large OPO networks, and a guide for device design and optimization.

Near threshold, we derived an eigenmode expansion that predicts the OPO threshold as a function of cavity dispersion and round-trip phase, and gives the correct steady-state pulse shape.  The pulse dynamics arise from competition between gain clipping, which shortens the pulse to maximize its overlap with the pump; and dispersion, which limits its bandwidth.  We noticed a smooth transition between degenerate and nondegenerate oscillation when the cavity dispersion is not compensated, which could be explained by a simple phase-matching argument.  In both the degenerate and nondegenerate regimes, we obtained analytic formulae for the pulse shape in terms of Airy and hypergeometric functions, which gave analytic expressions for the pulse shape and its threshold.  Moreover, pulse stability could be explained using bifurcation theory with a simple two-mode model.

Far from threshold, the steady-state pulse was found to have a narrow spectrum, and we obtained a box-like pulse shape by solving the equations without dispersion.  In the frequency domain, this ``box pulse'' appears as a sinc-shaped spectrum which grows narrower the higher the pump relative to threshold.  An analytic expression for the pulse width and amplitude was derived, which agrees with the numerics.

Working between these regimes, we obtained a reduced model based on projection onto a sech-shaped pulse.  This was physically motivated by the ``simulton'' solution in a $\chi^{(2)}$ waveguide, and we accounted for the effects of gain-clipping and dispersion as perturbations to this solution.  While only valid in the degenerate regime close to threshold, this model is helpful because it is fully analytic, and within its regime of validity, agrees with the both the eigenmode model and the numerics.

In future work, we hope to extend this analysis to systems in the ultrafast, dispersion-engineered limit, where second- and third-order dispersion are equally important in shaping the signal pulses.  In high-power systems, spatial effects will also play an important role \cite{Jankowski2016}.  In the nondegenerate regime, the eigenmode model in Sec.~\ref{sec:10-emshapes} could be used to quantify the offset between signal- and idler-comb carrier envelope frequencies, which can has been observed as an RF beatnote of the output power \cite{Wolf2013}.

Multi-OPO systems are another avenue for future study.  Studies of OPO-based Ising machines have shown convergence to the ground-state for small problems with very high probability \cite{Marandi2014, KentaThesis}, suggesting that a single-mode model may not be accurate to describe their dynamics \cite{Takata2016}.  The theory in Sec.~\ref{sec:10-linear}-\ref{sec:10-nonlinear} could easily be extended to OPO networks.  Beyond solving Ising problems, it is likely that such ``multimode'' OPO networks will have richer nonlinear dynamics, and may thus be a more useful resource for neuromorphic computing and machine learning \cite{Izhikevich2000}.

\begin{acknowledgements}

The authors would like to thank Chris Phillips for helpful discussions.  R.H.\ is funded by a seed grant from the Precourt Institute for Energy at Stanford University.  This research is supported by the DARPA DODOS and MTO programs, and the ImPACT Program of the Council of Science, Technology and Innovation (Cabinet Office, Government of Japan).

\end{acknowledgements}


\begin{thebibliography}{99}

\bibitem{Marandi2016}{A.~Marandi, K.~A.~Ingold, M.~Jankowski, and R.~L.~Byer, Optica {\bf 3}, 324 (2016).}
\bibitem{Parker2012}{F.~Parker, {\it Applications of infrared spectroscopy in biochemistry, biology, and medicine} (Springer Science \& Business Media, 2012).}
\bibitem{Popmintchev2012}{T.~Popmintchev, M.-C.~Chen, D.~Popmintchev, P.~Arpin, S.~Brown, S.~Ali{\v{s}}auskas, G.~Andriukaitis, T.~Bal{\v{c}}iunas, O.~D.~M{\"u}cke, A.~Pugzlys, et al. Science {\bf 336}, 1287 (2012).}
\bibitem{Peralta2013}{E.~Peralta, K.~Soong, R.~England, E.~Colby, Z.~Wu, B.~Montazeri, C.~McGuinness, J.~McNeur, K.~Leedle, D.~Walz, et al., Nature {\bf 503}, 91 (2013).}
\bibitem{Wang2013}{Z.~Wang, A.~Marandi, K.~Wen, R.~L.~Byer, and Y.~Yamamoto, Phys.\ Rev.\ A {\bf 88}, 063853 (2013).}
\bibitem{Marandi2014}{A.~Marandi, Z.~Wang, K.~Takata, R.~L.~Byer, and Y.~Yamamoto, Nature Photonics, {\bf 8}, 937 (2014).}
\bibitem{Tezak2015}{N.~Tezak and H.~Mabuchi, EPJ Quantum Technology {\bf 2}, 1 (2015).}
\bibitem{Jackel1991}{J.~Jackel and J.~Johnson, Electron.\ Lett.\ {\bf 27}, 1360 (1991).}
\bibitem{Korkishko1998}{Y.~N.~Korkishko, V.~Fedorov, T.~Morozova, F.~Caccavale, F.~Gonella, and F.~Segato, JOSA A {\bf 15}, 1838 (1998).}
\bibitem{Iwai2003}{M.~Iwai, T.~Yoshino, S.~Yamaguchi, M.~Imaeda, N.~Pavel, I.~Shoji, and T.~Taira, Appl.\ Phys.\ Lett.\ {\bf 83}, 3659 (2003).}
\bibitem{Roussev2004}{R.~V.~Roussev, C.~Langrock, J.~R.~Kurz, and M.~Fejer, Opt.\ Lett.\ {\bf 29}, 1518 (2004).}
\bibitem{Chang2016}{L.~Chang, Y.~Li, N.~Volet, L.~Wang, J.~Peters, and J.~E.~Bowers, Optica {\bf 3}, 531 (2016).}
\bibitem{Poberaj2012}{G.~Poberaj, H.~Hu, W.~Sohler, and P.~Guenter, Laser \& Photonics Reviews {\bf 6}, 488 (2012).}
\bibitem{Rabiei2014}{P.~Rabiei, J.~Ma, J.~Chiles, S.~Khan, and S.~Fathpour, in {\it 2014 IEEE Photonics Conference} (2014).}
\bibitem{Guarino2007}{A.~Guarino, G.~Poberaj, D.~Rezzonico, R.~DeglÕInnocenti, and P.~Gu\"{u}nter, Nature Photonics {\bf 1}, 407 (2007).}
\bibitem{Lin2015}{J.~Lin, Y.~Xu, Z.~Fang, M.~Wang, N.~Wang, L.~Qiao, W.~Fang, and Y.~Cheng, Science China Physics, Mechanics \& Astronomy {\bf 58}, 1 (2015).}
\bibitem{VanDriel1995}{H.~Van Driel, Applied Physics B 60, 411 (1995).}
\bibitem{Khaydarov1994}{J.~D.~Khaydarov, J.~H.~Andrews, and K.~D.~Singer, Opt.\ Lett.\ {\bf 19}, 831 (1994).}
\bibitem{Marandi2015}{A.~Marandi, C.~Langrock, M.~M.~Fejer, and R.~L.~Byer, in {\it Nonlinear Optics} (Optical Society of America, 2015), pp.~NM1AÐ2.}
\bibitem{Roslund2014}{J.~Roslund, R.~M.~De Araujo, S.~Jiang, C.~Fabre, and N.~Treps, Nature Photonics {\bf 8}, 109 (2014).}
\bibitem{Yokoyama2013}{S.~Yokoyama, R.~Ukai, S.~C.~Armstrong, C.~Sornphiphat- phong, T.~Kaji, S.~Suzuki, J.-i.~Yoshikawa, H.~Yonezawa, N.~C.~Menicucci, and A.~Furusawa, Nature Photonics {\bf 7}, 982 (2013).}
\bibitem{KentaThesis}{K.~Takata, Ph.D.~thesis, University of Tokyo (2014).}
\bibitem{Ikeda1979}{K.~Ikeda, Opt.\ Commun.\ {\bf 30}, 257 (1979).}
\bibitem{Takata2016}{K.~Takata, A.~Marandi, R.~Hamerly, Y.~Haribara, D.~Maruo, S.~Tamate, H.~Sakaguchi, S.~Utsunomiya, and Y.~Yamamoto, Scientific Reports {\bf 8}, 34082 (2016).}
\bibitem{Inagaki2016}{T.~Inagaki, K.~Inaba, R.~Hamerly, K.~Inoue, Y.~Yamamoto, and H.~Takesue, Nature Photonics {\bf 10}, 415-419 (2016).}
\bibitem{Hamerly2016-2}{R.~Hamerly, K.~Inaba, T.~Inagaki, H.~Takesue, Y.~Yamamoto, and H.~Mabuchi, arXiv:1605.08121 (2016)}
\bibitem{Kinsler1991}{P.~Kinsler and P.~D.~Drummond, Phys.\ Rev.\ A {\bf 43}, 6194 (1991).}
\bibitem{Akhmanov1968}{S.~Akhmanov, A.~Chirkin, K.~Drabovich, A.~Kovrigin, R.~Khokhlov, and A.~Sukhorukov, IEEE J.\ Quantum Electronics, {\bf 4}, 598 (1968).}
\bibitem{Trillo1996}{S.~Trillo, Opt.\ Lett.\ {\bf 21}, 1111 (1996).}
\bibitem{Jankowski2016}{M.~Jankowski, A.~Marandi, K.~Ingold, R.~Hamerly, et al., in preparation.}
\bibitem{Becker1974}{M.~Becker, D.~Kuizenga, D.~Phillion, and A.~Siegman, J.\ Appl.\ Phys.\ {\bf 45}, 3996 (1974).}
\bibitem{McMahon2016}{P.~McMahon, A.~Marandi, et al., in preparation.}
\bibitem{BoydBook}{R.~W.~Boyd, Nonlinear Optics (Academic press, 2003).}
\bibitem{AgrawalBook}{G.~P.~Agrawal, {\it Nonlinear Fiber Optics} (Academic press, 2007).}
\bibitem{Cheung1990}{E.~Cheung and J.~Liu, JOSA B {\bf 7}, 1385 (1990).}
\bibitem{Patera2010}{G.~Patera, N.~Treps, C.~Fabre, and G.~J.~de Valcarcel, EPJ D {\bf 56}, 123 (2010).}
\bibitem{DeValcarcel2006}{G.~J.~de Valcarcel, G.~Patera, N.~Treps, and C.~Fabre, Phys.\ Rev.\ A {\bf 74}, 061801 (2006).}
\bibitem{Raymer1991}{M.~G.~Raymer, P.~Drummond, and S.~Carter, Opt.\ Lett.\ {\bf 16}, 1189 (1991).}
\bibitem{Werner1997}{M.~Werner and P.~Drummond, J.\ Comput.\ Phys.\ {\bf 132}, 312 (1997).}
\bibitem{Drummond2001}{P.~Drummond and J.~F.~Corney, JOSA B {\bf 18}, 139 (2001).}
\bibitem{Phillips2011}{C.~Phillips, C.~Langrock, J.~Pelc, M.~Fejer, I.~Hartl, and M.~E.~Fermann, Optics Express {\bf 19}, 18754 (2011).}
\bibitem{PhillipsThesis}{C.~R.~Phillips, Ph.D.~thesis, Stanford University (2012).}
\bibitem{CUDAGuide}{Nvidia Corporation, ``Cuda C Programming Guide'', http://docs.nvidia.com/cuda/cuda-c-programming-guide/, accessed: May 2016-17-05.}
\bibitem{Moreland2003}{K.~Moreland and E.~Angel, in Proceedings of the {\it ACM SIGGRAPH/EUROGRAPHICS conference on Graphics hardware} (Eurographics Association, 2003), pp.~112Ð119.}
\bibitem{Sreehari2012}{A. Sreehari, ``Implementations of the FFT algorithm on GPU'', Masters Thesis, Link\"{o}pings Universitet (2012)}
\bibitem{Parameswaran2002}{K.~R.~Parameswaran, R.~K.~Route, J.~R.~Kurz, R.~V.~Roussev, M.~M.~Fejer, and M.~Fujimura, Opt.\ Lett.\ {\bf 27}, 179 (2002).}
\bibitem{Langrock2007}{C.~Langrock and M.~Fejer, Opt.\ Lett.\ {\bf 32}, 2263 (2007).}
\bibitem{Kuizenga1970}{D.~J.~Kuizenga and A.~Siegman, IEEE J.\ Quantum Electron.\ {\bf 6}, 694 (1970).}
\bibitem{Siegman1970}{A.~Siegman and D.~J.~Kuizenga, IEEE J.\ Quantum Electron.\ {\bf 6}, 803 (1970).}
\bibitem{Haus2000}{H.~A.~Haus, Selected Topics in IEEE J.\ Selected Topics in Quantum Electronics {\bf 6}, 1173 (2000).}
\bibitem{Khaydarov1995}{J.~D.~Khaydarov, J.~H.~Andrews, and K.~D.~Singer, JOSA B {\bf 12}, 2199 (1995).}
\bibitem{StrogatzBook}{S.~H.~Strogatz, {\it Nonlinear dynamics and chaos: with applications to physics, biology, chemistry, and engineering} (Westview press, 2014).}
\bibitem{Lugiato1988}{L.~A.~Lugiato, C.~Oldano, C.~Fabre, E.~Giacobino, and R.~J.~Horowicz, Nuovo Cimento {\bf 18}, 959 (1988)}
\bibitem{Jiang2012}{S.~Jiang, N.~Treps and C.~Fabre, New J.\ Phys.\ {\bf 14}, 043006 (2012).}
\bibitem{Wolf2013}{S.~J.~Wolf, C.~R.~Phillips, A.~Marandi, K.~L.~Vodopyanov, M.~M.~Fejer, and R.~L.~Byer, in {\it CLEO: Science and Innovations} (Optical Society of America, 2013), pp.CW1B}
\bibitem{Mabuchi2008b}{H.~Mabuchi, Phys.\ Rev.\ A {\bf 78}, 015801 (2008).}
\bibitem{VanHandel2005}{R.~Van Handel and H.~Mabuchi, Journal of Optics B: Quantum and Semiclassical Optics {\bf 7}, S226 (2005).}
\bibitem{Izhikevich2000}{E.~M.~Izhikevich, Int.\ J.\ Bifuraction and Chaos {\bf 10}, 1171 (2000)}

\end{thebibliography}
\end{document}